\documentclass[a4paper, 11pt]{article}
\usepackage{jheppub}
\usepackage{comment,subfigure,hyperref}
\usepackage{float}
\hypersetup{
colorlinks = true,
linkcolor=blue,
anchorcolor=black,
citecolor=black,
filecolor=cyan,
menucolor=black,
runcolor=cyan,
urlcolor=blue
}
\usepackage{romannum}
\usepackage{pst-node}
\usepackage{epsfig}

\title{Phases of a 10-D Holographic hard wall model}
\author[1]{Akash Singh}
\author[2]{K. P. Yogendran}
\affiliation{Department of Physics,\\
IISER Mohali\\Sector 81, Knowledge City\\Punjab, 140306, India}
\emailAdd{akashsingh@iisermohali.ac.in}
\emailAdd{yogendran@iisermohali.ac.in}

\abstract{
In this article, we study the finite temperature properties of a 10-D version of a hardwall model for QCD. Introducing fundamental matter via probe D7-branes and separate cutoffs $r_m$ and $r_g$ for the branes and the bulk, we present a detailed exploration of the phases for varying temperature and quark mass. Finite thermodynamic quantities are calculated using the procedure of holographic renormalization and used to characterize the phases. Finally, by fitting glueball and vector meson masses, we show how a unique phase diagram can be isolated.
}
\arxivnumber{}

\def\be{\begin{equation}}
\def\ee{\end{equation}}
\def\barr{\begin{array}{lr}}
\def\earr{\end{array}}
\def\bea{\begin{eqnarray}}
\def\eea{\end{eqnarray}}

\def\nn{\nonumber}

\def\del{\partial}

\def\o{{\cal O}}

\def\a{\alpha}
\def\b{\beta}

\def\e{\epsilon}

\def\f{\phi}

\def\g{\gamma}
\def\h{\eta}

\def\k{\kappa}
\def\l{\lambda}
\def\m{\mu}
\def\n{\nu}
\def\o{\omega}
\def\p{\pi}
\def\q{\theta}
\def\r{\rho}
\def\s{\sigma}

\def\x{\xi}

\def\L{\Lambda}
\def\O{\Omega}

\definecolor{maroon}{rgb}{0.5, 0.0, 0.0}	
\definecolor{arsenic}{rgb}{0.23, 0.27, 0.29}

\begin{document}
\pagenumbering{arabic}
\maketitle
\flushbottom
\section{Introduction}
The AdS/CFT correspondence or Gauge/Gravity duality \cite{Maldacena:1997re} sets up a radical equality \cite{Gubser:1998bc, Witten:1998qj} between quantum field theories and string theory in one higher dimension (some relevant reviews include \cite{Aharony:1999ti, Natsuume:2014sfa, Ammon:2015wua}). 
In particular, since it is a {\it duality}, we can obtain valuable insight into the strong coupling regime on the one side from the other. Much effort has gone into exploring whether such a duality can be useful to understand the phases and properties of QCD \cite{Gubser},  which is a strongly interacting field theory at low energies. Calculations from the gravity side seem to generically capture phenomenologically relevant features such as Regge behavior of Hadron spectra \cite{Afonin:2021cwo}, low shear viscosity \cite{kss, Baggioli:2021tzr} and consequent elliptic flow \cite{Noronha:2010zc} which are hard to obtain from first principles QCD. There is also a large body of work on the properties of strongly interacting condensed matter \cite{McGreevy:2016myw,Blake:2022uyo} using holographic methods of which QCD at finite density is but a special case. Openings into the large literature on QCD related explorations of the holographic correspondence can be found in the reviews/theses \cite{Gubser},\cite{Remes:2020hwo},\cite{Rebhan:2014rxa},\cite{Jarvinen:2021jbd},\cite{Evans:2021lrg}. A useful resource with many references is the \href{https://ncatlab.org/nlab/show/AdS-QCD+correspondence}{webpage at nLab}.

The major research directions in AdS/QCD can be divided into the following categories: a bottom-up approach based on phenomenology, a top-down approach based on stringy considerations and a first principles approach based on dualities which depend critically on supersymmetry.

The latter approaches, especially those in which both sides of the duality are clearly identified, have shown that many of the properties of strongly interacting gauge theories are indeed recoverable from holographic calculations such as confinement and condensates \cite{PolStrass}), hard scattering \cite{PolStrass2} and partonic substructure \cite{Bianchi:2021sug}, Regge features \cite{CarvalhoAmorimdeSousa:2021kxz} etc. 
including recent efforts \cite{Bena} to obtain a finite temperature phase diagram for this system. 

The former directions are more phenomenologically oriented and incorporate various heuristic features that seem to be implicated in obtaining the relevant features of 4D QCD like theories. For instance, inclusion of black holes to model finite temperature \cite{Witten:1998qj}, branes with various couplings to introduce fundamental matter \cite{Karch-Katz}, chiral symmetry \cite{ekss, DaRold:2005mxj}, radially varying dilaton \cite{KKSS, Gubser:2008ny} to capture running coupling and linear confinement,  etc.

Among the top-down approaches, a well developed set of studies are the Witten-Sakai-Sugimoto models \cite{Sakai:2004cn}.  This approach (reviewed in \cite{Rebhan:2014rxa}) works in the ten dimensional spacetime produced by D4-branes with probe D8 branes representing massless chiral quark degrees of freedom (masses can be introduced through \cite{Dhar:2007bz}, \cite{McNees:2008km}, \cite{Kovensky:2019bih}). For another set of studies approaching QCD, starting with the Klebanov-Strassler duals \cite{KS} and introducing branes, see \cite{Yadav:2020pmk}. On the other hand, the V-QCD model \cite{Jarvinen:2011qe} (reviewed in the thesis \cite{Remes:2020hwo}) operates in five spacetime dimensions, incorporates running coupling, space filling branes representing chiral quarks and a host of other features. In this theory, multi-parameter potentials are tuned to capture QCD physics and have been considerably developed and applied to model the interiors of neutron stars \cite{Jarvinen:2021jbd,Hoyos:2016zke}.  The above models fall into the soft-wall \cite{Herzog} class of models \cite{Kajantie:2006hv}-\cite{Ballon-Bayona:2020qpq}. 
In contrast, the hardwall models approximate QCD by a segment of AdS-type spaces with parameters, boundary conditions and other features chosen to best fit low energy phenomenology \cite{Herzog}. Some further explorations of the hardwall models can be found in \cite{Domokos:2012da}-\cite{Rebhan}. 
Nevertheless, it is perhaps fair to say that a fully consistent, predictive model is very much work in progress.

Even this very brief introduction gives a glimpse of the richness of the holographic correspondence and the optimism and excitement that underlies these studies. A fundamental understanding of holography can consolidate these studies and perhaps even make contact with QCD proper.

An important shortcoming of nearly all these approaches is a lack of methods to estimate the reliability of the predictions from these models - at least partly because both sides of the duality are not precisely known. These issues are generic to all 5D models. In the Witten-Sakai-Sugimoto models \cite{McNees:2008km} on the other hand, the high temperature phase is unreliable because for $T>\L_{QCD}\sim M_{KK}$, Kaluza-Klein modes must be taken into account. Further, the naive idea for quark masses gives $m_q\approx M_{KK}.$

A second reason is that, in many cases, for phenomenologically interesting values of the parameters, the bulk calculations can be expected to receive significant corrections. There are many sources for these corrections: higher derivative corrections, string loop effects, additional fields including a variety of scalar fields (moduli). 
Several of these have been explored in the literature giving rise to a rich variety of phenomena. This causes difficulty in identifying the precise dual of a QCD-like theory (with fundamental quarks, chiral symmetry breaking, etc). Hence it is important to have control systems to understand the limits of the present calculations.

A general issue at finite chemical potential is sources of Baryon number which are  non-Abelian configurations of D-branes. This is because while the non-Abelian DBI action is accurate only up to $O(\a'^4 F^2)$, these configurations involve fields for which higher order terms are as significant as the ones being retained. It is difficult to take into account the masses and polarization effects of these D-branes  (\cite{Kovensky:2021ddl} is a recent effort in this direction).

In the present work, we add to this exploration by studying a {\it hardwall} model that is ten dimensional, includes probe D7-branes \cite{Karch-Katz}, and takes into account the effect of the shape of the brane on hadron spectra \cite{Babington} as detailed in the following sections. The key difference in our work is  using separate hard cutoffs for the bulk and the branes, which leads to explicit breaking of scale invariance in the bulk. This idea, albeit in 5D has been explored recently by \cite{Rebhan}. Introducing quarks via D-branes can lead to large deformations of both the bulk geometry and of the D-branes, especially in regions where the branes change their shape significantly. The second cutoff for the D-brane allows us to excise these regions, thereby maintaining control over bulk calculations. However, we also take into account the effect of the excluded regions via modified boundary conditions {\it dictated by phenomenology} much in the way renormalization conditions fix parameters in terms of experimental data in the usual QFT. In particular, this could enable us to handle non-Abelian deformations of the DBI action by hiding them behind the IR-cutoff but nevertheless including their effects via boundary conditions.

A second advantage of working in 10-D with D7-branes is that the precise gauge theory is known and controlled. 
Further, several deformations of this parent theory have been studied - most notably the Polchinski-Strassler and Klebanov-Strassler \cite{KS} families.  This allows us to test these computations against field theory calculations, including those obtained from lattice gauge theory approaches \cite{Schaich:2022xgy} when those become available. 
On the gravity side, fully backreacted geometries have been proposed \cite{PolGrana}, including at finite temperature \cite{Bena}. Thus, corrections to the predictions obtained from the gravitational side are controllable and can be compared with predictions for the phases of the dual field theory.

After explaining how the hardwall idea can be made to work in a full ten dimensional AdS setting, we describe how cutoffs are to be modified in the presence of D7-branes. Because of branes, several new parameters are introduced into the hardwall model apart from the bulk IR-cutoff $r_g$: a dimensionless DBI normalization parameter $b$, a brane IR-cutoff $\r_m$ and the quark mass parameter $m_q.$  

Given this data,  we subsequently identify all the candidate classical solutions of the cutoff models in section III. Section IV discusses the finite temperature phases of this system at zero and nonzero quark masses. We also study how the phase diagram varies with the parameter $b=\frac{\l}{4\pi^2}\frac{N_f}{N_c}$ and the ratio $\frac{r_m}{r_g}.$ In obtaining the phase diagram, we only need to consider {\em differences} in Helmholtz free energies. In section V, we use counterterms prescribed by the holographic renormalization procedure to obtain finite free energies and other thermodynamic quantities in the various phases. This allows us to identify order parameters as well.  Section VI translates the phase diagram into QCD units by using phenomenology to fix the numerical parameters of our model. We conclude with a summary of the results and discuss future directions for exploration.

\section{The Hard Wall Model in 10D}

To begin with, we describe how the hard wall calculations of \cite{Herzog} can be uplifted to 10D $AdS_5\times S^5$. We may expect that the phase diagram is not affected but the way this occurs is somewhat interesting.

We start with the full 10-D \Romannum{2}B string theory low energy effective action in Lorentzian signature with only a 5-form field strength as appropriate to a situation containing only D3-branes:
\be\label{2b}
S_{\Romannum{2}B}=\frac{1}{2\k_{10}^2}\int d^{10}x \sqrt{|g|}\left(e^{-2\Phi}(R+4(\nabla\Phi)^2)-\frac{1}{2.5 !}|F_5|^2\right)
\ee
We add boundary terms and counter terms in an upcoming section, but for obtaining equations of motion, \eqref{2b} is sufficient.  Here $2\k_{10} ^2=(2\pi)^7g_s ^2 (\a')^4$.

We can see that 10-Dimensional $AdS_5 \times S^5$ metric written as:
\be
ds^2=\frac{r^2}{L^2}(-dt^2+d\Vec{x}^2)+\frac{L^2}{\r^2+y^2}\left(d\r^2+\r^2d\O_3^2 +dy^2+y^2\ d\q^2\right)\label{D7metric}
\ee
with the usual radial coordinate $r$ being obtained as $r^2=y^2+\r^2$,
and the $AdS_5$ black hole $\times S^5$
\be 
ds^2=\frac{r^2}{L^2}(-f dt^2 +d\Vec{x}^2)+\frac{L^2}{\xi^2}\left(d\tilde\r^2 +\tilde\r^2d\O_3^2+d\tilde y^2+\tilde y^2 d\tilde\q^2\right)
\ee
where 
$ r^2=\left(\xi^2+\frac{r_0^4}{4\xi^2}\right)$,  
$\xi^2 =\tilde\r^2+\tilde y^2$,
and the blackening factor 
\be
f=\frac{\left(1-\frac{r_0^4}{4\xi^4}\right)^2}{\left(1+\frac{r_0^4}{4\xi^4}\right)^2}
\ee 
together with a five form field strength,  
\be 
F_{\m\m_2\m_3\m_4\m_5}=\e_{\m\m_2\m_3\m_4\m_5}\, \frac{4r^3}{L^4},
\ee 
and a vanishing dilaton $\Phi=0$, are solutions of equations of motion. 
To avoid confusion with the AdS coordinates $y,\r$,  we label the black hole coordinates as $\tilde y,\tilde\r$. The horizon in these new coordinates is defined by $\x= r_0/\sqrt{2}$. The UV boundary is at $r \to \infty$ which, at fixed $y(\r_{UV})$, $\tilde y(\tilde \r_{UV})$ means $\r_{UV}$, $\tilde \r_{UV} \to \infty$. The AdS radius $L^4=4\pi g_s (\a')^2N_c$ is fixed from the quantization condition that $\int_{S^5} *F =2\pi \m_3 N_c.$

The hardwall approach requires us to compute the on-shell Euclidean action
\be \label{2bE}
S_{\Romannum{2}B}^{E}=-\frac{1}{2\k_{10}^2}\int d^{10}x\  \sqrt{|g|}\left(e^{-2\Phi}(R+4(\nabla\Phi)^2)-\frac{1}{2.5 !}|F_5^E|^2\right)=\frac{1}{2\k_{10}^2}\int d^{10}x\sqrt{g} \frac{8}{L^2}
\ee
with a truncated integration range for the AdS radial coordinate.
The Lagrangian density is evaluated to be {\it the same as in 5-D} with the entire contribution coming from the five form since the total curvature of $AdS_5 \times S^5$ vanishes.  
Thus, upon comparing the on-shell action for thermal AdS and AdS black hole geometries, we obtain a first order phase transition identical to the 5-D hardwall analysis of \cite{Herzog}.

We now introduce quark degrees of freedom by adding D7-branes represented by the DBI action to \eqref{2b} involving the pull back $K_{a b}=P[g]_{a b}$ of the background metric  to the world volume:
\be 
S_{DBI}=- N_f \m_7\int d^8\s \sqrt{-det (P[g])}
\ee 
where $N_f$ represents the number of quarks, $\m_7=\frac{1}{g_s (2\pi)^7 l_s ^8}$ is the tension of the D7-brane.

Therefore, the total Euclidean action we will work with is:
\be\label{actionEuclidean}
S_{E}=-\frac{1}{2\k_{10}^2}\int d^{10}x \sqrt{|g|}\left(R-\frac{1}{2.5 !}|F_5^E|^2\right)+ N_f \m_7\int d^8\s \sqrt{det (P[g])}.
\ee
Following \cite{Karch-Katz}, the D7-branes are chosen to wrap an $S^3$ in the $S^5$ part of the geometry and to be parallel to the boundary directions for Poincare invariance. Finally, these D-branes also extend along a radial direction of the entire spacetime. 

The low energy dynamics is that of an $\mathcal{N}=2$ supersymmetric gauge theory with $N_f$ hypermultiplets in the fundamental representation \cite{Karch-Katz}.
These models have been studied in \cite{Babington}-\cite{Mateos:2006nu} - the novelty in our work being the hardwall cutoffs. 

\subsection{Cut-offs}\label{cutoff}

The hardwall approach to modeling the physics of QCD requires us to impose cutoffs on the ranges of integration of the bulk radial coordinate. 
The total action involves two separate contributions from the bulk gravity fields and those fields which are restricted to the world volume of the D7-branes. The gravity part, which can be said to capture the contribution of pure glue to the free energy is defined with an IR-cutoff $r_g.$ This breaks the conformal symmetry of the AdS geometry, and so we can interpret $r_g$ as being a proxy for $\L_{QCD}.$
This cutoff can be translated into physical units by computing an observable such as the glueball mass \cite{rinaldi}.

The new feature in our work is the introduction of a second IR-cutoff $r_m$ for the brane world volume which restricts the extent $\r_m$ of the brane world volume in the radial AdS direction where  $r_m ^2=\r_{m} ^2+y(\r_m)^2$, a relation which is applicable if the spacetime is thermal AdS. This relation uses the solution $y(\r)$ of the DBI equations of motion and so the cutoff $\r_m$ varies dynamically with the shape of the brane.  Further, given the shape $y(\r)$, the IR-cutoff $r_m$
can be translated into, for instance a meson mass, by studying the fluctuations of the brane degrees of freedom. 
We will refer to $r_m$ loosely as meson mass, while $r_g$ will be similarly termed the glueball mass. 

Thus, we have a dimensionless ratio $\frac{r_m}{r_g}$, which can be tuned to bring the model closer to QCD. In this work, only those cases where $\frac{r_m}{r_g}>1$ are considered, i.e., where the brane is embedded in a {\it geometric} background; but we wish to emphasize that in our work, we keep $r_m$ fixed rather than the quantity $\r_m.$ If a cutoff $\r_m$ is imposed on the world volume coordinate, the shape of the D-brane is unaffected by the cutoff coordinate, and a completely different story ensues.

Once we fix an IR-cutoff $r_m$ in the AdS geometry, the appropriate cutoff for the D7-branes in the {\it black hole} geometry is determined from
\be
r_m ^2= \tilde{y}(\tilde\r_m)^2+\tilde{\r}^2 _m+\frac{r_0 ^4}{4(\tilde{y}(\tilde\r_m)^2+\tilde{\r}^2 _m)}. \label{CutoffIR}
\ee
by using the solution $\tilde{y}(\tilde{\r})$ for the shape of the D7-brane in the black hole background.

Determining the phases of the theory requires us to compare the various classical solutions. 
To ensure that the solutions being compared are candidate phases for the {\it same} underlying theory, we need to ensure that the non-normalizable modes of the various fields are equal on the UV-cutoff surface (because these translate into sources for various operators of the field theory). In addition to this, the periodicity of the thermal circle in the AdS geometry $\b'$ is determined by the Hawking temperature $\b_H$ \cite{Herzog}. We can understand this as follows. We can rescale the time coordinates of both geometries so that the time circle has the same periodicity, say unity. In this case, the $g_{tt}$ component of the  two metrics become
\be
g_{tt} ^{AdS}= \frac{{\b'}^2}{z^2} \quad g_{tt} ^{BH}= \frac{{\b_H}^2}{z^2}f(z).
\ee
The AdS/CFT dictionary requires us to equate the non-normalizable mode of {\em all} the bulk fields at the UV-cutoff surface. Thus, we get 
the condition 
\be
\b'=\b_H\sqrt{1-\frac{r_0 ^4}{\L^4}} \label{Trelation}
\ee
where we perform a Fefferman-Graham expansion on the metric to identify these modes. 
The insight in \cite{Herzog} is the observation that the term subleading in the cutoff can play a significant role. We will drop the suffix on the Hawking temperature in what follows.

Imposing the IR-cutoffs leads to the model breaking conformal invariance. However, the absence of a radially varying dilaton suggests that the $\b$-function of QCD still vanishes.

In thermal AdS geometry, the relationship between the UV-cutoff $\L$ on the radial coordinate and the UV-cutoff on $\r$ is 
\be
\r^2_{UV}=\L^2-m_q ^2
\ee
In the black hole geometry however, the UV-cutoff $\tilde\r_{UV}$ is determined by the equation 
\bea
\label{uv}
\L ^2 &=& m_q ^2+\tilde{\r}^2 _{UV}+\frac{r_0 ^4}{4(m_q^2+\tilde{\r}^2 _{UV})}. 
\label{CutoffUV}\\
\implies \tilde\r_{UV} ^2&\approx&\L^2-m_q^2-\frac{r_0 ^4}{4\L^2}
\eea
It turns out that the subleading term above plays an important role similar to the subleading term in the temperature relation .

At the UV-cutoff surface, we demand that, in the two geometries, the D-brane shapes match $y(\L)=\tilde{y}(\L)=m_q.$

\section{Solutions/Saddle Points}
\label{solutions}
In this section, we describe various classical solutions that could represent the phases of the dual theory at finite temperatures. Most of these solutions have already been considered in the literature (\cite{Karch-Katz, Babington} for instance) - we present them in detail for completeness and ease of reference. 
\subsection{Thermal AdS}
Since the D7-branes are point-like in the $y_{1,2}$ plane, the action has a rotational U(1) R-symmetry in this plane. We assume that the D7-brane is located at $y_2=0$ and in what follows, we will denote $y_1$ as simply $y$. 
The pull back metric on the D7-brane embedded in the Thermal AdS geometry is :
\be\label{pull_back_ads}
ds^2=\frac{r^2}{L^2}(-dt^2+d\Vec{x}^2)+\frac{L^2}{\r^2+y^2}\left(d\r^2(1+y'^2)+\r^2d\O_3^2\right)
\ee
where $\r, y$ have dimensions of length. $y$ is a function of $\r$ and determines the shape of the D7-brane. Its non-normalizable mode at the UV boundary determines the mass of quarks and the normalizable mode accounts for a quark condensate related to the spontaneous breaking of the U(1) R-symmetry by the solution ansatz \cite{Babington}. This will lead to a Goldstone boson in the boundary theory.

The action \eqref{actionEuclidean} in this geometry can be evaluated to be:
\be\label{SadsE} 
S_{AdS}^{E}= a\frac{\b'}{\b}\left[4\int_{r_g} ^\L dr r^3 +b\int_{\r_m} ^{\r_{UV}} d\r \r^3\sqrt{1+y'^2}\right]
\ee
where we have defined boundary theory parameters
\bea 
a &=& \frac{\O_5 V_3 }{\k_{10}^2} = \frac{N_c^2 V_3}{4\p^2 L^8}\label{a} \\
b &=&\frac{\m_7 N_f \O_3 \k_{10} ^2 }{ \O_5} = \frac{\l}{4\pi^2}\frac{N_f}{N_c}\label{b}
\eea 
and $\l=4\pi g^2 _{YM} N_c=4\pi g_s N_c.$

In thermal AdS background, the shapes of the branes are given by the equation
\be\label{y'}
y'=\frac{c^3}{\sqrt{\r^6-c^6}}
\ee
where $c^3$ is an integration constant proportional to the quark condensate $\langle\bar{q}q\rangle$. Since a string stretching from the D7 at $y=m_q, r=\L$ to the D3-brane at $y=0,r=\L$ has length $y(\L)=m_q$, the mass of this string is 
\be\label{M}
M_q=\frac{m_q}{2\pi\a'}=\frac{m_q\sqrt{\l}}{2\pi L^2}.
\ee
which we identify with the physical quark mass. Here  $\frac1{2\pi\a'}$ is the tension of a fundamental string. We take the variation of the action \eqref{SadsE} with the physical quark mass to find the condensate $\s$ as
\be\label{con}
\s=\m_7\O_32\p \a' c^3
\ee

\subsection{Cutoff D-branes}
Once we introduce a cutoff $r_m$, we have new possibilities. The first is a family of D-branes which end on the cutoff surface with varying values of $y_m=y(\r_m)$ given by
\bea\nn
y(\r)=m_q-\frac{c}{2\, 3^{\frac14}} F \left(\cos
   ^{-1}\left(2-\sqrt{3}\right),\frac{1}{4} \left(2+\sqrt{3}\right)\right)\\ +\frac{c}{2\, 3^{\frac14}} F\left(\cos
   ^{-1}\left(\frac{\left(\sqrt{3}-1\right) \r^2+c^2}{\left(1+\sqrt{3}\right)
   \r^2-c^2}\right), \frac{1}{4} \left(2+\sqrt{3}\right)\right)
   \label{ads-cut-emb}
\eea
where $F$ is the elliptic integral defined by
\be
F(\f,\b)=\int_0 ^\f \frac{d\q}{\sqrt{1-\b^2\sin^2\q}}
\ee

\begin{figure}[ht]
    \centering
    \includegraphics[scale=0.23]{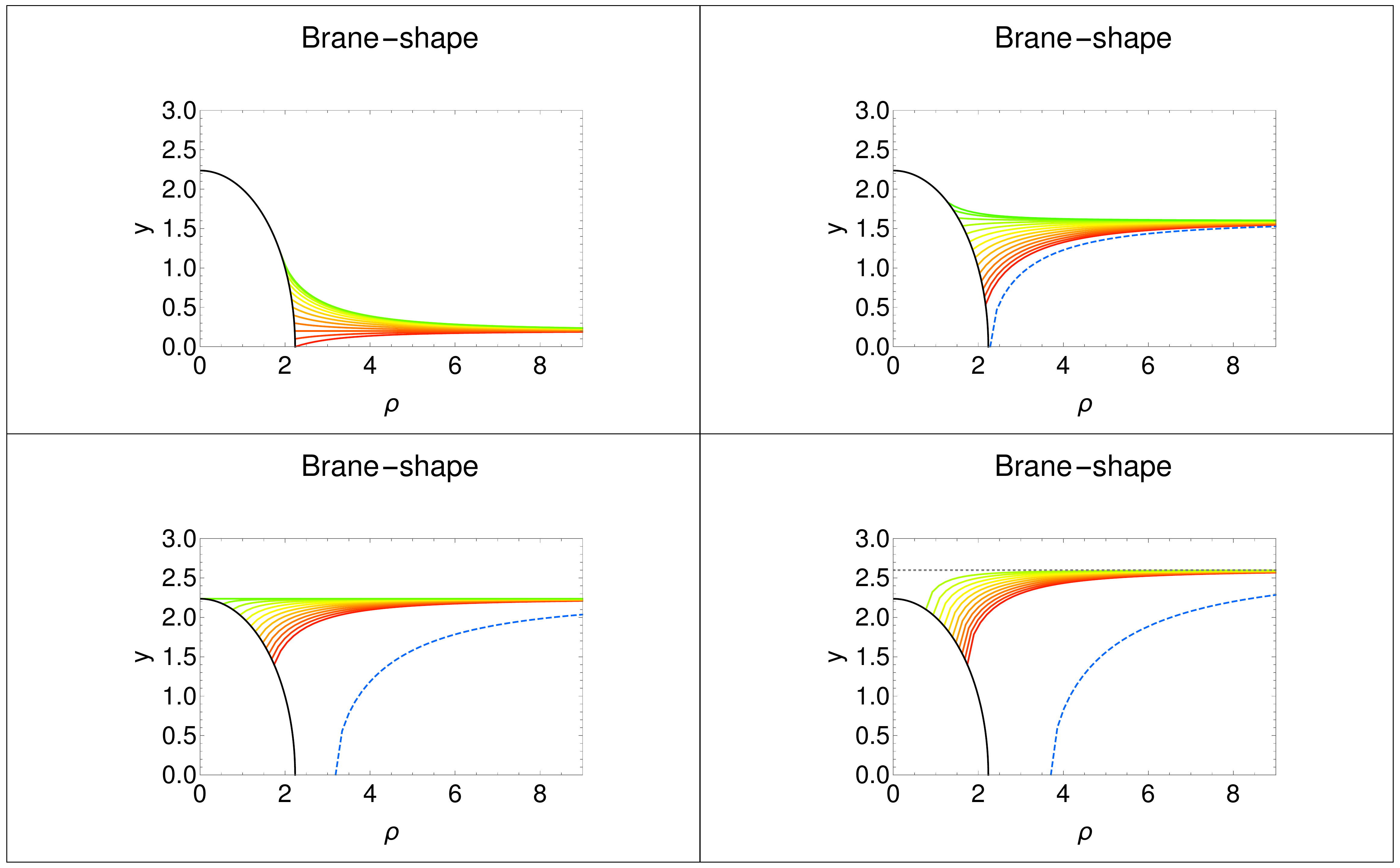}
    \caption{D-branes embeddings in AdS}
    \label{fig:ads_branes}
\end{figure}

We show the various possible shapes of the branes in Fig:[\ref{fig:ads_branes}] where the black solid arc represents the cutoff and
the red, yellow and green curves represent the above solutions.
The value of $c$ is fixed by the geometrical IR-cutoff relation $y^2(\r_m)+\r_m ^2=r_m ^2$ and can be negative. 
For large enough $m_q$, we can see that there are no D7-branes that end on the cutoff surface. The limiting value of $m_q$ occurs when $\r_m=c$, whence we get a quadratic equation for $c$. The discriminant of this equation should be positive which gives us an upper bound
\be
\frac{m_q^2}{r_m ^2} < 1+\frac{\sqrt{3}}{12}\left[F\left(\cos
   ^{-1}(2-\sqrt{3}), \frac{1}{4}(2+\sqrt{3})\right)\right]^2\sim 1.49153
\label{critmass}
\ee
When this condition is satisfied, we have brane solutions for a range of $c$ - at either ends of this range $\r_m=c$ (and hence 
$y'(\r_m)=\infty$), but in any case - all these branes terminate on the cutoff surface.

Secondly, we have the ``Hairpin branes" that bend over before the cutoff surface, that is to say, where $c>r_m$, as shown by blue dotted curve in Fig:[\ref{fig:ads_branes}]. For such branes, we must have $y'(\r_{min})=\infty$ in order that there is no conical singularity in the $y_1,y_2$ plane. Then, a pair of such branes can be smoothly joined producing a configuration which is symmetric under $y_1\to -y_1$. In this case we have only one possible brane since the quark masses fixes the condensate $m_q\sim 0.7 c$ uniquely and thereby the turning point $\r_m=c.$

Finally, we can also have D7-branes which end at $\r=0.$ At this point, the $S^3$ shrinks to zero size and the D7 effectively terminates. In 10-D, this occurs at $r=y_m>r_m.$ However, when the branes end at $\r=0$, they must do so horizontally, i.e., $y'(0)=0$ - since otherwise the induced metric on the D7-branes will have a conical defect. In this case, higher derivative terms in the brane action can be expected to play a significant role and must be included from the get go. However, if we start at $\r=0,$ with $y'(0)=0$ - then we can have only straight (or flat) branes in AdS. Thus, for a given quark mass, we have a single D7-brane configuration of this type which is shown as the straight dotted line in Fig:[\ref{fig:ads_branes}].

In summary, we have three possibilities at a given $m_q$. A straight brane $y(\r)=m_q,$ a Hairpin brane with the turning point fixed by $m_q\sim0.7 c$ and a family of branes ending on the cutoff surface with varying $y_m$ with possibly negative and positive values of $c.$ 

In the absence of the cutoff scale $r_m$, the scale of the D7-world volume fluctuations is set by the quark mass $m_q$ independent of $\L_{QCD}$ i.e., $r_g$. In QCD, the meson masses are determined by chiral symmetry breaking and $m_q$, whereas in all probe-brane situations without a cutoff, the masses are of the form $\frac{m_q}{g_{YM}(m_q)}$ \cite{Myers:2006qr}. 
In our model for the D7-branes which end on the cutoff, the condensate $c^3$ is not entirely determined by the quark mass $m_q$. For a given $m_q$ there are several solutions and the minimum energy configuration {\it dynamically} picks out the value of $c$ (as a function of $r_m$). In section \ref{PhysicalUnits}, we will fix the value of $r_m$ by comparing with actual meson masses - thus phenomenology will relate $\L_{QCD}$ and $r_m.$
In a more complete scenario, the bulk geometry and brane configurations behind the cutoff will determine the relation between $\L_{QCD}$ and $r_m$.

\subsection{Chiral Symmetry}

As seen from the above figure, the branes which end on $y_1=0$ or the branes which end on the cutoff surface have a non-trivial profile and hence the condensate $c^3\neq0.$ 

\begin{figure}[h]
\centering
\includegraphics[width=0.5\textwidth]{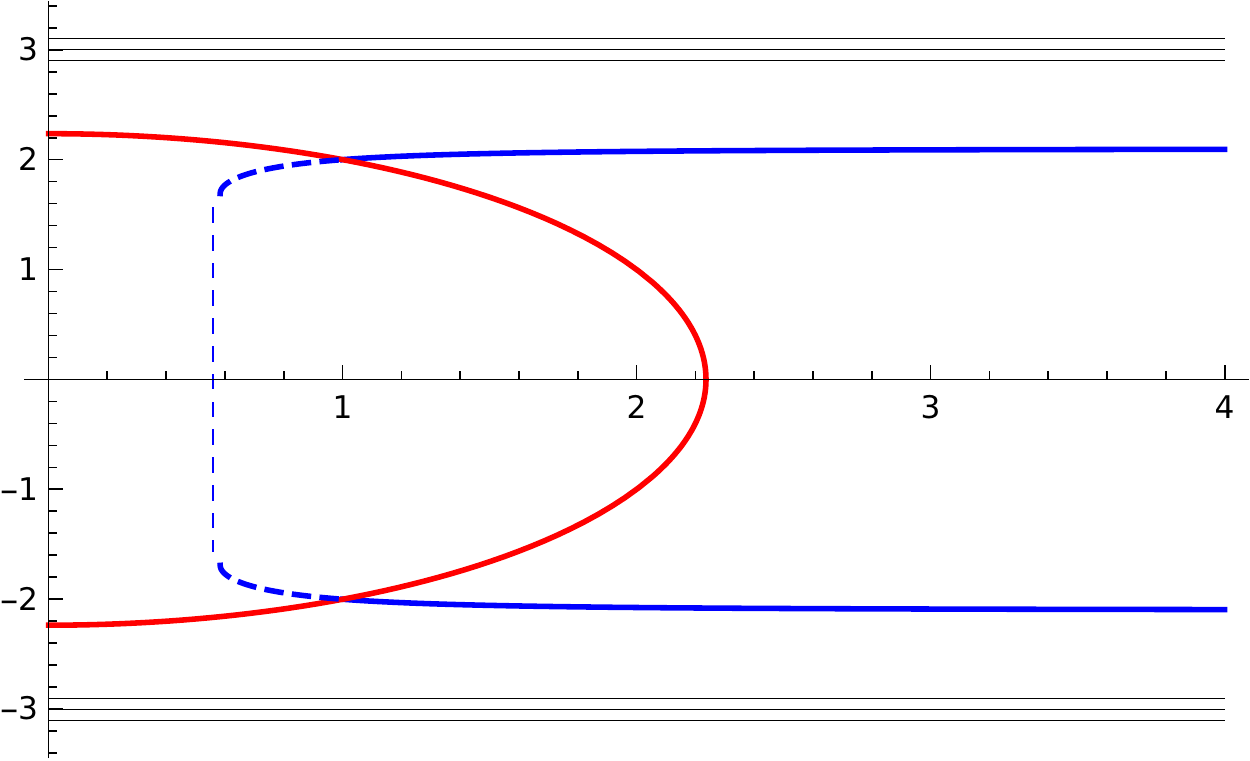}
\caption{Extendend brane shape}
\label{ext-brane}
\end{figure}

We can introduce a second set of D7-branes antipodally in the $y_1-y_2$ plane which will lead to a $SU(N_f)_U\times SU(N_f)_D$ flavour symmetry arising from the two sets of D7-branes. In the Fig:\eqref{ext-brane}, these embeddings are shown as straight black lines, in the presence of the IR-cutoff $r_m$ shown as a red arc. The world volume of the lower brane is taken to be oppositely oriented to the upper one (i.e., these form a brane antibrane pair). 

On the other hand, the configurations of D7-branes which end on the cutoff surface (shown in blue) can be interpreted in a manner very similar to the $D8-$branes of the Sakai Sugimoto model \cite{Sakai:2004cn}, provided we ``complete" the brane embedding by attaching a second D7-brane as shown. We suggest that in a more complete treatment, these branes form a hairpin like configuration (shown as the dotted line), and the separate flavor symmetries $SU(N_f)_U\times SU(N_f)_D$ are broken to a diagonal $SU(N_f)$ subgroup which is suggestive of chiral symmetry breaking in QCD if we identify $U\to L$ and $D\to R$ where $L,R$ are chirality labels. This is supported by the orientation reversal on the lower half arising from the joining.

Developing this idea further, an important question relevant to QCD, is whether we have {\em spontaneous} breaking of chiral symmetry as represented by a nonzero condensate in the chiral limit $m_q\to 0$. This does not occur in our model. At $m_q=0$, the lowest energy D7-brane embedding is a straight brane in the AdS background which ends on the cutoff surface without any condensate. Thus, even if we regard $SU(N_f)_U\times SU(N_f)_D$ as the chiral symmetry, it does not break spontaneously in this model.

We also note that this model has a rotation symmetry in the $y_1-y_2$ plane, which defines an $R$-symmetry of the field theory. A nonzero quark mass leads to each set of D7-branes {\it spontaneously} breaking this rotation symmetry. When the two branes are conjoined, this leads to a single Goldstone boson, whereas when the two branes are distinct, we will obtain a pair of Goldstone bosons. Unfortunately, this symmetry is also not broken spontaneously in the massless limit.

In the forthcoming sections, rather than refer to these two classes of branes as chirally symmetric and broken phases, we will speak instead of quarks being confined or deconfined. This is because as long as there is no horizon on the brane world volume, the meson masses computed from the quasiparticle energy eigenvalues will be real. On the other hand, when the branes have a worldvolume horizon, the mesons will become worldvolume quasinormal modes and hence acquire a temperature dependent width. We also note that for {\it chiral} quarks transforming differently under the flavour symmetries $SU(N_f)_U\times SU(N_f)_D$, additional ingredients are necessary. 

\subsection{D7-branes in the AdS Black Hole Geometry}

Similarly, we can embed the D7-brane in the AdS-black hole background
with the pull back metric:
\be
ds^2=\frac{r^2}{L^2}(-f dt^2 + d\Vec{x}^2) + \frac{L^2}{\xi^2}\left(d\tilde\r^2(1+\tilde y'^2)+\tilde\r^2d\O_3^2\right)
\ee
where $\xi^2={\tilde y}^2+\tilde\r^2.$
The on-shell action in black hole geometry is:
\be\label{SbhE}
S_{BH}^{E}=a\b\left[4\int_{r_g} ^{\L} dr r^3 +b\int_{\tilde\r_m} ^{\tilde\r_{UV}} d\tilde\r \tilde\r^3\left(1-\frac{r_0^8}{16\xi^8}\right)\sqrt{1+\tilde y'^2}\right]
\ee
Because of the presence of potential terms, straight $y={\rm constant}$ D7-branes are not possible in the black hole geometry. If the hard wall cutoff on the bulk geometry is such that $r_m>r_g>r_0$, the cutoff surface hides the horizon. Secondly, if $r_m>r_0>r_g$, a black hole horizon is visible in the cutoff bulk geometry though the D7-branes do not see the horizon. Finally, if $r_0>r_m>r_g$, D7-branes can end on the horizon.  

In any case, near the UV-boundary $\tilde\r\sim \tilde\L$, the branes approximately satisfy 
\be
\tilde y'\sim \frac{c^3}{\tilde{\r}^3}
\ee
defining the condensate $c^3$. The full solution and, thus, the value of the condensate is determined by the IR boundary conditions, as we discuss below. 
The cutoff $\tilde \r_m$ is dynamically determined by the cutoff relation \eqref{CutoffIR} which we reproduce below
\be
r_m ^2=\tilde y_m^2+\tilde{\r}^2 _m+\frac{r_0 ^4}{4(\tilde{y}^2_m+\tilde{\r}^2 _m)}.
\ee
The above relation implies that 
$2\tilde \r_m ^2=r_m ^2+\sqrt{r_m ^4-r_0 ^4}-2{\tilde y}_m ^2$
which cannot be satisfied for large quark mass $r_m ^2+\sqrt{r_m ^4-r_0 ^4}-2{\tilde y}_m ^2<0$ or for high temperature $r_0>r_m.$
In the former case, the branes in the black hole geometry end at $\tilde\r=0.$ In this case, we need that $\tilde y'(0)=0$ from the requirement that there will be no world volume conical singularity for the D7-brane.  
In the latter case, when $r_0>r_m$, the branes either bend over to end up at ${\tilde y}_m=0,$ end on the horizon of the black hole or if the mass is large enough, terminate at $\tilde\r=0$. 
\subsubsection{Low Temperature}
In this subsection, we consider the case when the temperature is smaller than the IR-cutoff $r_m>r_0$. 
The equations of motion show that the trivial solution $y(\r)=0$ still exists. 
We find non-trivial solutions for $y(\r)$ numerically, for different initial conditions ${\tilde y}_m={\tilde y}(\tilde{\r}_m)$, by using a Newton-Raphson routine which adjusts the initial slope ${\tilde y}'(\tilde\r_m)$ so that we obtain the correct quark mass ${\tilde y}(\L)=m_q.$ 
\begin{figure}[ht]
    \centering
    \includegraphics[scale=0.23]{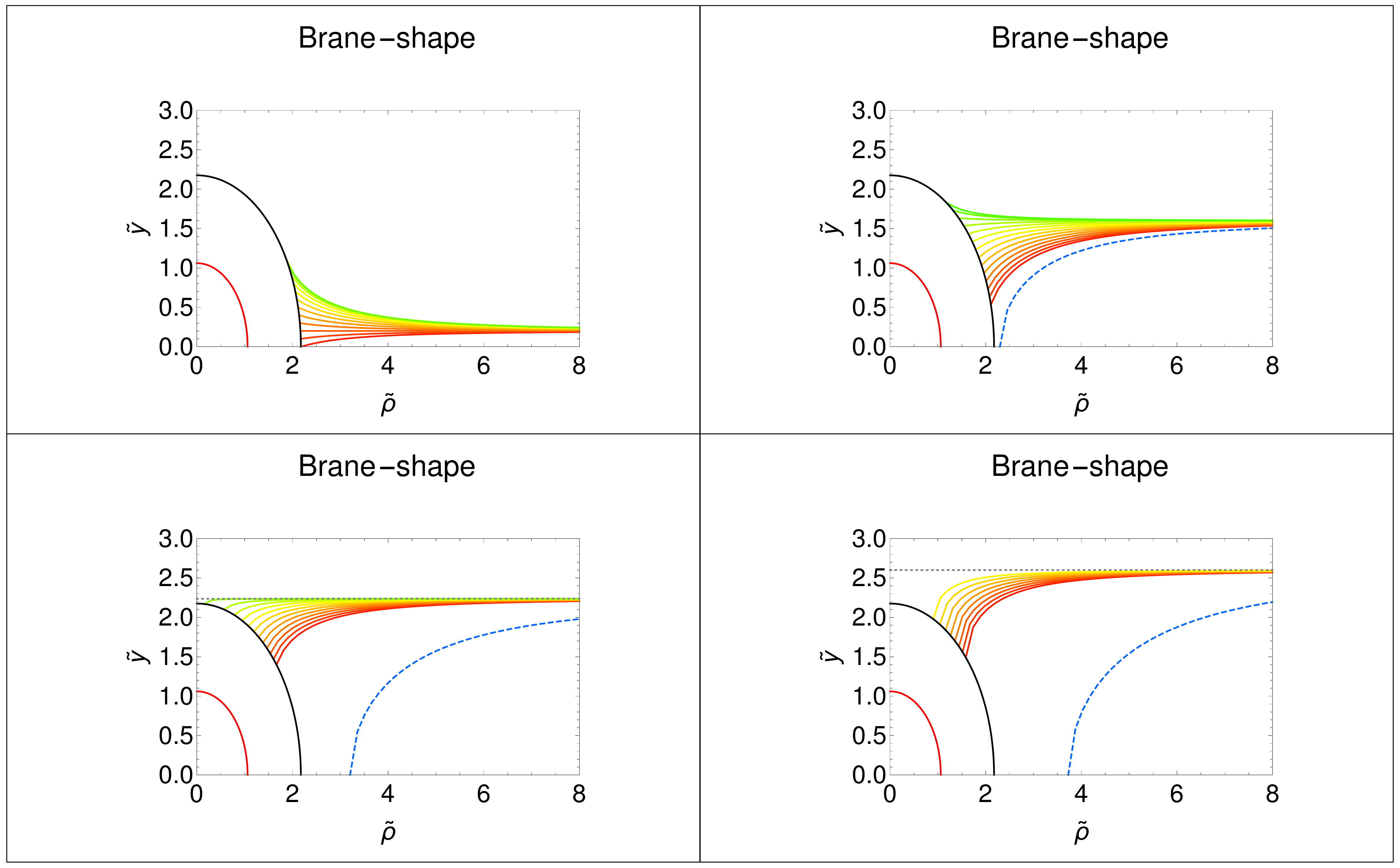}
    \caption{D-branes in BH geometry for $r_0<r_m$.}
    \label{fig:bh_branes}
\end{figure}
Fig:\eqref{fig:bh_branes} shows all possible D7-brane for a fixed IR cut-off $r_m$ and horizon $r_0$, and varying the quark masses $m_q=y(\L)$. The black arc is the IR-cutoff, and the red arc represents the horizon which is covered by the cutoff. 
The Hairpin D-branes (blue dashed lines) which bend over before the cutoff occurs for relatively larger values of the quark mass. 
For larger quark masses, we also have ``straightish" branes in the black hole geometry, which end on the $y-$axis at $\tilde\r=0.$
Note that there are solutions which start at the IR-cutoff with a value greater than the quark mass (green curves).

\subsubsection{High Temperatures}

When the temperature $r_0>r_m$, the D7-branes in the black hole geometry are somewhat different since in this case the cutoff radius $r_m$ is behind a horizon. 

\begin{figure}[ht]
    \centering
    \includegraphics[scale=0.25]{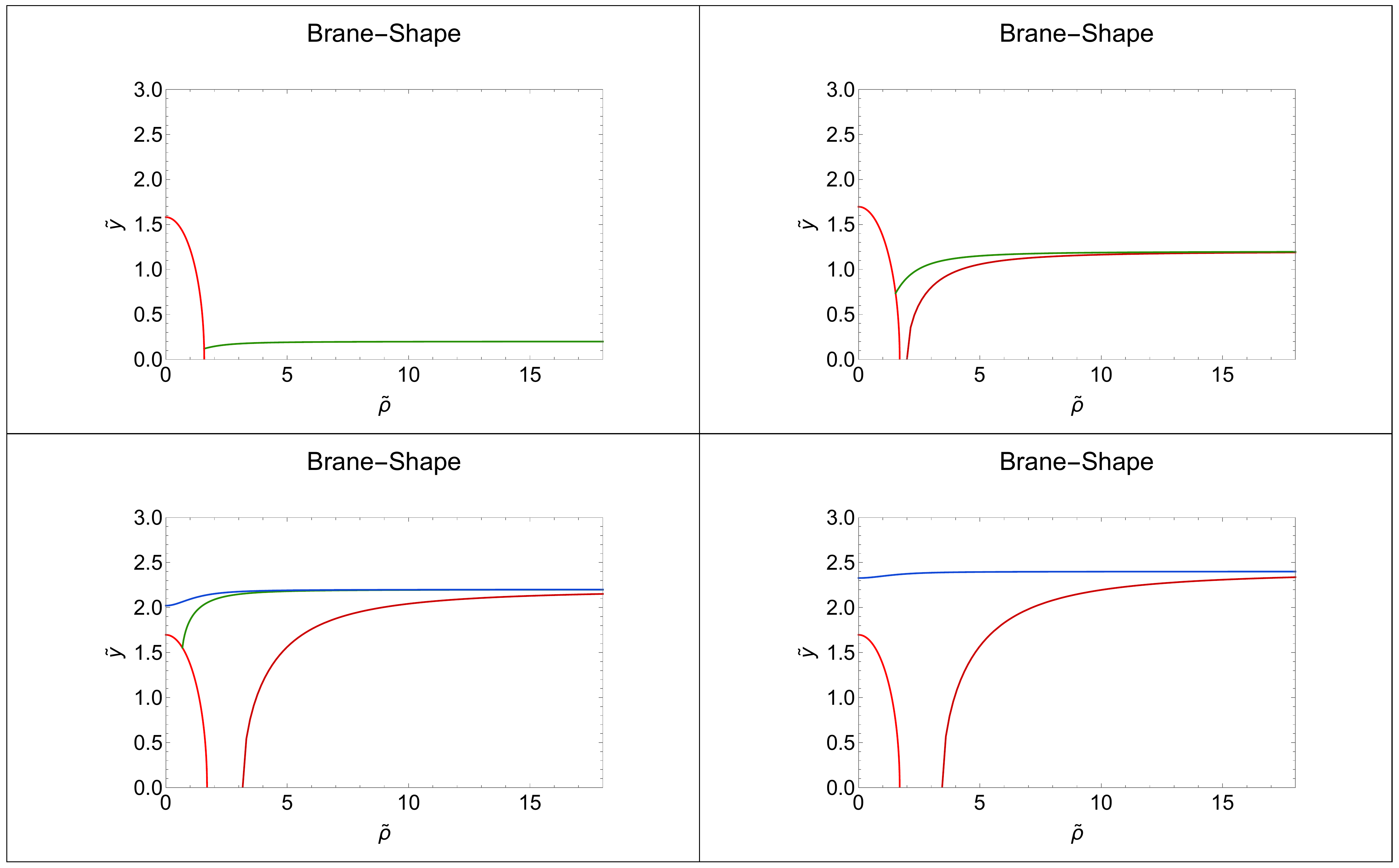}
    \caption{D7-brane in BH geometry for $r_0>r_m$}
    \label{fig:D7-shapes}
\end{figure}

These different possibilities are shown in Fig:\eqref{fig:D7-shapes}.
It is now possible for the D7-brane to end on the horizon of the black hole at ${\tilde y}^2+{\tilde\r}^2=r_0 ^2/2$. From the equations of motion, we can obtain a regularity condition for the embedding at the horizon, ${\tilde y}'=\frac{\tilde y}{\tilde\r}.$ This implies that we have only one such D7-brane configuration for a given quark mass. These are shown in green in the figure. In this case, we find that the Hawking temperature of the world volume metric on the D7-brane matches with that of the bulk. Thus, we can say that the quark and the gluon degrees of freedom are in thermal equilibrium. 

We can also have solutions that terminate on the $\r=0$ surface - and thus the world volume $S^3$ collapses to zero size. As before, we will require ${\tilde y}'(0)=0$ for such branes and hence we have a single solution of this type. These are the blue colored curves in the figure and we refer to them as the ``straightish" branes. 

Finally, we can also have Hairpin branes which bend over before the horizon and intersect the x-axis and are shown in red in the figure. 

It is important to point out that for the D7-branes which end  on the horizon of the black hole, the worldvolume fluctuations will be quasinormal modes, that is to say, the mode energies will acquire imaginary parts. Stated in boundary terms, the `mesons' now acquire a lifetime which depends on the temperature. For convenience, we will refer to this situation as meson 'melting' although that term is more properly used when the quasiparticle peak in the spectral function disappears. 

\section{Phase transitions at finite temperature}\label{PT}

In this section, we will discuss the phase diagram of this system in detail. This requires us to compare the Helmholtz free energy of the various configurations we have discussed. 

The action when evaluated on the solutions described earlier, has UV-divergences coming from the upper limit of integration. Thus, it is not immediately interpretable as free energy. However, for the purpose of computing the phase diagram, we can take differences in the on-shell action between various solutions to determine the thermodynamically preferred configuration. The differences are guaranteed to be finite because UV-divergences in field theories are independent of temperature. On the gravity side, this translates into identical cutoff dependence of the on-shell action when evaluated on the various solutions. 

Recall that we have two dimensionless parameters: the ratio
 $\frac{r_m}{r_g}\geq 1$ and the coefficient $b=\frac{\l N_f}{4\pi^2N_c}.$ In this section, we will present our study of the phase diagram for a few values of these parameters.

\subsection{Zero quark mass}
\label{section0qm}
The equations of motion of the D-branes (in the probe limit) admit a solution $y(\r)=0$ in both thermal AdS and black hole geometries. These solutions represent the introduction of zero mass quarks into the dual boundary field theory since the D3-D7 strings have zero length.

The possible configurations, in this case, are the trivial solution and the nontrivial solutions of the form \eqref{ads-cut-emb}, 
with $c<0$  and $m_q=0.$
However, the latter solutions turn out to have larger free energy and are ignored in the following. 

Given the two cutoffs, we have three possible scenarios in the black hole geometry
\begin{itemize}
    \item $r_0<r_g<r_m:$ In this case, the bulk has $r_g$ as IR-cutoff while the IR-cutoff is $r_m$ for the DBI action.
    \item $r_g<r_0<r_m:$ In this case, the bulk has $r_0$ as its IR-cutoff. The IR-cutoff for the DBI  is $r_m$. On the brane coordinate, this can be written as  $\tilde\r_m ^2=(r_m^2+\sqrt{r_m^4-r_0^4})/2.$
    \item $r_g<r_m<r_0:$ In this case, the black hole horizon acts as the IR-cutoff for both systems. 
\end{itemize}
We now consider these in turn. 

When the bulk geometry is that of thermal AdS, the total Euclidean on-shell action \eqref{SadsE} is
\bea
S_{AdS}^{E}&=&a\b'\left( \left[\L^4-r_g^4\right] + \frac{b}{4}\left[\L^4-r_m^4\right]\right)\\
&=&a\b\left(\L^4-r_g^4-\frac{r_0^4}{2}+\frac{b}{4}\left(\L^4-r_m^4-\frac{r_0^4}{2}\right)\right) 
\eea
where we have used the relation \eqref{Trelation} between $\b'$ and $\b.$ We observe that if we interpret that  on-shell action as free energy, the $r_0^4$ temperature dependence indicates a nonzero entropy. This is at odds with the traditional association of a horizon with entropy.

For the first case $r_0<r_g<r_m$, the on-shell action \eqref{SbhE} for branes embedded in the black hole geometry evaluates to:
\bea
S_1&=&a\b\left(\L^4-r_g^4+\frac{b}{4}\left(\tilde\L^4+\frac{r_0^8}{16\tilde\L^4}-(\tilde\r_m^4+\frac{r_0^8}{16\tilde\r_m^4})\right)\right)\\
&=&a\b\left(\L^4-r_g^4+\frac{b}{4}(\L^4-r_m^4)\right)
\eea
For the range $r_g< r_0<r_m$, we obtain $S_2$ as:
\be
S_2=a\b\left(\L^4-r_0^4+\frac{b}{4}(\L^4-r_m^4)\right)
\ee
Finally, when the black hole horizon become larger than $r_m$, we get $S_3$ :
\be
S_3=a\b\left(\L^4-r_0^4+\frac{b}{4}\left(\L^4-r_0^4\right)\right)
\ee
For a given $r_0$, we take the differences in the on-shell action of the AdS and Black hole configurations. 
\begin{align}
S_1-S&= a\b\left(1+\frac{b}{4}\right)\frac{r_0^4}{2}\: ;\:\qquad\qquad\qquad\qquad r_0<r_g\\
S_2-S&=a\b\left(r_g^4+r_0^4\left(\frac{b}{8}-\frac{1}{2}\right)\right) \: ;\:\qquad\qquad r_g< r_0<r_m\\
S_3-S&=a\b\left(r_g^4+\frac{b}{4}r_m^4-\frac{r_0^4}{2}(1+\frac{b}{4})\right) \: ;\:\qquad r_m<r_0
\end{align}
We see that if $0<b\leq 4\left(1-2\frac{r_g^4}{r_m^4}\right)$, we obtain a possible phase transition in the region $r_g < r_0 < r_m$. Positivity of $b$ requires  $\frac{r_m}{r_g}>2^{\frac14}\sim1.19$.
When these two conditions are satisfied, the phase transition occurs at a temperature
\be\label{Tcg} 
T_{cg}=\frac{2^{3/4}r_g}{\p L^2 (4-b)^{1/4}}
\ee
This transition which depends only on $r_g$ deconfines the gluons while the quarks remaining bound in mesons. The critical temperature is independent of the brane cutoff $r_m$. However, the possibility of this transition does depend on the ratio $\frac{r_m}{r_g}.$
As we increase the temperature further ($r_0\geq r_m$), we see a horizon appearing in the D7-brane world volume beyond a temperature
\be \label{Tq}
T_{cq}= \frac{(8 r_g^4 +2 b r_m^4)^{1/4}}{\p L^2 (4+b)^{1/4}}
\ee
depending on the brane cutoff $r_m$. For $r_0=r_m$ (when case II and III both become possible) and if $b=4\left(1-2\frac{r_g^4}{r_m^4}\right)$, both critical temperatures become equal.

However, even if the conditions for the first transition are not satisfied, as we increase the temperature for any $b>0$, we always obtain a transition in the region $r_0>r_m>r_g$, when $S_3$ becomes the lowest Free energy configuration (compared to the AdS embeddings). The temperature is given by the same formula \eqref{Tq}.
This is a simultaneous deconfinement/melting transition of the quarks and gluons - since the branes now end on the horizon. Therefore, we denote this temperature by $T_c$.

\subsection{Finite Quark mass transitions}

We now move to non-zero current quark mass, that is to say, for non-zero $y$ field profiles. We will find the solutions numerically and determine the on-shell action using numerical integration techniques.
For a given temperature and quark mass, we then determine which of the various branes and backgrounds represents the least action configuration (after including the gravity terms that represent the glue contribution). 

In the thermal AdS background, the total action is
\bea
S&=&a\b'\left(\L^4-r_g ^4 +b \int_{\r_m} ^{\sqrt{\L^2-m_q ^2}} d\r \r^3\sqrt{1+{y'}^2}\right) \\
&=& a\b \left(-r_g ^4-\frac{r_0^4}{2}+b \left(-\frac{r_0^4}{8}+\int_{\r_m} ^{\sqrt{\L^2-m_q ^2}} d\r \r^3\sqrt{1+{y'}^2}\right) \right)
\label{ADSaction}
\eea
where, in the second line, we have to drop  divergent terms coming from the bulk as in the minimal subtraction scheme. This integral can be evaluated in terms of Gauss Hypergeometric functions, and we get the following expression for the action after taking the limit $\L\to\infty$
\be
S=a\left( -r_g ^4- \frac{r_0^4}{2}+b \left(
-\frac{\rho_m^4}{4}\, {}_2F_1(-\frac{2}{3},\frac{1}{2};
\frac{1}{3};\frac{k^6}{\r_m ^6}) -\frac{r_0^4}{8}\right)\right)\label{ADSFreeExact}
\ee
For this on-shell action to be interpreted in boundary terms, we need to use renormalization conditions to translate the various parameters appearing above into physical quantities. 

\subsubsection{Ground state as a function of quark mass}

Before proceeding to finite temperature, we discuss the properties of this system at zero temperature as we vary the quark mass. At zero temperature, the on-shell gravitational action, regarded as the ground state energy of the boundary theory, involves two contributions. One comes from the bulk gravitational degrees of freedom and depends on $r_g$, while the other is the D7-brane contribution depending on the cutoff $r_m$. However, since the background remains the same, the difference is independent of $r_g$, and the nature of the ground state is controlled only by $r_m$. Secondly, in this probe approximation, the contribution of the D-brane involves the coupling constant $b$, only as an overall multiplication factor. Thus, we could say that the properties of the ground state are entirely controlled by the meson mass via its proxy $r_m.$ In particular, if we associate the bulk cutoff $r_g$ with $\L_{QCD}$, the phase changes in the quark sector are not affected by $r_g$ so long as $r_g<r_m.$

For a given value of $m_q$ and cutoff $r_m$, the shape of the brane maybe characterized by the condensate parameter $c.$ The minimum energy condition picks out a particular value of $c(m_q,r_m)$ for a given value of $m_q$ and $r_m$. In QCD, the chiral condensate defined as $\langle \bar{\psi}\psi\rangle\equiv C^3$ is a function $C(m_q,\L_{QCD})$ with the property that $C(0,\L_{QCD})\neq 0.$ If we relate $c\sim C$, then the minimum energy condition determines the chiral condensate as a function of $\L_{QCD}$ and $m_q.$

However, we find that $m_q=0$, we find that the {\it minimum energy configuration} is a straight brane ending on the cutoff surface. This means that $c(0,r_m)=0$ which is translated, in boundary terms, as the absence of spontaneous breaking of the chiral symmetry. This is in contrast to other holographic QCD models \cite{Babington,Kruczenski,Bak:2004nt}, where we see that the minimum energy, stable branes acquire a nontrivial profile leading to a condensate. It is interesting to note that in all the cases with symmetry breaking, we have a nontrivial dilaton profile as well suggesting that conformal symmetry breaking alone does not trigger chiral symmetry breaking in the massless limit.  

\begin{figure}[t]
    \centering
    \includegraphics[scale=0.25]{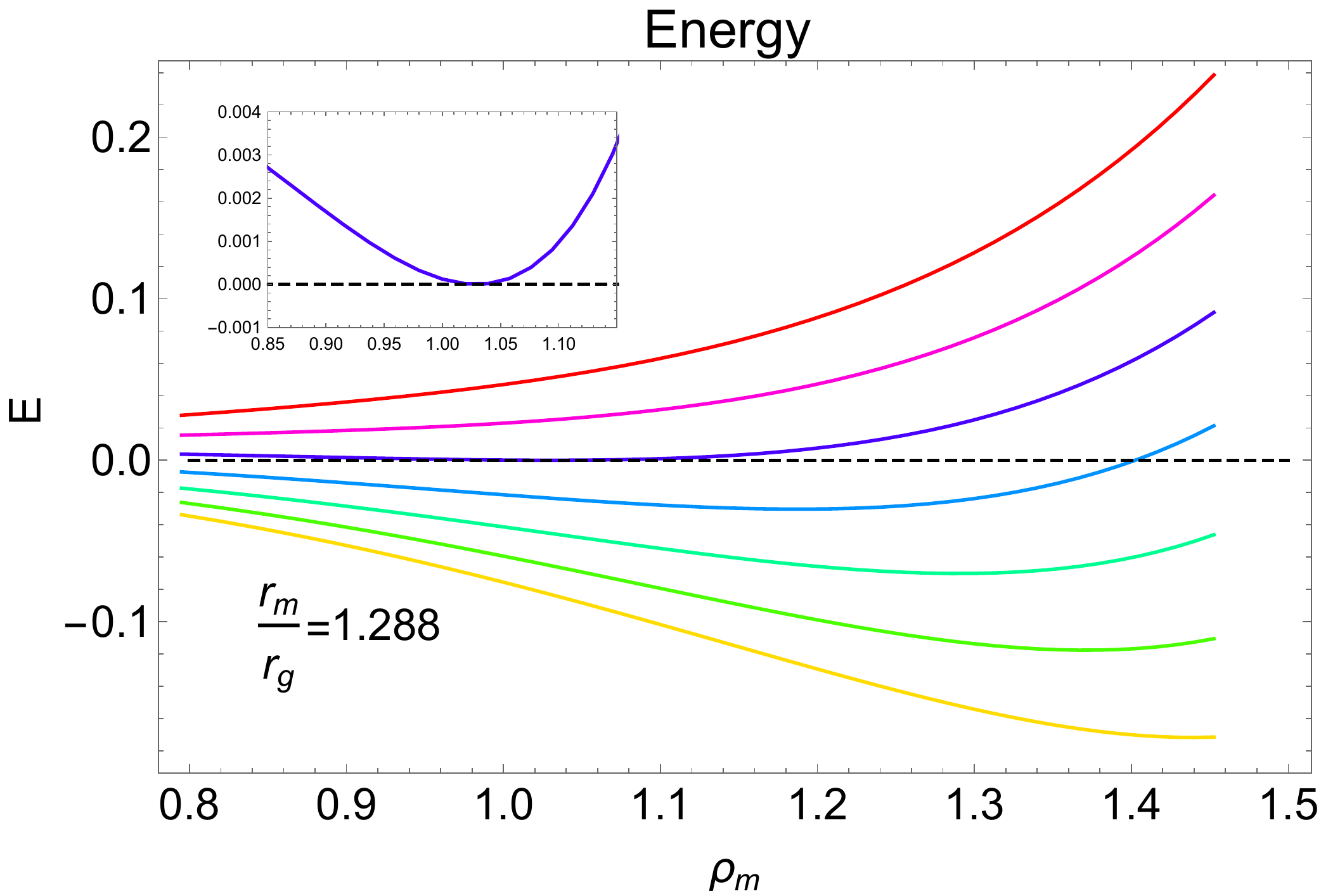}
    \caption{The black dotted line is a visual representation of the energy of the straight brane (it should be a point at $\r_m=0$) while each colored line represents all brane configurations ending on the $r_m$ surface. Different colors are shown for different quark masses (with yellow being the smallest and red being the largest).}
    \label{fig:qmass}
\end{figure}

For a given $m_q<r_m$, as shown in Fig:\eqref{fig:ads_branes}, we have a family of solutions ending on the cutoff surface (including a straight brane $y=m_q$). In Fig:\eqref{fig:qmass}, we plot the DBI energies of the curved branes in AdS for various quark masses shown by the colored curves. The horizontal axis is the IR-cutoff which varies between $0<\r_m<r_m.$
From the expression for the action \eqref{ADSaction}, it is obvious that any curvature in the shape of the brane caused by the condensate $c^3$ increases the free energy. However, somewhat surprisingly, we find that the lowest energy configuration in AdS is a curved brane that ends on the cutoff surface at an angle. This is because a curved brane has a smaller integration range along the $\r$ coordinate. Therefore, for low values of $m_q/r_m$, we have a minimum energy configuration (see Fig:\eqref{fig:qmass}) representing a brane that ends on the cutoff surface. However, for large values of $m_q$ (at fixed $r_m$), we see that there is no minimum. Thus, there is a value $m_q ^1$ when the minimum disappears (the magenta curve, say). 

If $m_q>r_m$, apart from branes that end on the cutoff surface, there is also a straight brane ending at $\r=0$ whose energy is visually represented as the black dotted line in the figure. In this case, we should compare the energy of the D-branes at the minima with the straight brane to determine the nature of the ground state. 

From the figure, we can make the following observations. 
\begin{itemize}
\item As we increase $m_q$,  the value of the minimum energy becomes equal to the energy of the straight brane. The inset zooms in on the comparison, clearly showing that at the value of $m_q$ corresponding to the blue curve, the energy of the curved brane equals that of the straight brane. For larger masses than this critical value, which occurs at $m_q^*/r_m= 1.12644$, the straight brane is the lower energy configuration. Thus, as a function of the quark mass, the ground state undergoes a qualitative change - while the energy itself changes smoothly.
\item The order parameter for this transition can be taken to be the condensate $c^3$ that appears as the normalizable mode of $y$. For the straight branes, $c=0$, while the curved branes that end on the cutoff surface all have non zero values for $c$.
\item
For large values of $m_q>m^1 _q$, the minimum in free energy disappears. However, $m_q^1$ is always greater than $m_q^*$. This is also clearly visible from the inset - which shows a clear minimum which will persist for larger values of $m_q.$ 
\end{itemize}
Thus, beyond the critical value $m_q ^*$, the minimum energy configuration changes abruptly from a curved brane ending on the cutoff to a straight brane. However, we note that even though the value $m_q ^1$ occurs well above this transition value $m_q ^*,$  the presence of a dynamical scale like $m_q ^1$ could be relevant in the phase diagram of other holographic models for QCD. 

It may be appropriate to point out here that if we were to impose the cutoff on the world volume $\r$ coordinate, then for a given quark mass, in AdS space, the straight branes are always the lowest free energy configurations.

\subsubsection{Nonzero temperatures}

In order to identify the phases at nonzero temperature, we use numerical techniques. As discussed in Section \ref{solutions}, we have various possible brane configurations in the thermal AdS as well as in the black hole geometry. Therefore, to find the lowest free energy configuration among all the possibilities, we first determine the lowest free energy brane in each background. We then compare the two least energy configurations for various values of the temperature. These phase transitions depend on quark mass as well as the parameters $b,r_m/r_g$ and will be studied in detail in what follows. We first analyze the phase structure for a fixed value of the IR-cutoff of the D7-brane $r_m=\sqrt{5}$ and determine the bulk IR-cutoff $r_g$ according to the ratio $\frac{r_m}{r_g}$ $(=1.288)$. 

\begin{figure}[ht]
    \centering
    \subfigure{\includegraphics[scale=0.2]{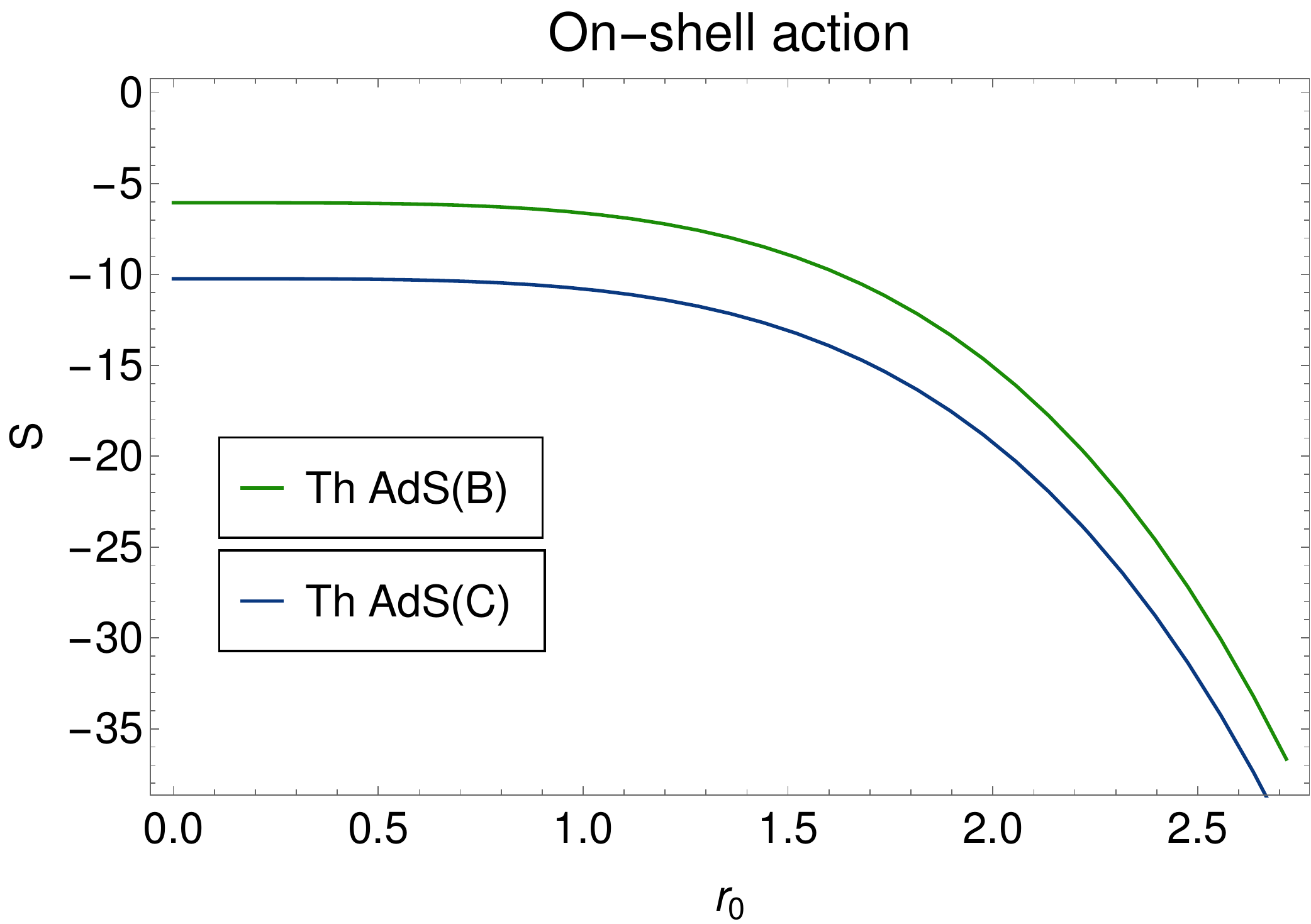}\label{fig:small_mass_action_ads}}
    \subfigure{\includegraphics[scale=0.2]{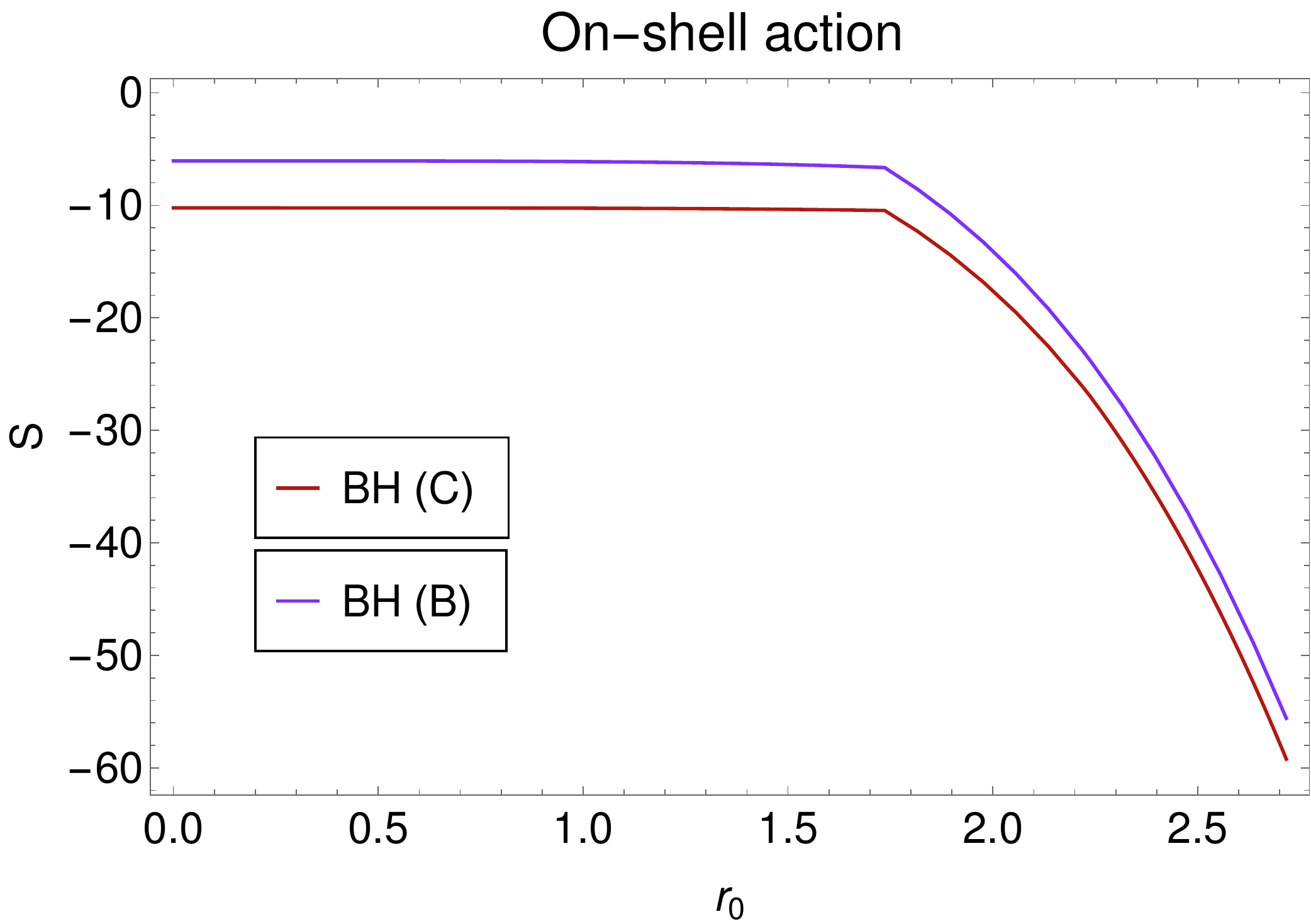}\label{fig:small_mass_action_bh}}
    \caption{Energies of various branes: $b=0.5, m_q=1.7$}
\end{figure}

The story of branes in the AdS background continues to nonzero temperatures as well. This is because the only effect of temperature on the free energy of these solutions is an overall factor that arises from $\b'$. Thus, as shown in Fig:\eqref{fig:small_mass_action_ads}, for low values of $m_q$, the curved branes (labeled Th AdS(C)) have lower free energy compared to the hairpin branes (labeled Th AdS(B)). 

However, at a larger temperature, we have other configurations with D7-branes embedded in the {\it black hole} background. In the black hole background, the total action is
\be
S_{BH}^{E}=a \left[-r_m ^4+ b\int_{\tilde\r_m} d\tilde\r \tilde\r^3\left(1-\frac{r_0^8}{16\xi^8}\right)\sqrt{1+\tilde y'^2} -b\frac{r_0^4}{8}\right]
\ee
where we have to drop UV divergent terms. 

\begin{figure}[ht]
    \centering
    \subfigure[]{\includegraphics[scale=0.2]{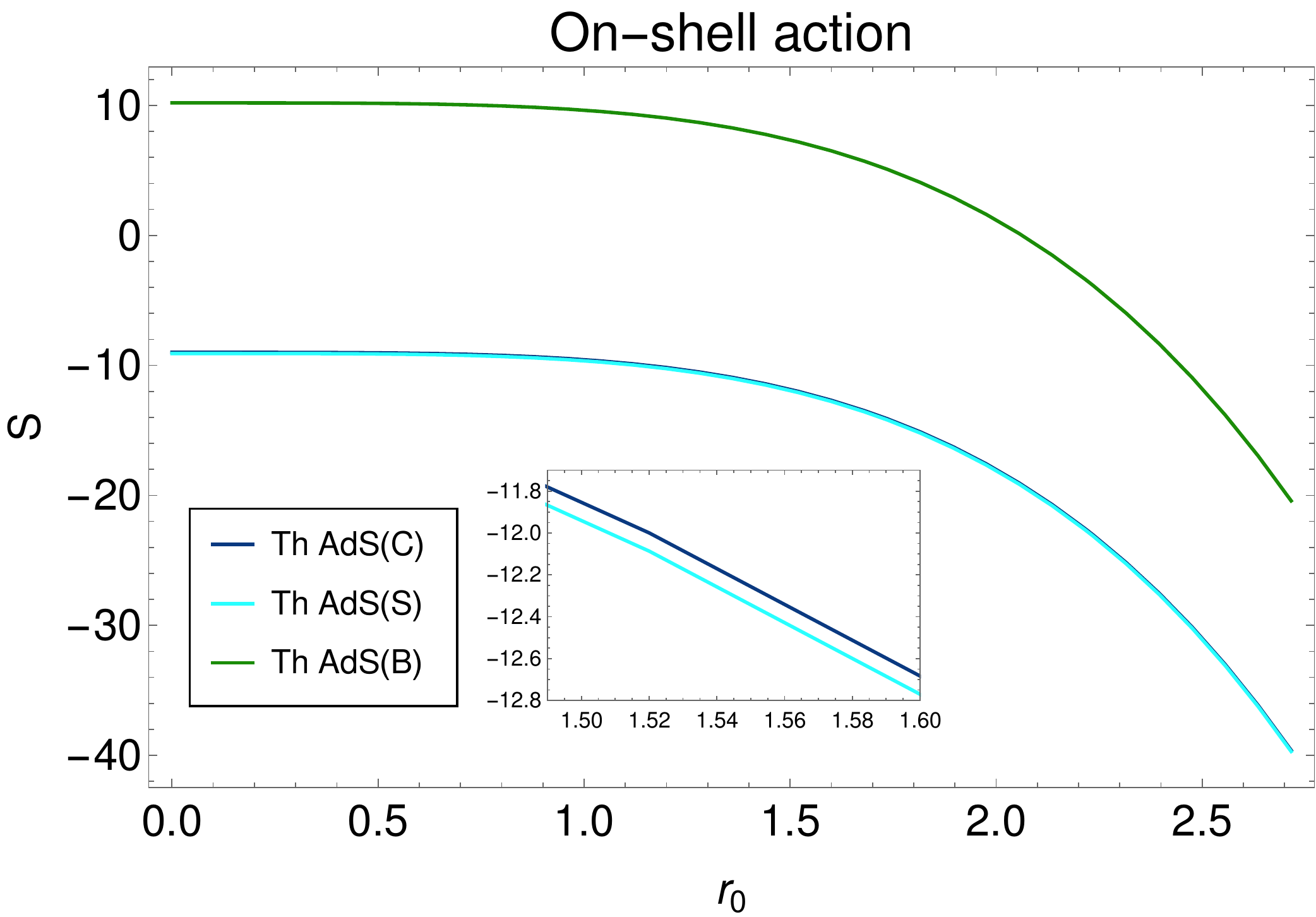}\label{fig:large_mass_action_ads}}
    \subfigure[]{\includegraphics[scale=0.2]{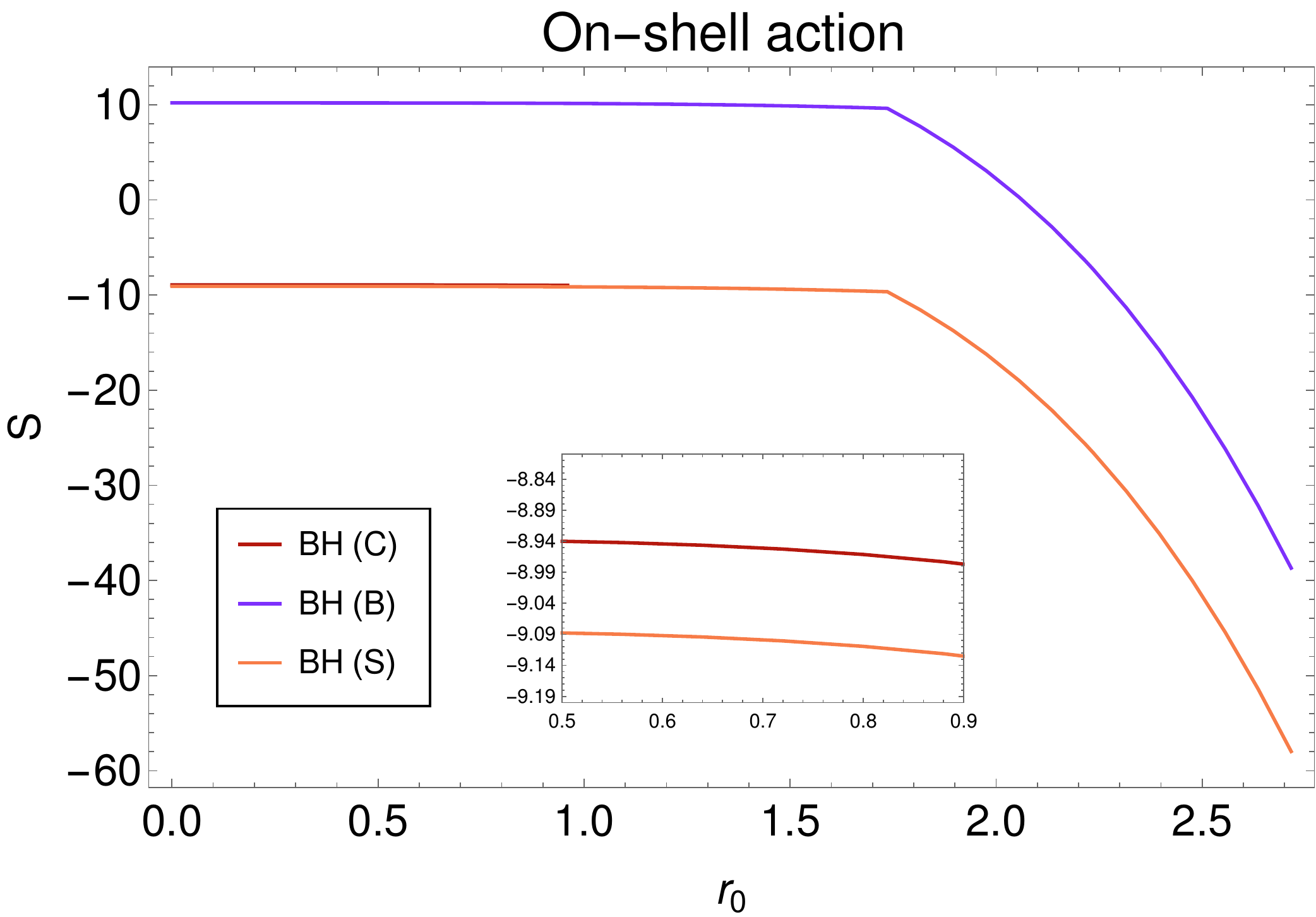}\label{fig:large_mass_action_bh}}
    \caption{Energies of various branes: $b=0.5, m_q=2.7$}
    \label{ONShellA}
\end{figure}

For low quark masses, the minimum free energy occurs for a positive condensate and for $y_m<m_q$ Fig:\eqref{fig:small_mass_action_bh}. The hairpin branes BH(B) have higher energy in this case as well. 
For large quark masses $m_q=2.7$, as shown in the Fig:\eqref{fig:large_mass_action_ads}, the straight branes are the lowest free energy configurations in both thermal AdS and in the black hole geometry Fig:\eqref{fig:large_mass_action_bh} where the labels $B,S,C$ refer to the hairpin, straight, and cutoff branes, respectively. 

\begin{figure}[ht]
    \centering
    \subfigure[$b=0.5$]{\includegraphics[scale=0.2]{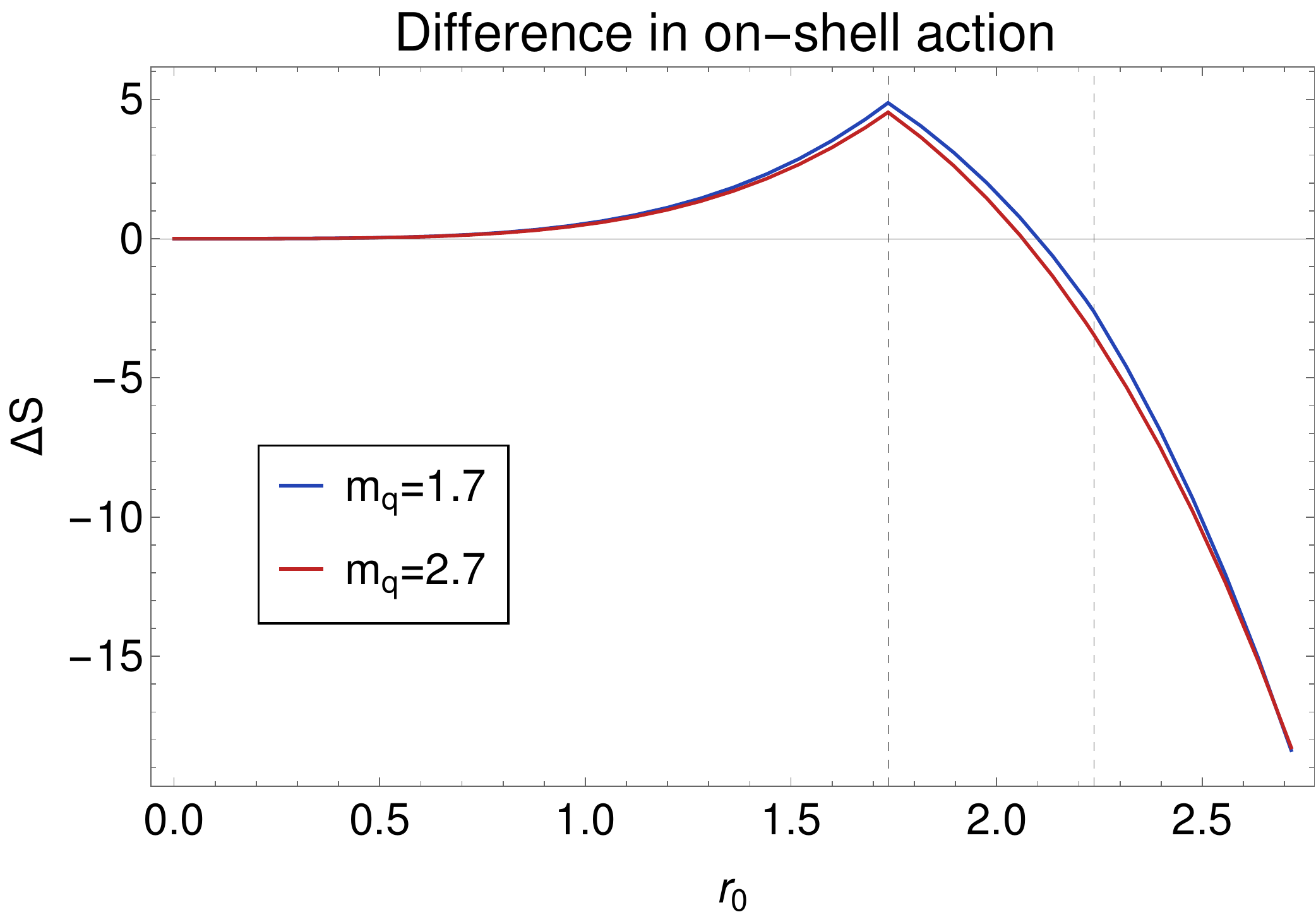}\label{fig:diff_FE_small_b}}
    \subfigure[$b=5$]{\includegraphics[scale=0.2]{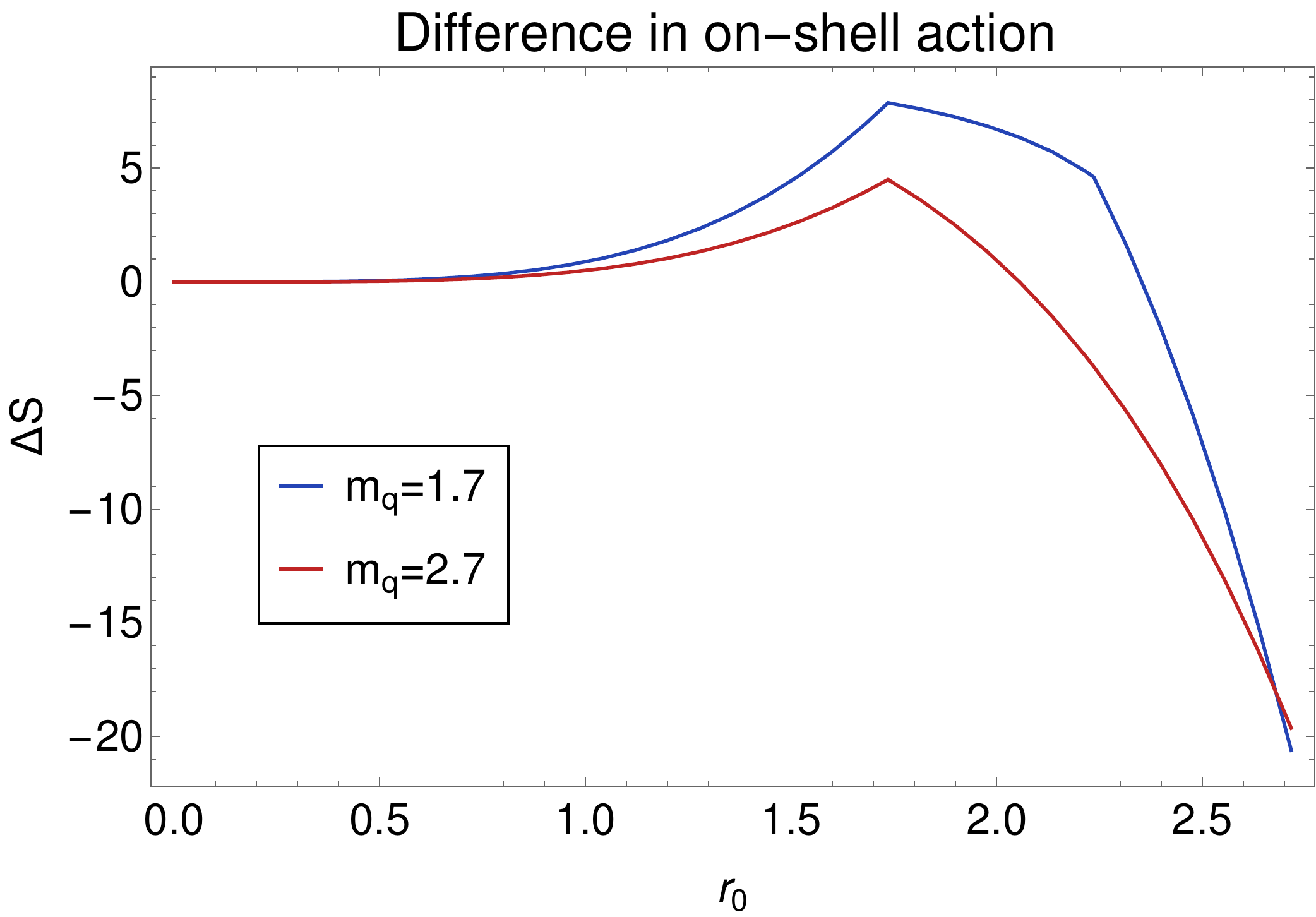}\label{fig:diff_FE_large_b}}
    \caption{Phase transition}
    \label{fig:diff_FE}
\end{figure}

The zero quark mass study suggests that we should study the system for at least one value of b in the range $b>4\left(1-2\frac{r_g^4}{r_m^4}\right)$ and one in the range $b<4\left(1-2\frac{r_g^4}{r_m^4}\right).$ Therefore, in Fig:\eqref{fig:diff_FE}, we plot the difference in free energy, now including the bulk background contribution as well, between the minimum energy black hole embedding and the corresponding minimum free energy thermal AdS embedding.  
 The two panels show the effect of varying the parameter $b$ which multiplies the DBI term in the free energy. The first panel in Fig:\eqref{fig:diff_FE_small_b} shows that the AdS embedding describes the phase in the range $r_0<r_g$. But once $r_0>r_g$, there is a critical temperature $T_{cg}$ after which the difference turns negative, showing that the cutoff branes in the black hole background have lower free energy. Therefore, as pointed out in Section \ref{section0qm}, the gluons deconfine even though quarks are bound in mesons. The mesons `melt' freeing the quarks at the higher temperature $T_{cq}$ as found in zero mass case. This latter transition does not appear in the figure. 
 
 In the second panel Fig:\eqref{fig:diff_FE_large_b}, the difference in the on-shell action increases until $r_0=r_g$, indicating that the AdS embedding has lower free energy. As soon as $r_0>r_g$, the difference starts to decrease and eventually becomes negative in the region $r_0>r_m$, stating that the black hole embedding with a horizon brane is preferred over the thermal AdS. This suggests that the deconfinement transition for gluons is simultaneous with the mesons `melting' temperature $T_c$. Note that the presence of the kink in the blue curve indicates a transition within the black hole geometry between curved branes ending on the cutoff and branes ending on the horizon, but this has no effect on the phase diagram which is determined by the lowest free configuration alone. 

\subsection{Phase diagrams}
\label{PD}
\begin{figure}[h]
    \centering  \subfigure[]{\includegraphics[scale=0.25]{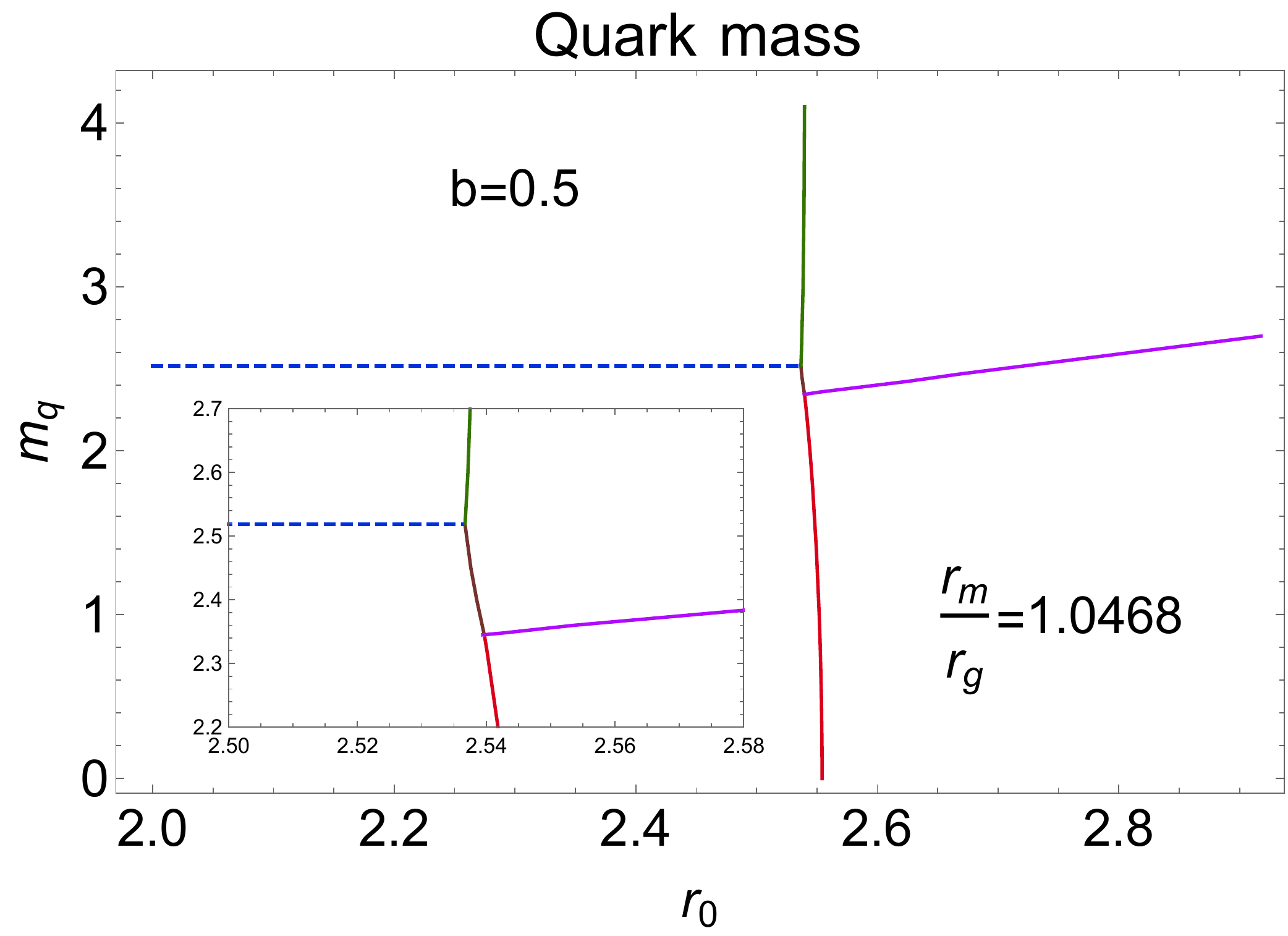}\label{PhaseDiagram1_a}}
    \subfigure[]{\includegraphics[scale=0.25]{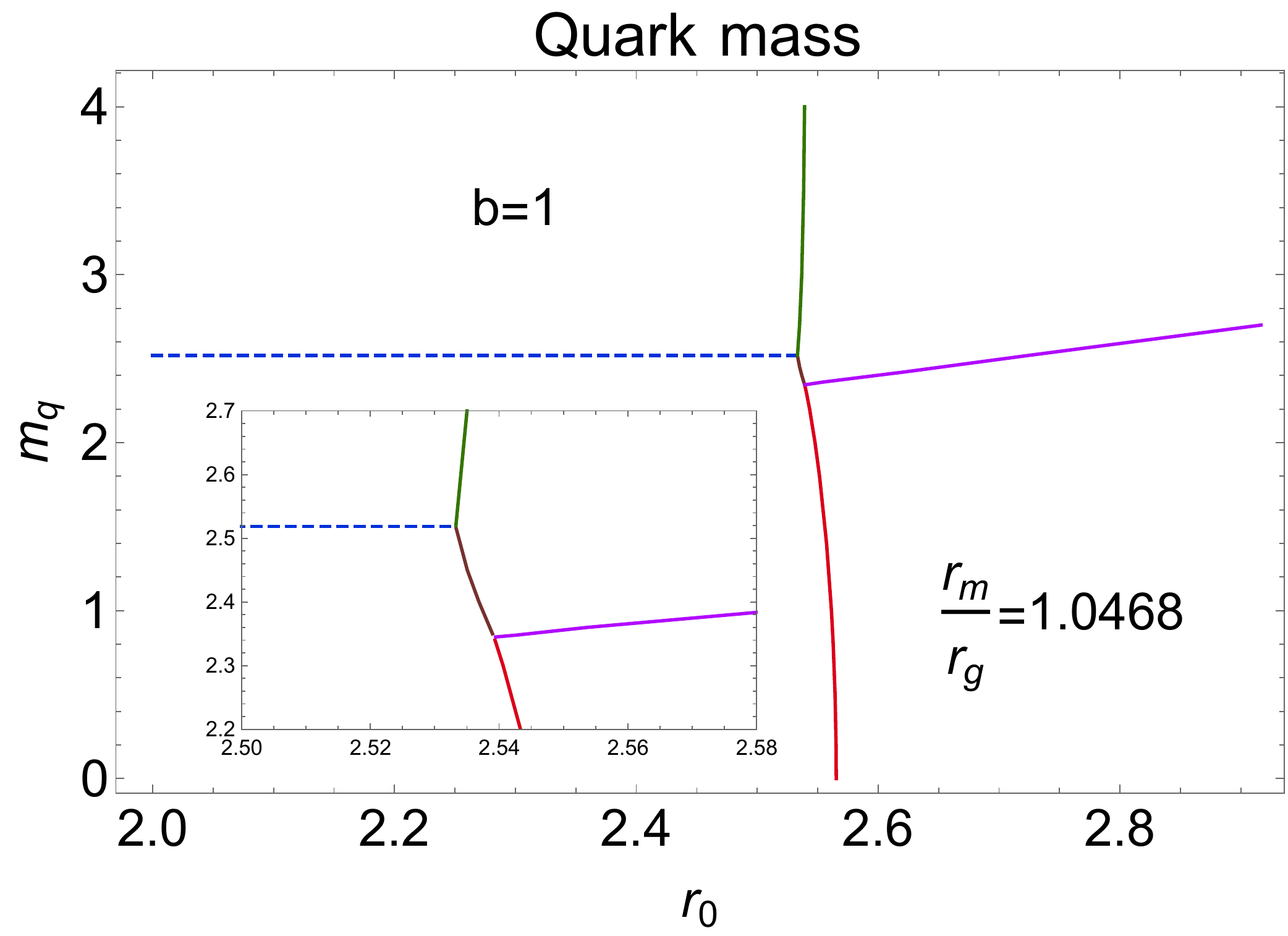}\label{PhaseDiagram1_b}}\\
    \subfigure[]{\includegraphics[scale=0.25]{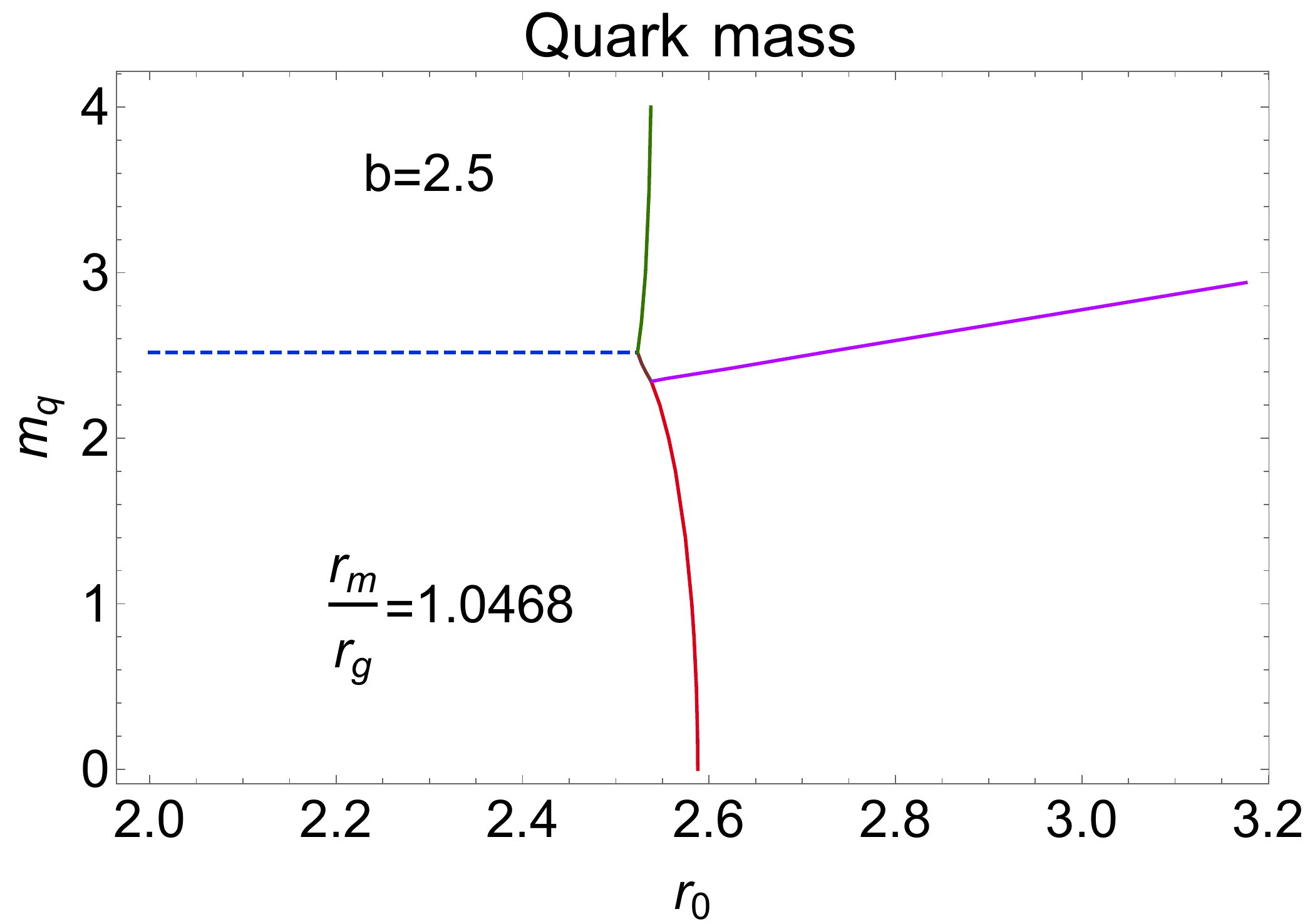}\label{PhaseDiagram1_c}}
    \subfigure[]{\includegraphics[scale=0.25]{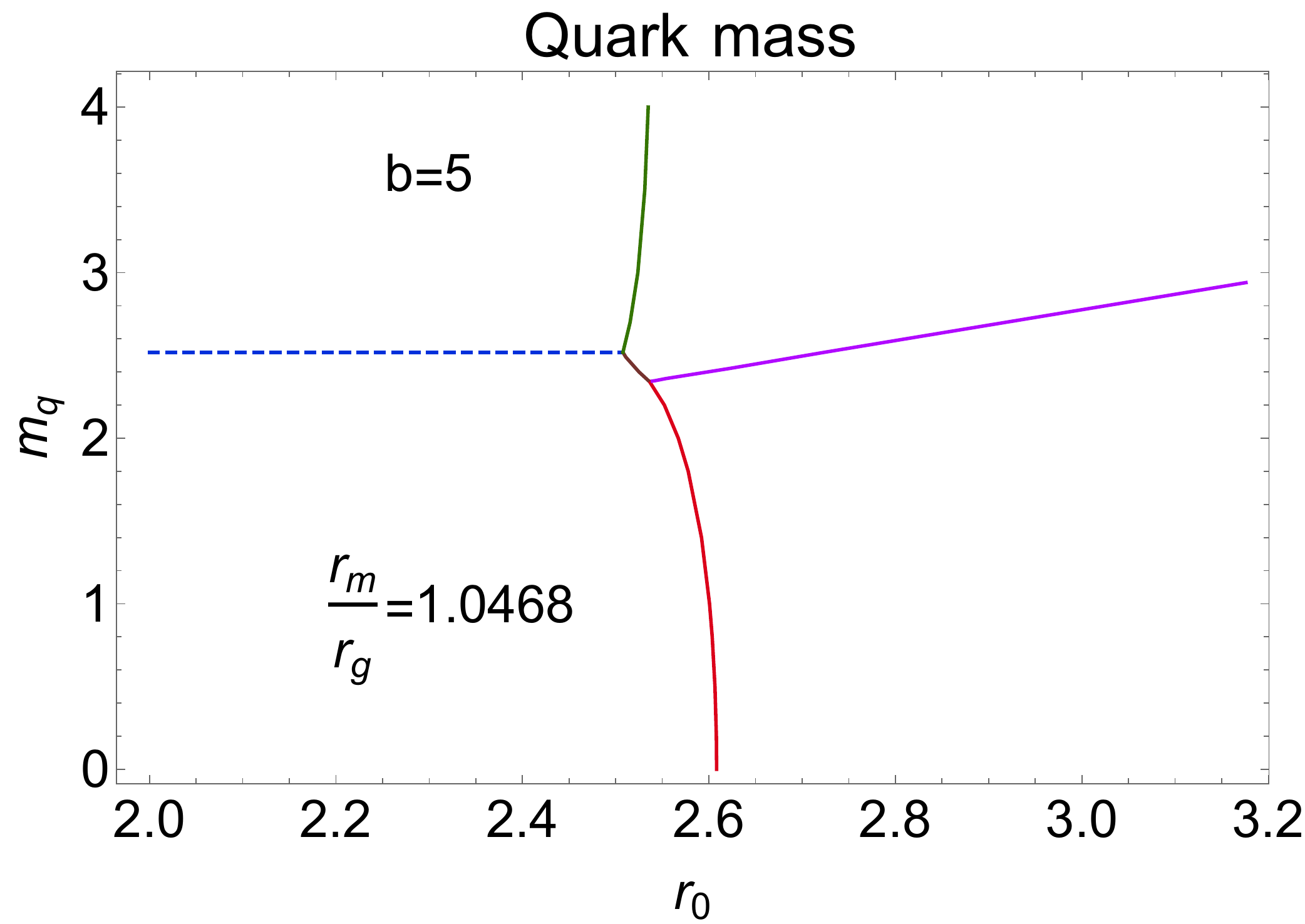}\label{PhaseDiagram1_d}}
    \caption{Phase Diagram for $\frac{r_m}{r_g}=1.0468$}
    \label{PhaseDiagram1}
\end{figure}

The considerations of the previous section assemble into the phase diagram shown in Fig:\eqref{PhaseDiagram1}. 
For small values of $m_q$ well below the horizontal blue dotted line, as the temperature is increased, we see that a transition occurs (red curve) from curved branes in the AdS background to a curved brane with a horizon on its world volume.  
For large values of $m_q$ though, full deconfinement occurs in two stages. The gluons first deconfine, represented by the changed background, but the quarks remain bound in mesons as indicated by the straightish branes in the black hole geometry. This transition is indicated by the green line in the phase diagram. 
At much higher temperatures, the brane ends on the horizon of the black hole which means that the brane modes will now become quasinormal with a temperature dependent imaginary part, as indicated by the magenta line. 
However, as shown in the inset, for values of $m_q$ just below the blue line, we have a new possibility where the curved branes in AdS give way to straight branes in the black hole background (brown curve). 

This produces two triple points (vertices in the figure). 
The change in the ground state from curved branes at low $m_q$ to straight branes at large $m_q$ is independent of temperature and the value of $b$ as it involves the comparison of two branes in the same background. Likewise, the magenta line is associated with the comparison of the branes in the black hole background and therefore is neither affected by $b$ nor by the ratio $\frac{r_m}{r_g}$.
As we vary $b$, the phase diagram is qualitatively unchanged, but for the changes in slopes, as shown in the various panels in Fig:\eqref{PhaseDiagram1}. We point out that there is no phase transition in the $r_g<r_0<r_m$ region as the ratio is $\frac{r_m}{r_g}<2^{\frac{1}{4}}$.

\begin{figure}[h]
    \centering
   \subfigure[]{\includegraphics[scale=0.25]{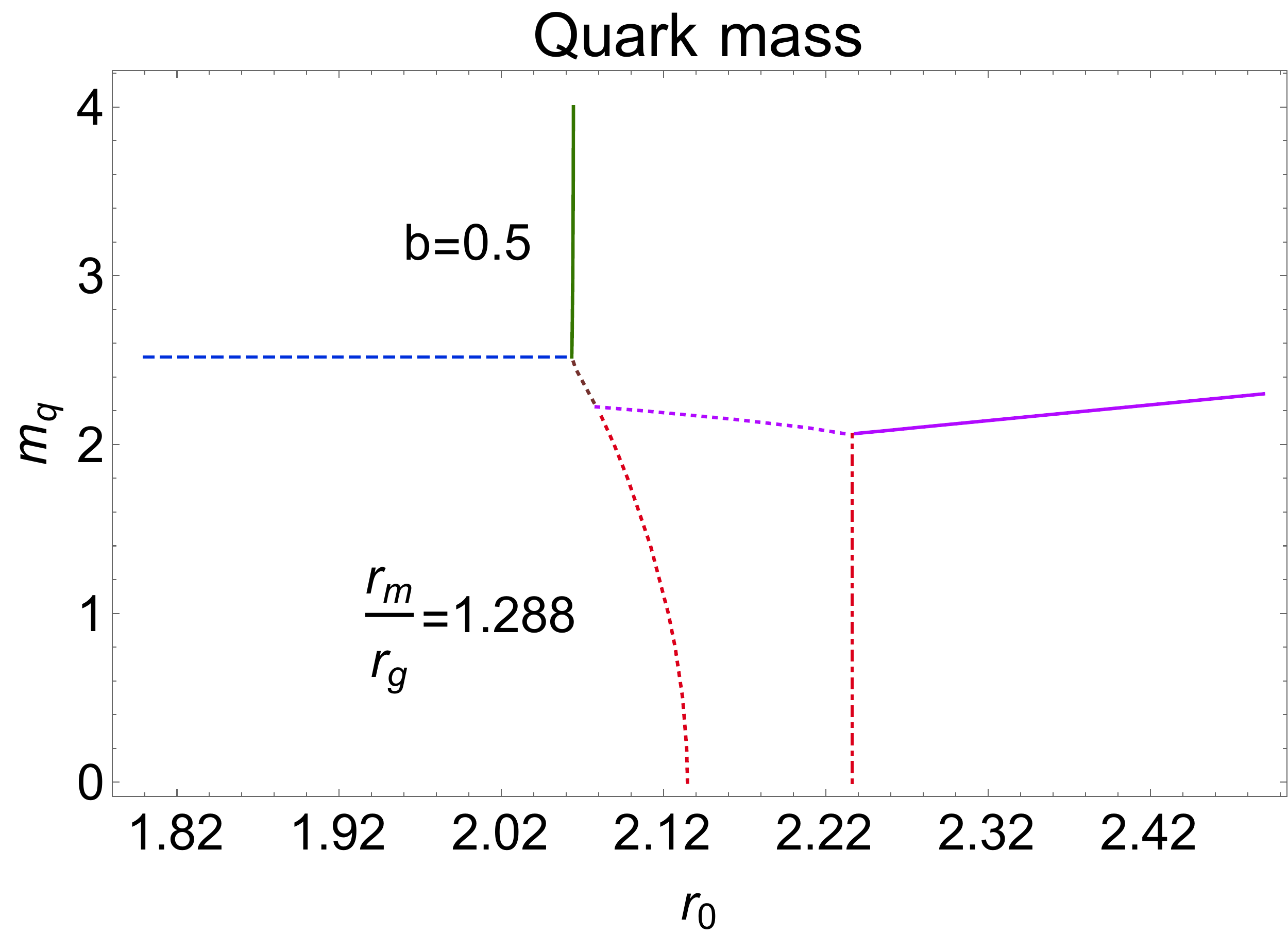}\label{fig:Intermediate_a}}
    \subfigure[]{\includegraphics[scale=0.25]{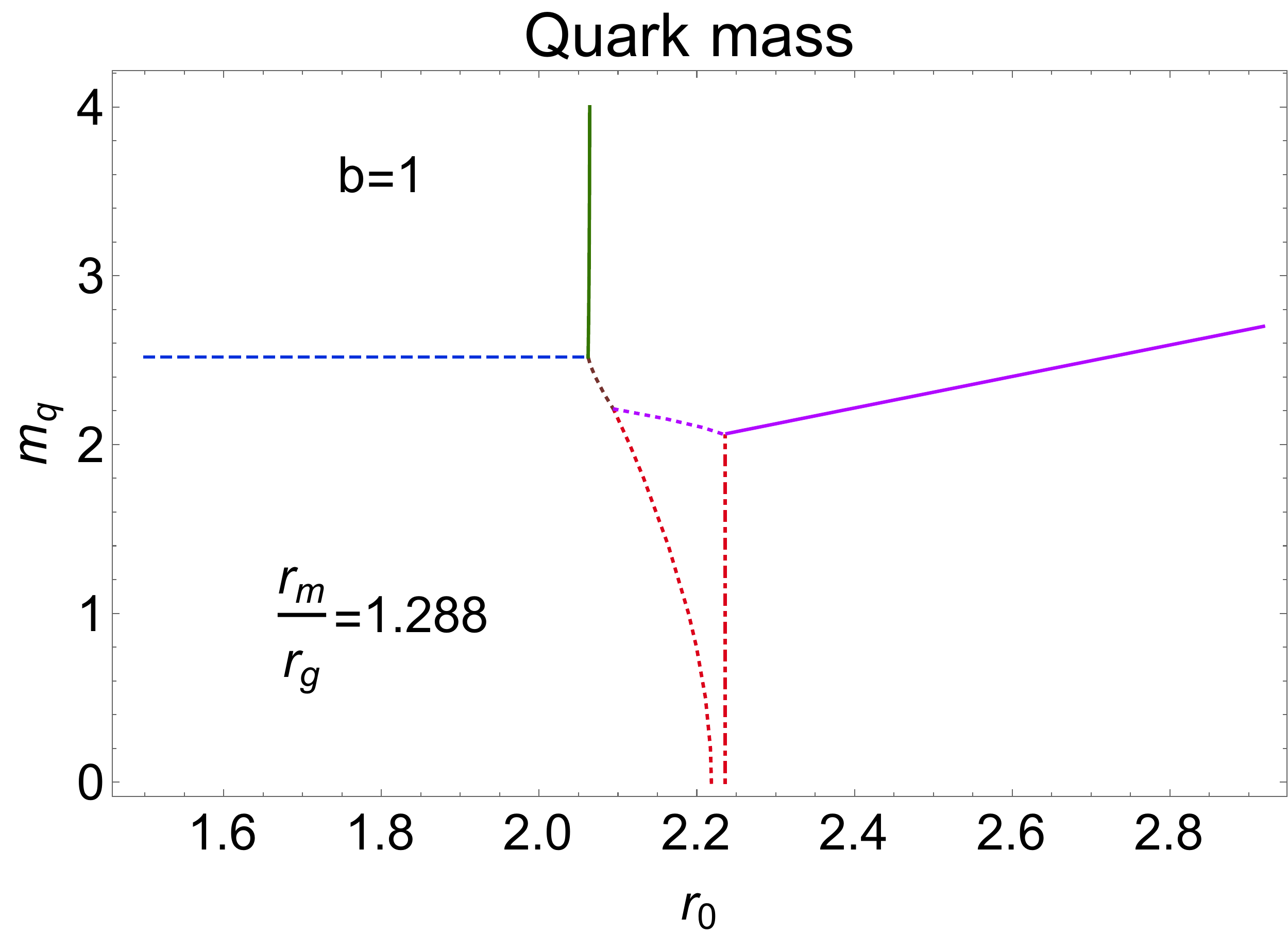}\label{fig:Intermediate_b}}\\
    \subfigure[]{\includegraphics[scale=0.25]{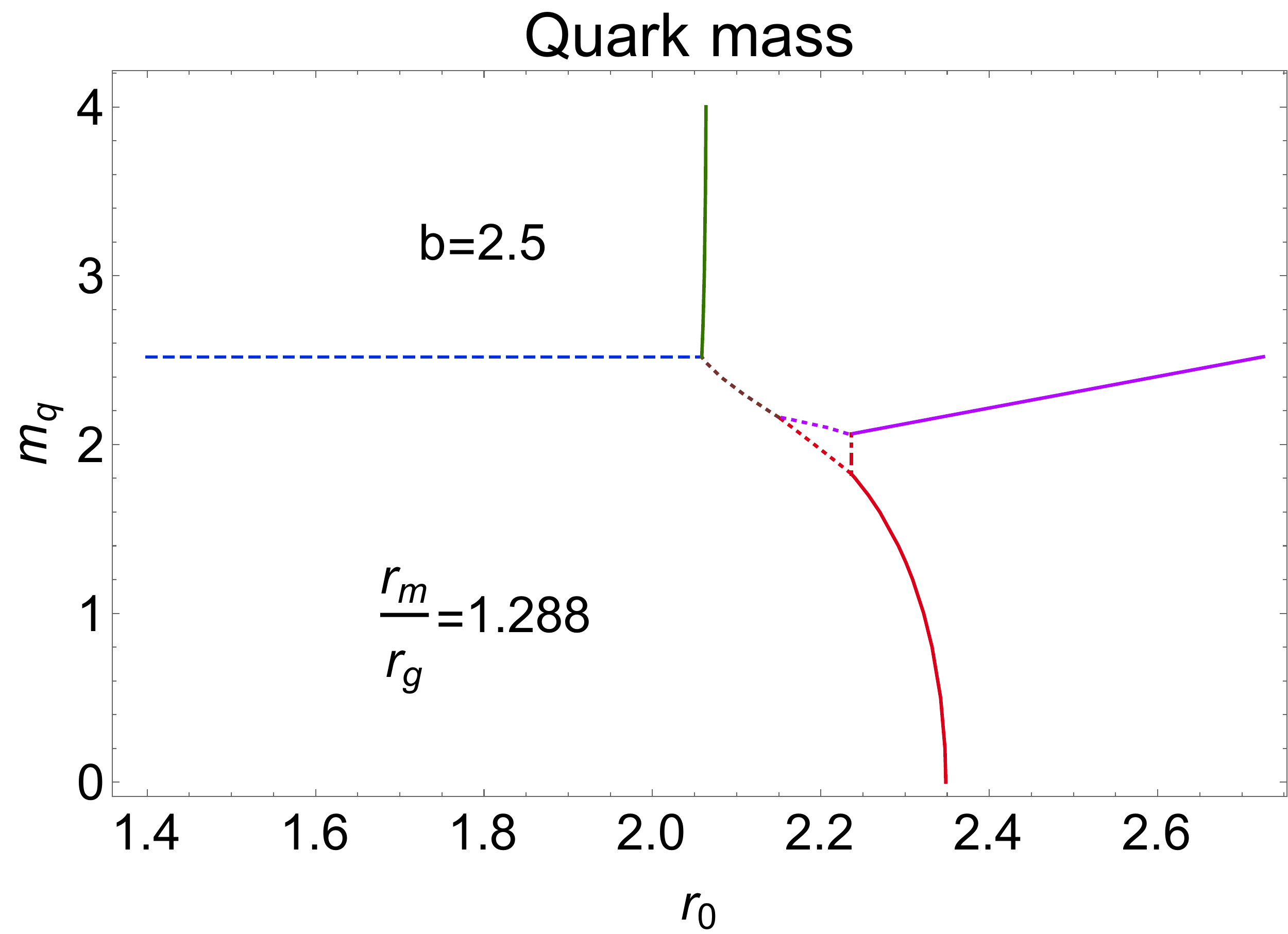}\label{fig:Intermediate_c}}
    \subfigure[]{\includegraphics[scale=0.25]{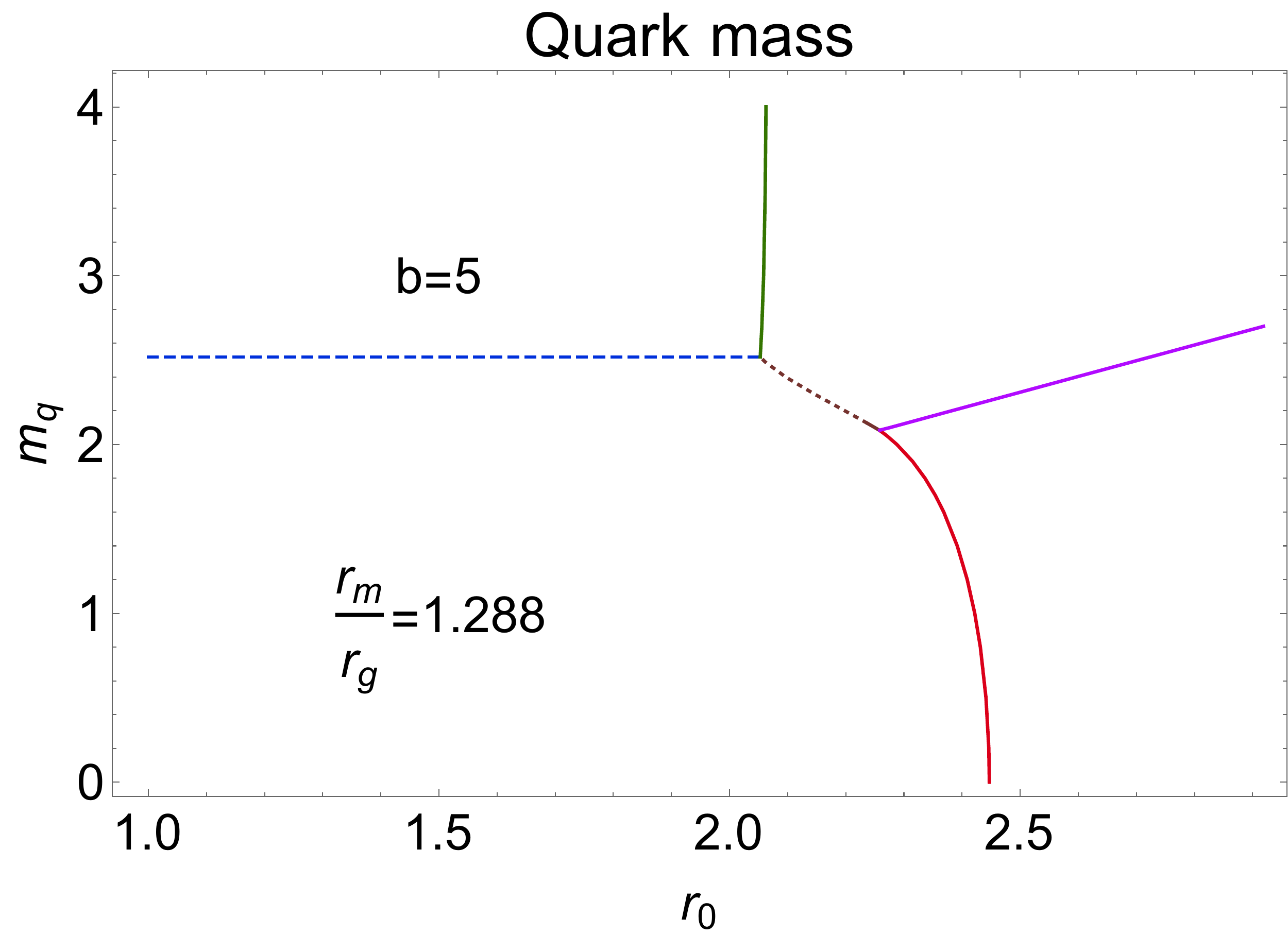}\label{fig:Intermediate_d}}
    \caption{Intermediate value $\frac{r_m}{r_g}=1.288$}
    \label{fig:Intermediate}
\end{figure}

For an intermediate value of the ratio ($\frac{r_m}{r_g}=1.288$), and for smaller $b$, we see new possibilities represented by dotted lines in the first three panels of Fig:\eqref{fig:Intermediate} in agreement with the analysis presented in Section \ref{section0qm} for zero quark mass. 
Firstly, for low quark masses, we have new transitions in the $r_g<r_0<r_m$ region. The red dotted line represents a transition between the branes ending on the cutoff surface in the AdS geometry and the branes ending on the cutoff surface in the black hole geometry (with the horizon behind the cutoff). Increasing the temperature further then leads to another first order transition, shown as the vertical dotted dashed red line (present only in the first two panels of Fig:\eqref{fig:Intermediate}). Once the horizon $r_0$ becomes greater than or equal to the brane IR-cutoff $r_m$, then a horizon can appear in the world volume of the D7-brane as well. The brown dotted line joining the first pair of vertices is a coexistence curve between branes ending on the cutoff surface in AdS and straightish branes in the black hole background. The dotted magenta line joining the second and third vertices represents a transition between curved branes ending on the cutoff and straightish branes - both in the black hole geometry. As $b$ increases, we see that the red lines draw closer and eventually merge (partially)  (third panel of Fig:\eqref{fig:Intermediate}) leading to the appearance of a new vertex (triple point). Increasing $b$ further removes the region in the middle range of temperature due to the coalescing of the two triple points.

We observe that the positions of the green line shift
towards lower temperatures as we increase $\frac{r_m}{r_g}.$ The locations of the vertices change with variations in both $b$ and $\frac{r_m}{r_g}$. 

\begin{figure}[h]
    \centering
    \subfigure[]{\includegraphics[scale=0.25]{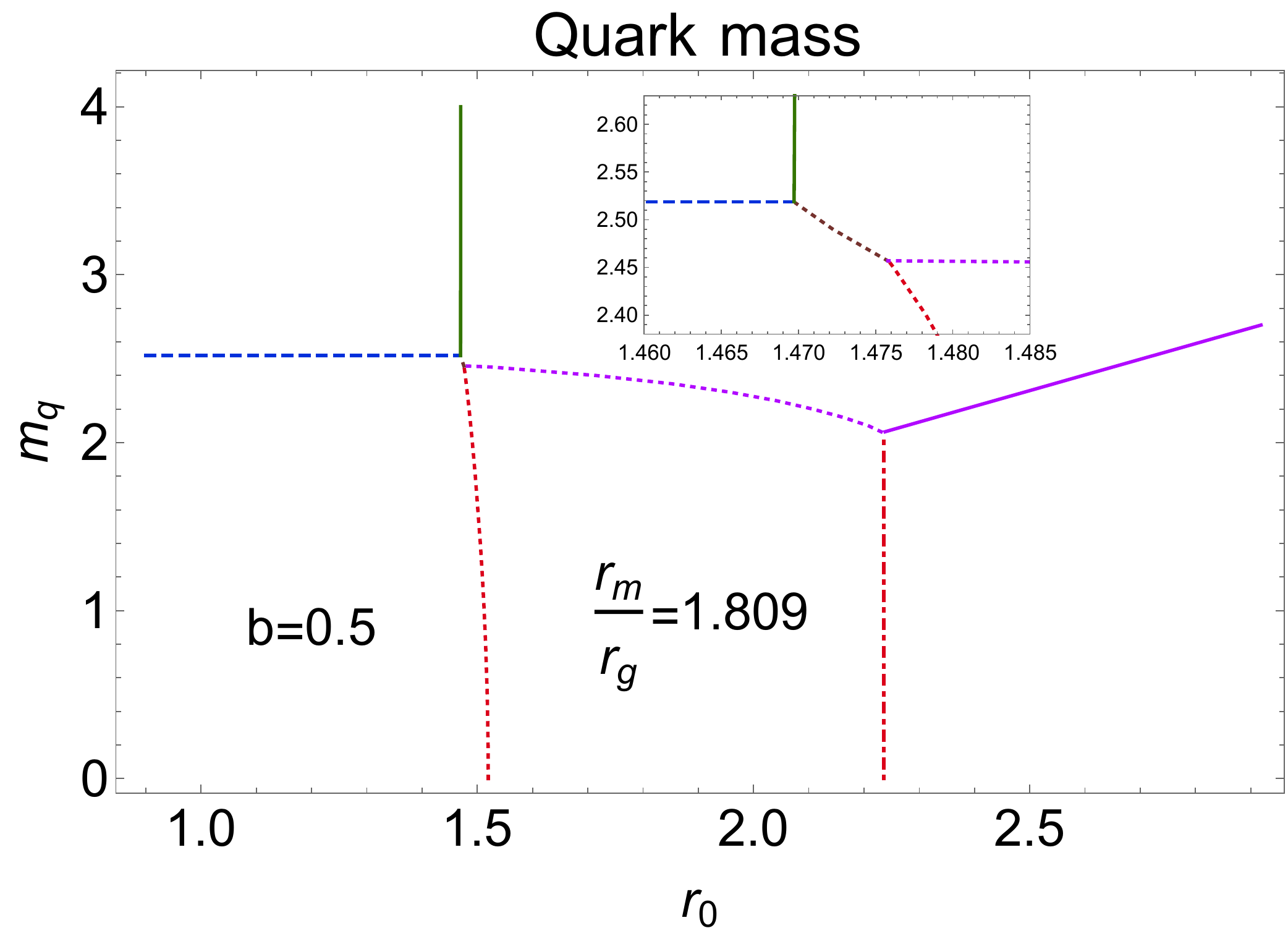}}
    \subfigure[]{\includegraphics[scale=0.25]{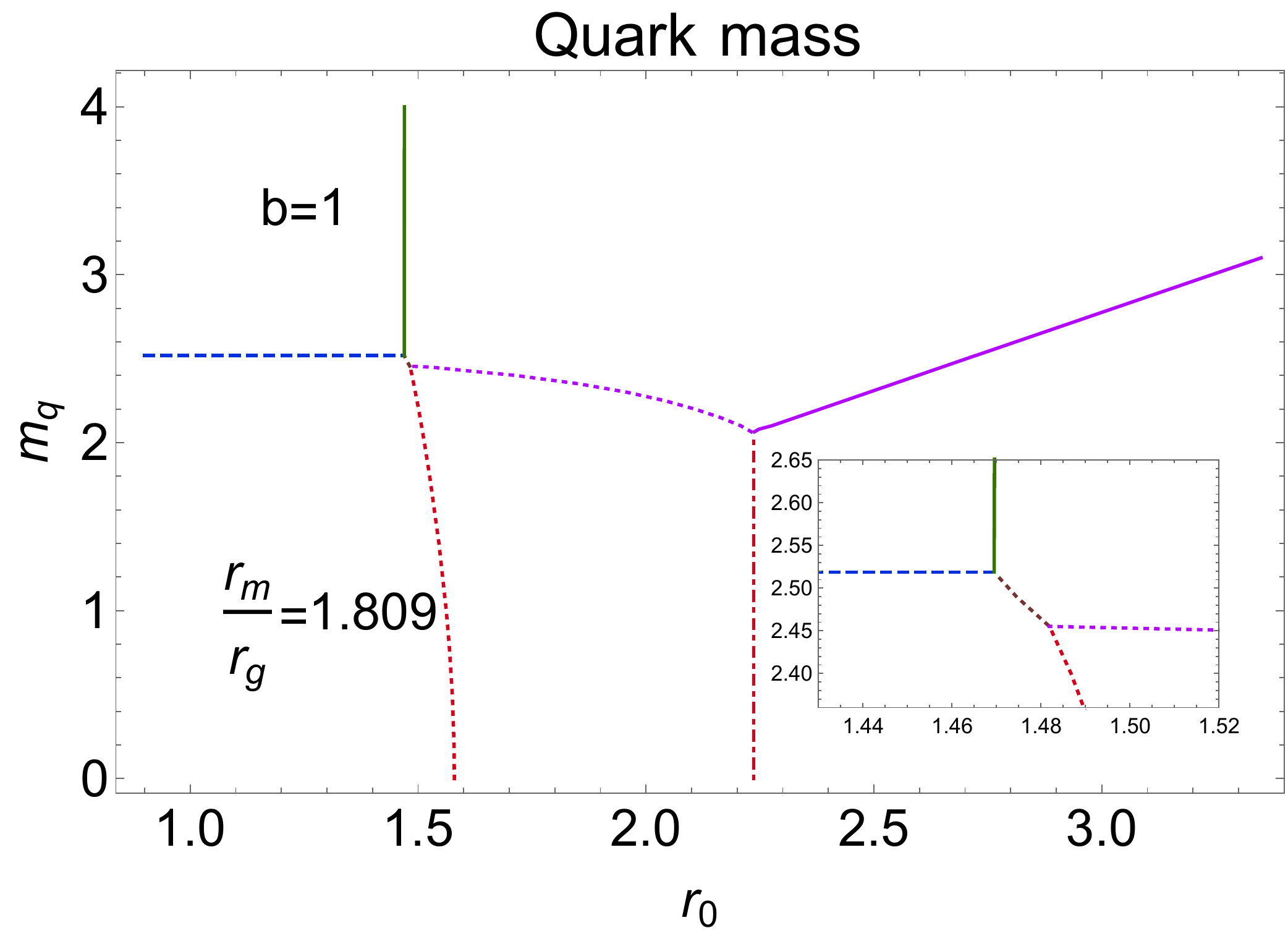}}\\
    \subfigure[]{\includegraphics[scale=0.25]{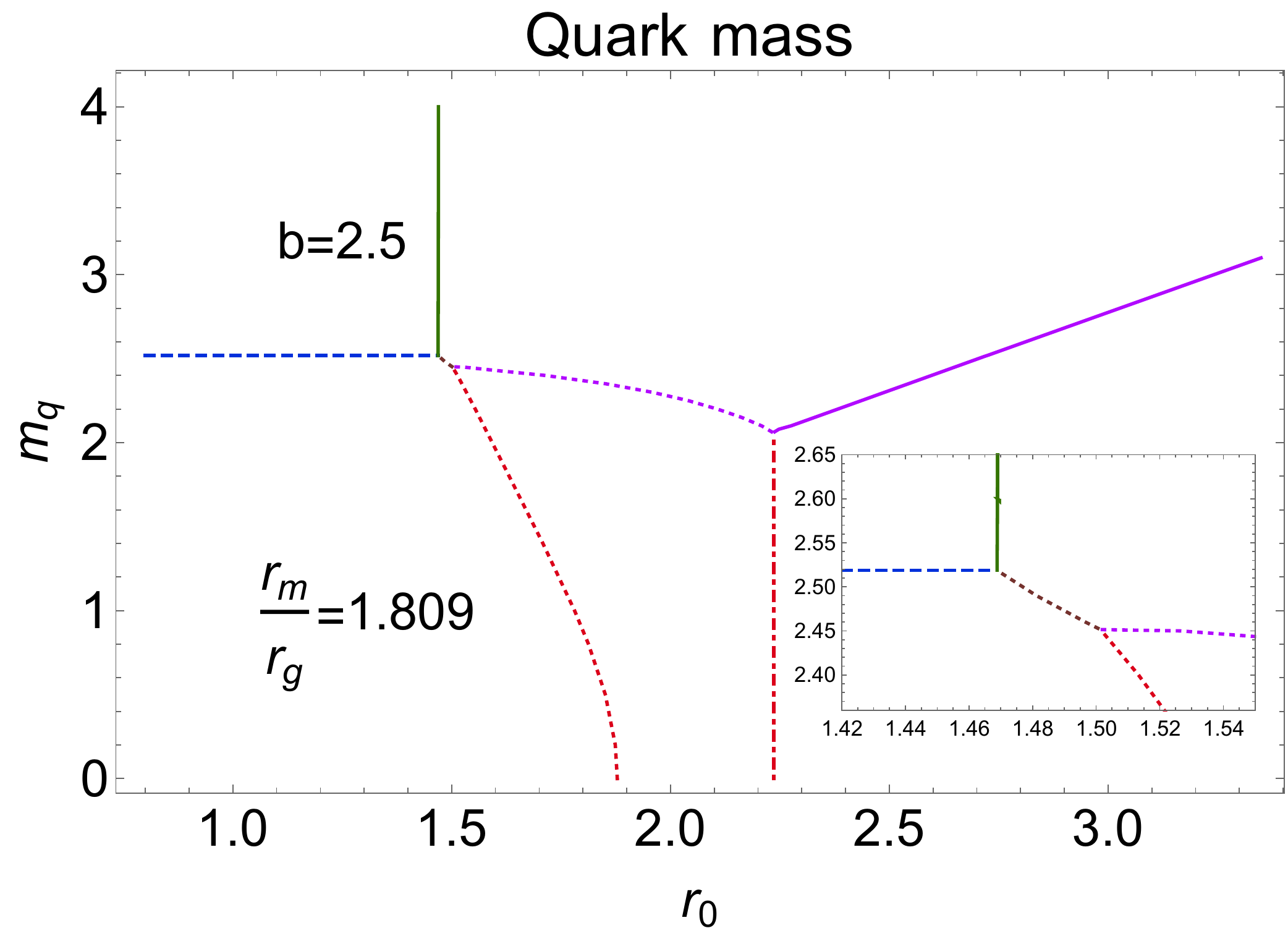}}
    \subfigure[]{\includegraphics[scale=0.25]{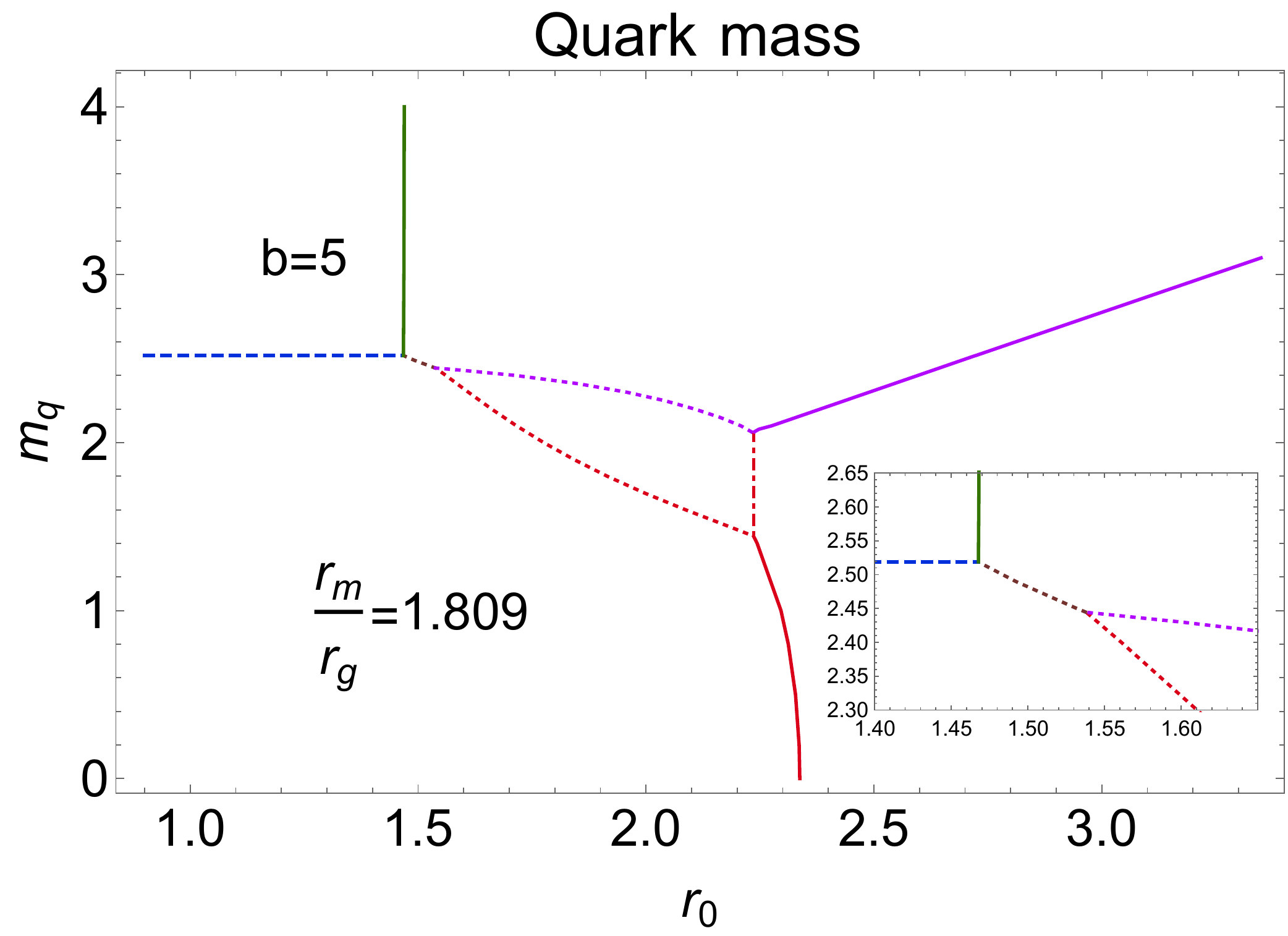}}
    \caption{Large Ratio $\frac{r_m}{r_g}=1.809$ }
    \label{fig:largeratio}
\end{figure}
Increasing the ratio ($\frac{r_m}{r_g}=1.809$) further leads to a qualitatively similar phase diagram as shown in Fig:\eqref{fig:largeratio}. For $b=0.5$ and $b=1$, we see that there are the same first order phase transitions in the region $r_g<r_0<r_m$ as they were in the intermediate value of the ratio. However, the temperature and masses at the phase transition have changed.

It can be checked that the phase diagrams are {\it invariant} when we rescale $r_g$, $r_m$, and $m_q$ by a common factor.

\section{Counterterms and Order Parameters} 

In the preceding sections, we obtained the phase diagram by considering the difference in the free energies of various classical solutions of the gravity equations. In considering differences, the divergent terms were canceled, and this was adequate for the purpose of finding minimum free energy configurations. 

The procedure of holographic renormalization gives a systematic method to remove divergences by adding specific counterterms to the gravitational action. The counterterms are constructed using various tensors made out of the bulk fields and have to be local in the bulk gravitational theory. This results in finite values for the free energy and other thermodynamic quantities (for a detailed introduction to the procedure of holographic renormalization, see \cite{deHaro:2000vlm}). 

If we view the holographic renormalization procedure merely as a renormalization scheme in the boundary theory, then we expect that the phase diagram is not modified.  However, we can then expect to obtain sensible  thermodynamic quantities and equations of state from the finite free energy so obtained. Further, the phase diagrams of the previous section can be restated in field theory terms provided we identify suitable order parameters. One obvious order parameter is the entropy. For a second order parameter, we will use the normalizable mode of the $y$ field which is the condensate. 

The total bulk action including the various counterterms in a full holographic treatment takes the form 
\be\label{S}
S_{ren}^{E}=S_{\Romannum{2}B}^E + S_{DBI}^E + S_{GH}^E + S_{ct1}^E + S_{ct2}^E+ S_{ct3}^E
\ee
where the individual terms are described briefly below.
\bea 
S_{\Romannum{2}B}^E&=&-\frac{1}{2\k_{10}^2}\int d^{10}x \sqrt{|g|}\left(R-\frac{1}{2.5 !}|F_5^E|^2\right)\\
S_{DBI}^E&=&N_f\m_7\int d^8\s\sqrt{det(P[g])}\\
S_{GH}^E&=&-\frac{1}{\k_{10}^2}\int d^{9}x \sqrt{\g}K \\
S_{ct1}^E&=&-\frac{1}{\k_{10}^2}\int d^{9}x \frac{\sqrt{\g}}{L}(1-p)\\
S_{ct2}^E&=&-N_f\m_7\int d^{7}\s \frac{\sqrt{\g_2}L}{4}\\
S_{ct3}^E&=&N_f\m_7\int d^{7}\s \frac{\sqrt{\g_2}y^2}{2r^2}
\eea 
The action has many terms which require some discussion, especially in this new context of cutoff geometries. For details pertaining to the D-brane counterterms, see \cite{Karch:2005ms}. 

\begin{itemize}

\item The Gibbons-Hawking term $S_{GH}$ is needed for the well-definedness of the variation principle because, at the UV-cutoff $r=\L$, we should not restrict derivatives of the bulk metric. 

This term is evaluated on the surface $r=\L$. $\g$ is the induced metric on the surface, $K$ is the trace of extrinsic curvature. The relative sign between $S_{\Romannum{2}B}$ and $S_{GH}$ is fixed using the direction of normal at UV boundary. 

\item 
Both $S_{\Romannum{2}B}$ and  $S_{GH}$ are separately divergent when evaluated on solutions. $S_{ct1}$ is a counterterm which cancels the above divergences as we take the limit $\L\to\infty$, $p=4$ is the $AdS_5$ boundary dimension.  

\item The D-brane action $S_{DBI}$ also produces divergences near UV boundary which are canceled by $S_{ct2}$ and $S_{ct3}$. In these terms, $\g_2$ is the brane metric on the cutoff surface. We can evaluate these terms either in terms of the $\r$ coordinates or in terms of the $r$ coordinates. They remove, entirely, the divergent term in the DBI action.

In this case, while the AdS action turns out to be independent of temperature, the black hole action includes up a finite piece coming from the counterterms.

\item
There is the possibility of including a finite counterterm $S_{ct4}$ that is required to ensure supersymmetry at zero temperature in AdS space. In our case, we are using a cutoff geometry and therefore the vanishing ground state energy is not a requirement. 
\be S_{ct4}=-\a\ y(\L)^4\ee
However, even if we include this term, it will not affect the phase diagram since it cancels when we compute differences at fixed quark mass $m_q=y(\L).$
\end{itemize}
Note that the counterterms do not depend on the details of the solutions in the interior of AdS, only the boundary values matter. The holographic renormalization scheme will differ from a minimal subtraction scheme by finite parts. 
Since there are two boundaries in the cut-off geometry, the IR cut-off can also contribute boundary terms with a sign opposite to the boundary term at UV. However, we propose that the IR-cutoff should not be treated as a boundary. Experimentally, we find that including the boundary contributions from the IR-cutoff removes even the high temperature phase transition of \cite{Herzog}. Thus, we propose that the IR-cutoff should not be regarded as a boundary. Rather, this surface should be joined smoothly to another geometry to produce a complete spacetime.  The IR-cutoff can be visualized as a surface which hides strong curvature/coupling effects. 
For our purposes, we imagine that the effects of this interior region are so as to determine the IR boundary values of the bulk fields in terms of physical quantities such as masses of the mesons/glueballs.

Before we discuss the situation, including D-branes, it is of some interest to ask what happens to the phase transition of \cite{Herzog} in the holographic renormalization scheme.
The various terms in the full action evaluate to be :
\bea 
\sqrt{-g}&=&L^2r^3\o^5\qquad 
R=0\qquad 
\frac{|F_5^E|^2}{2.5!}=\frac{8}{L^2}\\
\sqrt{\g}&=&r^4 L\sqrt{f}\o^5\qquad 
K=\frac{1}{2L\sqrt{f}}(8f+rf')
\eea 
As a result, we find that the free energy of the AdS bulk is $F=-a r_g ^4$ while the black hole free energy evaluates to $F=-a(r_g^4-\frac{r_0 ^4}{2})$ when $r_g>r_0$ and $F=-a\frac{r_0 ^4}{2}$ for $r_g<r_0$. The second term in the latter case arises from the temperature dependence in the counterterms.  Note that, in this scheme, the subleading terms in $\b'$ which are important in the hardwall model do not change anything since the total action is finite.  
Nevertheless, it is easy to see that the phase transition observed in \cite{Herzog} occurs at the same temperature $r_0=r_g2^{\frac14}.$

In the upcoming sections, using
\be 
\sqrt{-det(P[g])} =\o^3\r^3\sqrt{1+ y'^2} \: ;\quad
\sqrt{\g_2}=\frac{r\o^3\r^3}{L}
\ee 
\be
\sqrt{-det(P[g])} =\o^3\tilde\r^3\left(1-\frac{r_0^8}{16\xi^8}\right)\sqrt{1+\tilde y'^2} \: ;\quad
\sqrt{\g_2}=\frac{\o^3\tilde\r^3\xi}{L}\left(1-\frac{r_0^8}{16\xi^8}\right)
\ee
we obtain a finite contribution  to the free energy from the D-brane in the AdS and black hole geometry, respectively. It is to be noted that the $\sqrt{\g_2}$ terms are evaluated on the cutoff surface $r=\L.$

\subsection{Zero quark mass case}\label{0qmFreeEnergy}

Having discussed the various parts of our action \eqref{S}, we first evaluate the free energies for the simple case, $y=0$. In this case, all quantities can be determined analytically. However, it is necessary to carefully consider the subleading terms that come from the relation between $\L$ and $\tilde\r_{UV}$ \eqref{uv}. The UV-finite on-shell action can then be {\it directly} interpreted as the Helmholtz free energy $S=\b F$ by dropping the integral over the time circle. 

{\bf Thermal AdS:}\\
With $p=4$ since $r_m=\r_m$ for $y=0$, we get our free energy as:
\be\label{FEAds}
F_{AdS}=-a\left[r_g^4 +\frac{br_m ^4}{4}\right]
\ee
is actually independent of temperature. Thus, the entropy vanishes consistent with the absence of any horizons which could have been the repositories of said entropy. 

\textbf{Black hole:}

In the black hole background, for $r_g<r_0<r_m$, the bulk terminates at $r_0$ and the brane cuts off at $r_m$. We study three cases as mentioned in section \ref{section0qm}. These three cases are in the same order as in that section and we will verify that we get the same phases by using the renormalized free energy. 

For $r_0<r_g<r_m$, the bulk and brane cutoff at $r_g$ and $r_m$, respectively. The free energy is:
\be \label{FBH1} 
F_{BH1}=a\left[\frac{r_0^4}{2}-r_g^4 -\frac{b}{4}\left(\tilde\r_m^4+\frac{r_0^8}{16\tilde\r_m^4}
\right)\right].
\ee
Here the IR-cutoff evaluates to $\tilde \r_m=\sqrt{\frac{r_m^2+\sqrt{r_m^4-r_0^4}}{2}}$ for $y=0$. 
\be\label{FEBHD1} 
F_{BH1}=a\left[\frac{r_0^4}{2}-r_g^4 -\frac{b}{4}\left(r_m^4-\frac{r_0^4}{2}\right)\right].
\ee
The difference \eqref{FEBHD1}-\eqref{FEAds},
\be
F_{BH1}-F_{AdS}=a\frac{r_0^4}{2}\left[1+\frac{b}{4}\right]\\
\ee
is always positive and we see that there is no phase transition. In this region, the Thermal AdS embedding has a lower free energy.

For $r_g<r_0<r_m$, the bulk and the brane have cutoff at the black horizon $r_0$ and $r_m$, respectively. Thus, the free energy for branes in the black hole background evaluates to: 
\be\label{FBHD2} 
F_{BH2}=-a\left[\frac{r_0^4}{2}+\frac{b}{4}\left(r_m^4-\frac{r_0^4}{2}\right)\right]
\ee  
Taking the difference \eqref{FBHD2}-\eqref{FEAds},
\be
F_{BH2}-F_{AdS}=a\left[r_g^4+r_0^4\left(\frac{b}{8}-\frac{1}{2}\right)\right]
\ee 
In this region, there is the phase transition as described in Section \ref{section0qm}. The conditions for the phase transition to happen are also the same. When conditions are satisfied and phase transitions occur, we get the same critical temperature $T_{cg}$, at which the gluons deconfine while the quarks remain bound.

For $r_g<r_m<r_0$, the bulk and the brane has a cutoff at the black horizon. Thus, the action evaluates to: 
\be\label{FEBHD3} 
F_{BH3}=-a\frac{r_0^4}{2}\left[1+\frac{b}{4}\right]
\ee  
Taking the difference \eqref{FEBHD3}-\eqref{FEAds},
\be
F_{BH3}-F_{AdS}=a\left[r_g^4+\frac{b}{4}r_m^4-\frac{r_0^4}{2}\left(1+\frac{b}{4}\right)\right]
\ee  
For $ b<4(1-2\frac{r_g^4}{r_m^4})$, this difference is always negative. The black hole has a lower free energy density. Thus, as soon as $r_0\geq r_m$, the quarks are freed. The temperature is denoted as $T_{cq}$. Somewhat remarkably, we see that the differences are identical to the analysis in Section \ref{section0qm} provided we omit the overall $\b$ factor.

For zero quark mass, we conclude that if the conditions given in Section \ref{section0qm} are satisfied we see the phase transition in the region $r_g<r_0<r_m$. The critical temperature is $T_{cg}$ for the deconfinement of the gluon and $T_{cq}$ for quark deconfinement ($T_{cg}<T_{cq}$). If the conditions are not satisfied, we see that there is no phase transition in the $r_g<r_0<r_m$ region. However, in the $r_g<r_m<r_0$ region, the quarks and gluons deconfine together, and the phase transition temperature is $T_c$.

\subsection{Finite quark mass}
For finite quark masses, we have analytical solutions for the free energy in the case of AdS embeddings of the D-brane. 
\be
F=-a\left( r_g ^4-b\left(
-\frac{\rho_m^4}{4}\, {}_2F_1(-\frac{2}{3},\frac{1}{2};
\frac{1}{3};\frac{c^6}{\r_m ^6})\right)\right)\label{ADSFree}
\ee
In this expression, $c$ is determined in terms of the quark mass and the cutoff $r_m$ by the condition
$y^2(\r_m)+\r_m^2=r_m^2$. It is clear that the free energy is independent of the temperature $r_0$. Special cases are obtained when the condensate $c^3$ vanishes. 

For black hole embeddings with $r_g<r_0$, the total free energy is dominated by a $r_0^4$ term coming from the bulk
\be
F=-\frac{r_0 ^4}{2} - b\  f(r_0,m_q, r_m)
\ee
where the first term gives the contribution of the bulk gravity (or gluons).
We focus our attention on the second term that arises from the DBI part of the action. As we shall see, it allows us to make qualitative distinctions between the phases in terms of quark degrees of freedom. 

In the case where the branes end on the cutoff surface within the black hole background (i.e. $r_0>r_g$), we obtain a family of solutions with different starting slopes on the cutoff surface. Therefore, following the thermal AdS computations, the {\it condensate} is dynamically determined by identifying the minimum free energy solution. For such branes, based on the observations of Section \ref{solutions}, it is natural to consider the ratios $\frac{m_q}{r_m}$ and $\frac{r_0}{r_m}.$  This is because the latter is always less than unity for these branes, and further we have an upper limit for $\frac{m_q}{r_m}$ for such branes to exist. Thus, we may write a series for the Free energy 
\be 
F= F_0 \left(b,\frac{m_q}{r_m}\right)\ r_m ^4+F_2 \left(b,\frac{m_q}{r_m}\right)\ r_0 ^2\  r_m ^2+F_4 \left(b,\frac{m_q}{r_m}\right)\ r_0 ^4 + F_6 \frac{r_0 ^6}{r_m^2}+...
\ee
\begin{figure}[ht]
    \centering
    \subfigure[D7-brane contribution]{\includegraphics[scale=0.2]{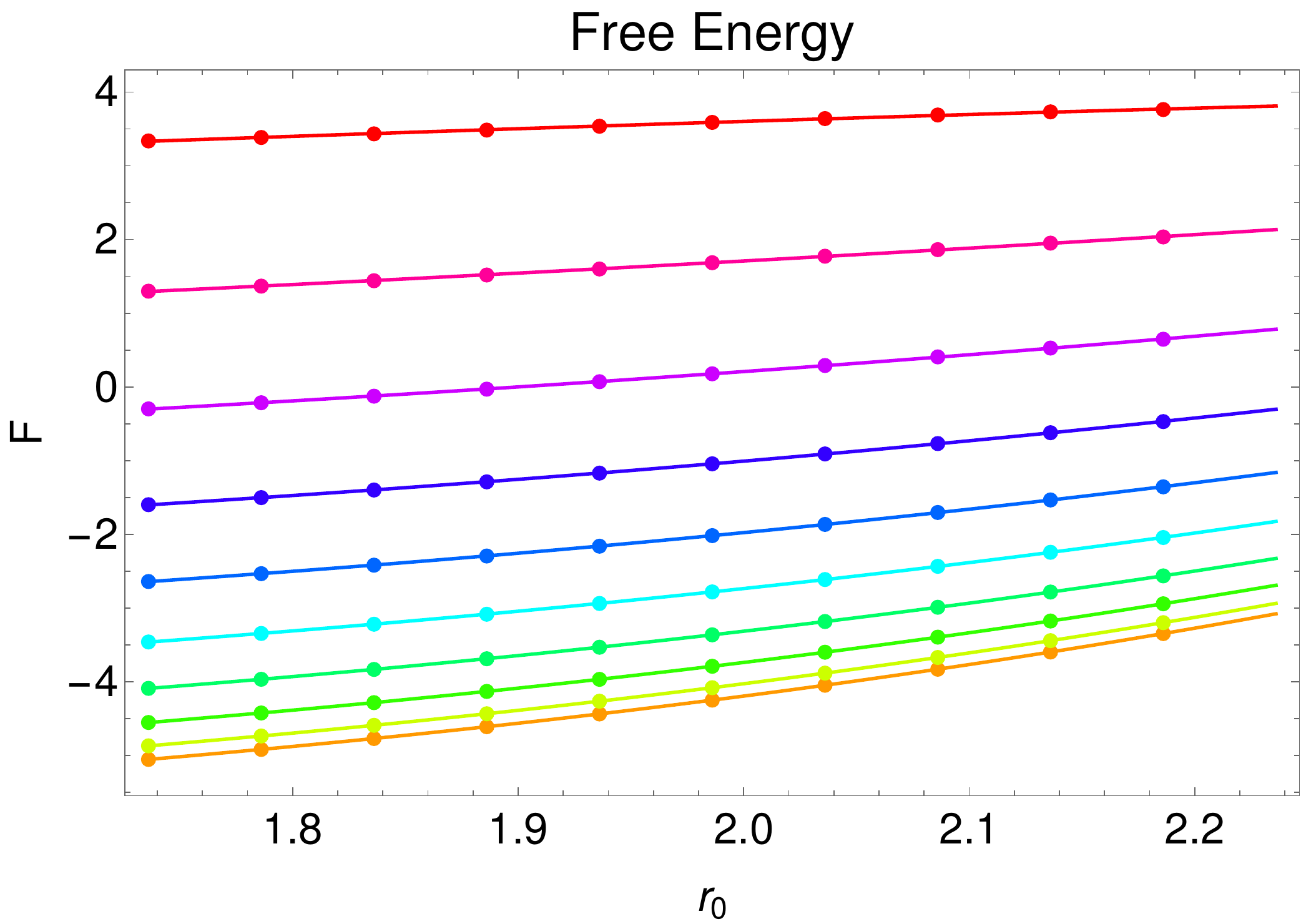}}
    \subfigure[Total free energy]{\includegraphics[scale=0.2]{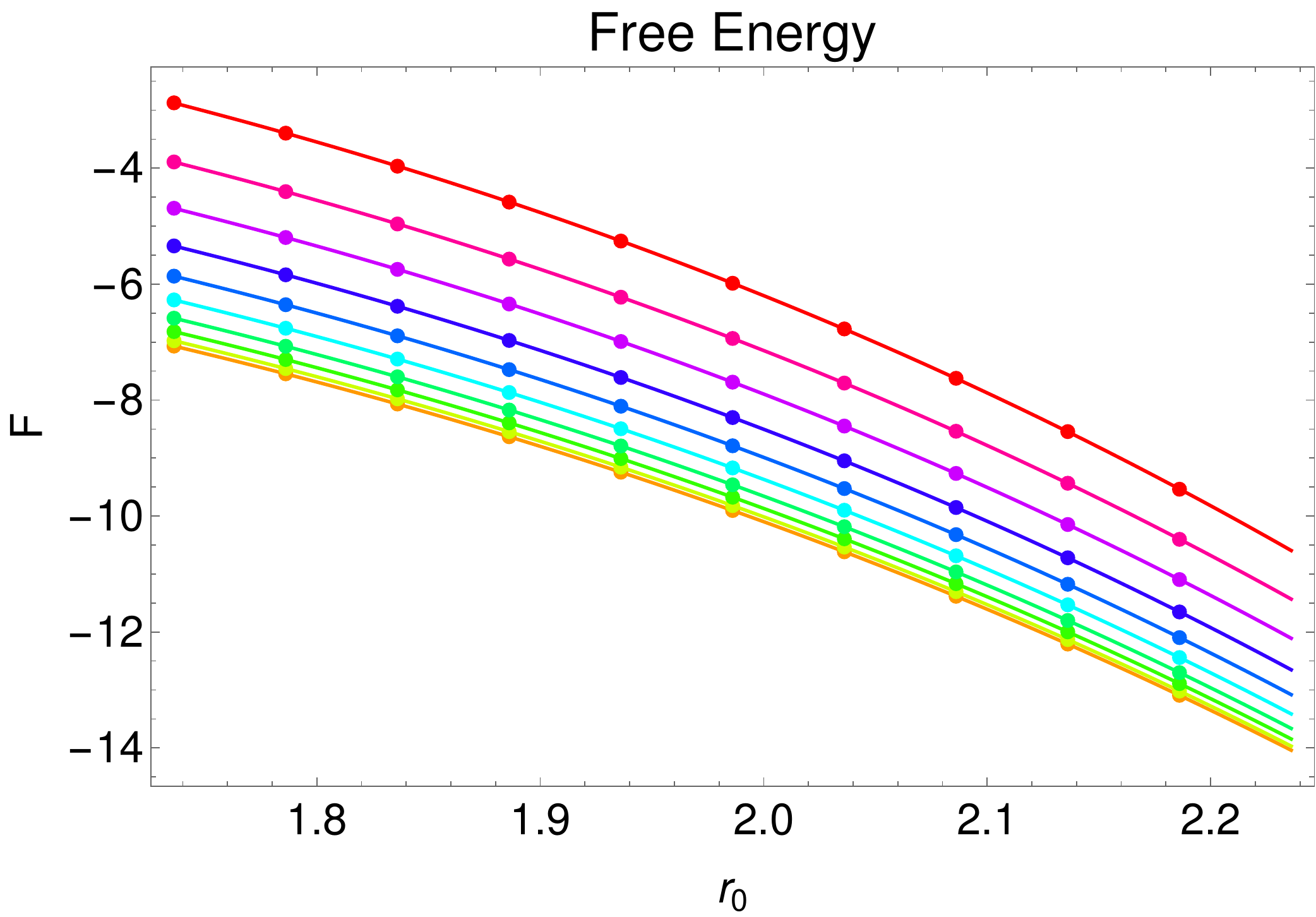}}
    \caption{Free Energy of cutoff branes in black hole background.}
    \label{fig:freeBHC}
\end{figure}
We have not studied these fitting functions in detail in this work. However, Fig:\eqref{fig:freeBHC} shows the numerically determined free energies (dots) shown with the best fit polynomials determined by the above logic. For these cutoff branes, the DBI contribution is seen to be increasing in temperature. However, the total free energy decreases with increasing temperature since the bulk contribution contributes negatively.

Since the branes that end on the horizon are not straight, we have a nonzero condensate, but it is not an independent parameter being determined by the condition $\tilde\r y'=y$ at the horizon. Consequently, the only independent dimensionful parameters are $m_q$ and $r_0$.  Hence, the free energy can be fitted to a high temperature series of the form 
\be 
F = F(b,m_q/r_0) r_0 ^4=F_4(b)r_0 ^4+F_2(b)r_0 ^2m_q ^2+F_0(b)m_q ^4+F_{-2}(b) \frac{m_q ^6}{r_0 ^2}...
\ee
Odd powers have been omitted by noting that the action has a discrete symmetry under $y\to -y$.  
\begin{figure}[ht]
    \centering
    \includegraphics[scale=0.2]{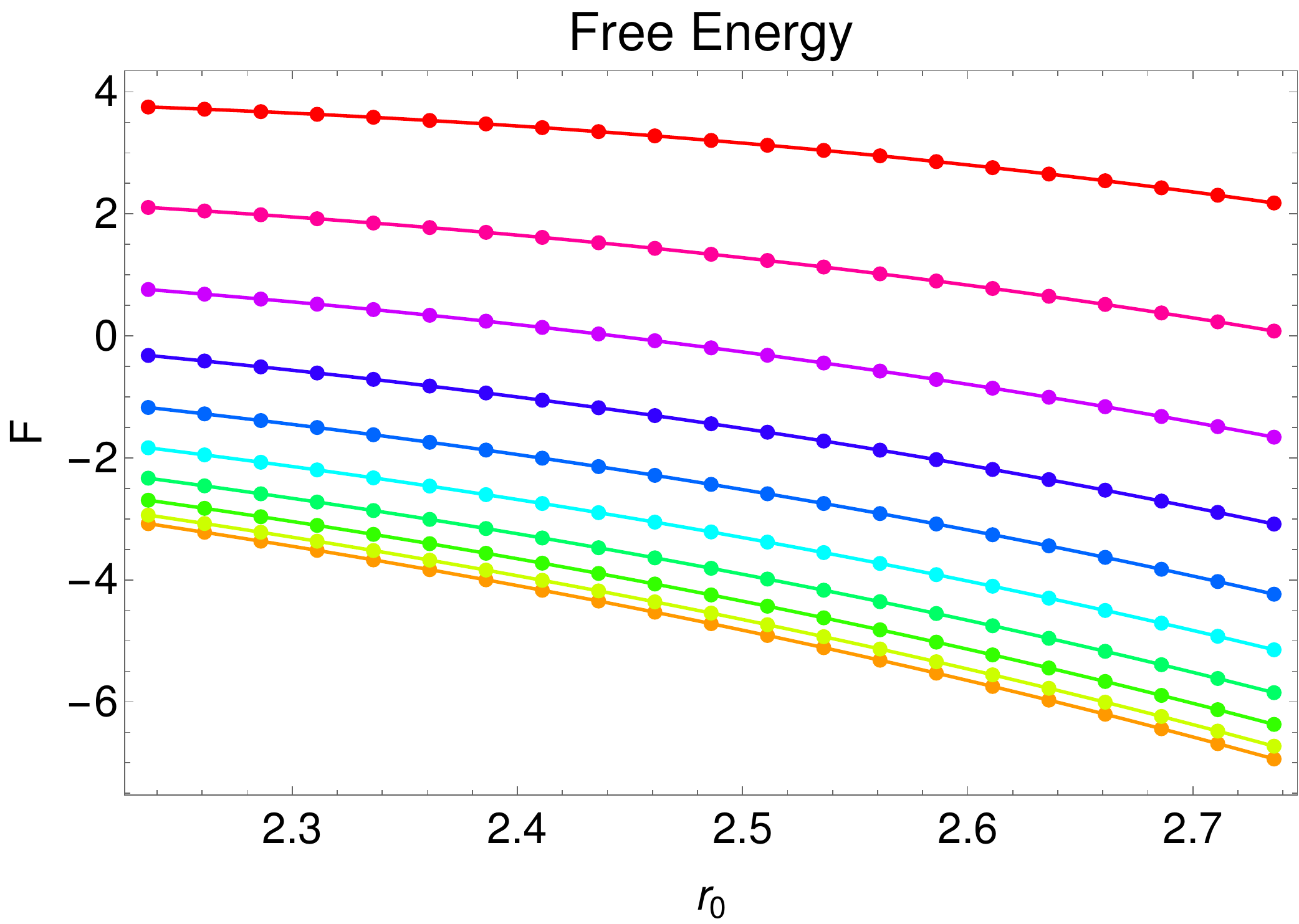}
    \caption{The free energy of D7-branes ending on the horizon}
    \label{fig:freebhh}
\end{figure}
Using the numerically determined free energy, we find that retaining up to quartic order produces an excellent fit, with the coefficients $F_0\approx \frac{b}{8}(\frac{5}{4}m_q^4+r_m^2 m_q^2)$,  $F_{2}=\frac{b}{8}$ and $F_4=-\frac{1}{2}(1+\frac{b}{4})$. In contrast to the cutoff branes, the free energy of these decrease with increasing temperature. We draw attention to the observation that the quadratic term in temperature is also proportional to $m_q^2$.

Finally, we have the straightish branes where the $S_3$ in the world volume shrinks to zero size. In this case, we have $m_q,r_0$ as the independent dimensionful parameters. Due to the gravitational potential, these branes are also not straight and therefore we have a condensate. However, the condensate is determined in terms of $m_q,r_0$ by the zero slope condition at $\tilde\r=0.$ 
\begin{figure}[ht]
    \centering
    \includegraphics[scale=0.2]{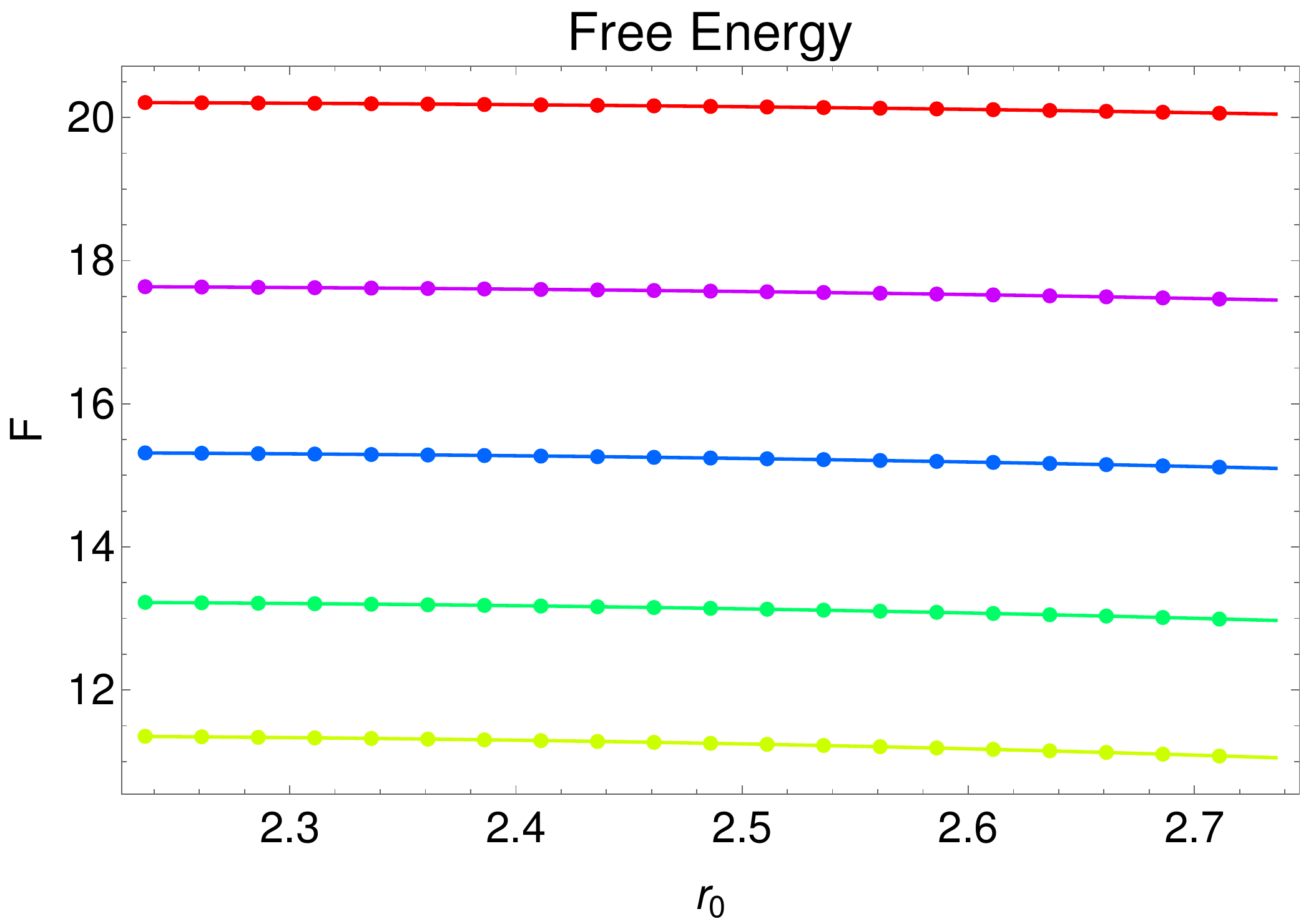}
    \caption{Free Energy of straightish branes in black hole background}
    \label{fig:freeBHS}
\end{figure}
As can be seen from Fig:\eqref{fig:freeBHS} the free energy of the DBI part depends only mildly on the temperature. For large enough quark masses the free energy contribution from the DBI is similar to that of the straight branes in the AdS-background. 

\subsection{Entropy}

In this section, we study the entropy of the various phases. There are two questions of interest. Firstly, we are interested in using entropy as an order parameter in the phase diagrams. Secondly, in holographic descriptions entropy is associated with the appearance of horizons. In our case, we have the possibilities of horizons appearing on the world volume of D-branes as well as in the bulk geometry. Therefore, a second point of interest is whether world volume horizons of branes contribute to the entropy. 

We start by noting that once we include counterterms, the free energies in the AdS background are finite and independent of temperature. Therefore, the entropy which will be determined as 
$S=-\frac{\del F}{\del T}$ vanishes for these configurations. 

\begin{figure}[ht]
    \centering
    \includegraphics[scale=0.2]{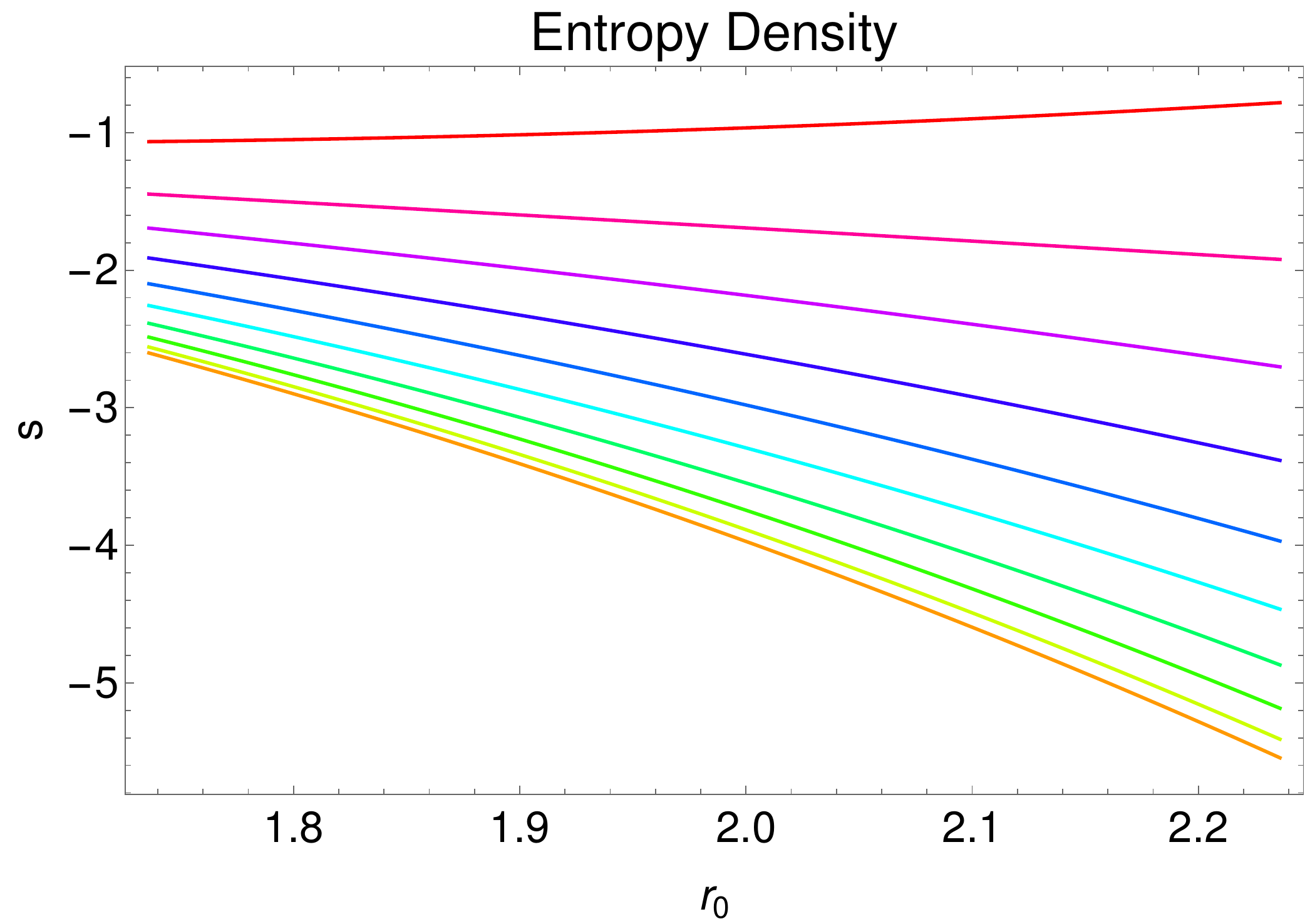}
    \caption{Entropy of cutoff branes in black hole geometry}
    \label{fig:SbhC}
\end{figure}

In the black hole geometry, at zero quark mass, we can have branes ending on the cutoff surface (when $r_0<r_m$). For these branes, the entropy density is obtained by differentiating the free energy of the preceding section 
\bea
s_2 &=& \frac{N_c^2\p^2 T^3}{2}\left(1-\frac{\l N_f}{16\p^2 N_c}\right)\qquad ;\quad T_{cg}<T<T_{cq}
\eea 
Since $\frac{\l N_f}{16\p^2 N_c}<1$ entropy density is always positive and entropy is an increasing function of temperature. However, surprisingly, we have a negative brane contribution to the entropy even though the branes do not have a horizon in their world volume. 
This negative contribution continues to non-zero quark masses as shown in Fig:\eqref{fig:SbhC}. Perhaps in this situation, the D7-brane behaves like a gas of mesons instead of a gas of fermions.  
We hasten to point out that even though the DBI contribution to the total entropy is negative, the total entropy (including the bulk) does turn out to be positive (this is easily seen from the total free energy curve \eqref{fig:freeBHC}). This suppression of the quarks occurs in the phase between the dotted red line and the quark deconfinement in the phase diagrams of the previous section. These phases occur only for sufficiently small $b$ which prevents the {\it total} entropy from turning negative. 

\begin{figure}[ht]
    \centering
    \includegraphics[scale=0.2]{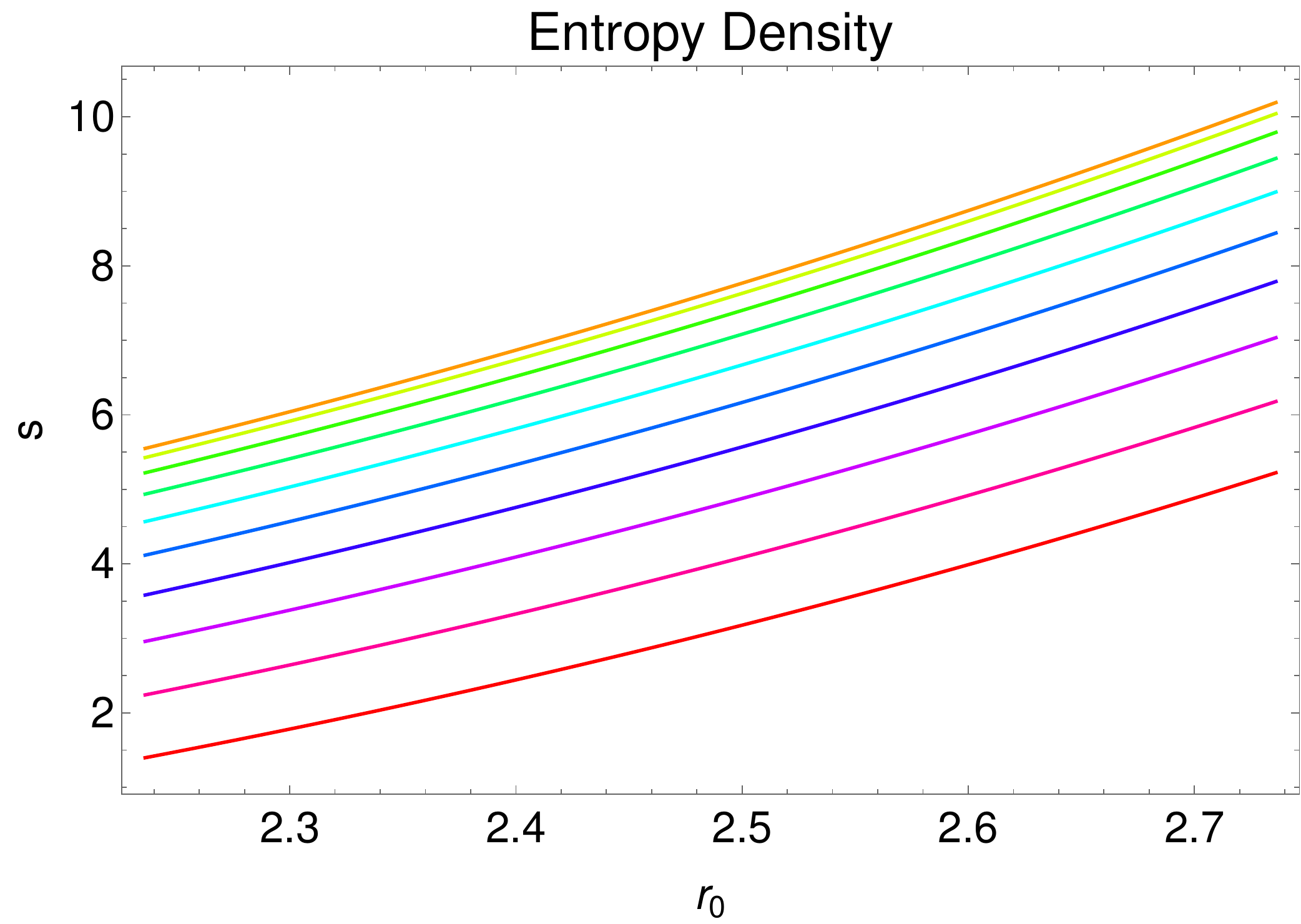}
    \caption{Entropy of branes ending on the horizon}
    \label{fig:sBHH}
\end{figure}

From our free energy computation, we know that at high temperature, branes that end on the horizon are preferred. For these branes, the entropy density is 
\bea 
s_3 &=& \frac{N_c^2 \p^2 T^3}{2}\left(1+\frac{\l N_f}{16\p^2 N_c}\right) \qquad ;\quad T_{c}<T
\eea
In a similar manner, we can find the entropy density for the finite quark masses. Fig:\eqref{fig:sBHH} shows entropy density as a function of temperature for various quark masses. These curves are presented in increasing order of quark mass, with yellow being the lowest.    
From the fitting form of the free energy given in the previous section, the entropy density is evaluated to be 
\be
s_{BH}=\frac{N_c^2 \p^2 T^3}{2}\left(1+\frac{ N_f}{4 N_c}\left(\frac{\l}{4\p^2}-\frac{M_q^2}{T^2}\right)\right) 
\ee
which includes both bulk and brane contributions. Different from the zero mass case is the presence of the linear term in entropy which will translate into a linear term in specific heat. Note that this term is also proportional to $M_q^2$ and that the entire free energy is independent of $r_m$, which we interpret to mean the absence of any meson contribution.

For the straightish branes, the contribution coming from the DBI to the entropy is negligible, as is clear from Fig:\eqref{fig:freeBHS}.  

In summary, the AdS phases are characterized by zero entropy. In the black hole background, we always have net positive positive entropy. However, the straightish branes for large quark masses hardly contribute to the entropy. 

\subsection{Condensate}

From the viewpoint of the open strings, the field $y$ is on par with the gauge fields $A_\m$ that live on the brane world volume. For instance, under T-duality, these two transmute into each other. Thus, we can identify the normalizable and non-normalizable modes of the y-field by comparing with those of the gauge field. For a gauge field, near the boundary $\r=\L$ of AdS, a Frobenius series analysis gives
\be
A_0(\r,x)=\m(x)+...+q(x) \r ^{2-d}+...
\ee
leading to the interpretation of $\m$ as the chemical potential and $q$ as the associated expectation value of the charge density operator. The Frobenius series for $y$ \eqref{y'}, takes the form 
\be
y(\r,x)=m_q(x)+...-\frac{c^3}{2\r^2}
\ee
and so by analogy, we see that the constant value of $y$ - the quark mass $m_q$ is the non-normalizable mode and the constant $c^3$ multiplying $\r^{-2}$ is the vacuum expectation value of a dimension 3-operator in the field theory. This can be regarded as the expectation value of a fermion bilinear \cite{Babington} in 3+1-dimensions. However, this should not be regarded as being related to the {\it chiral} condensate because the D7-branes do not introduce {\it chiral} fermions.

For AdS embeddings, identifying the lowest free energy branes fixes the slope $y'$ on the cutoff surface $r_m>r_0$. However, this lowest free energy configuration is the same for any temperature since upon adding counterterms, free energy is independent of temperature. Thus, for a given value of the quark mass $m_q$, the condensate is constant in the AdS phases. 

\begin{figure}[ht]
    \centering
    \subfigure{\includegraphics[scale=0.2]{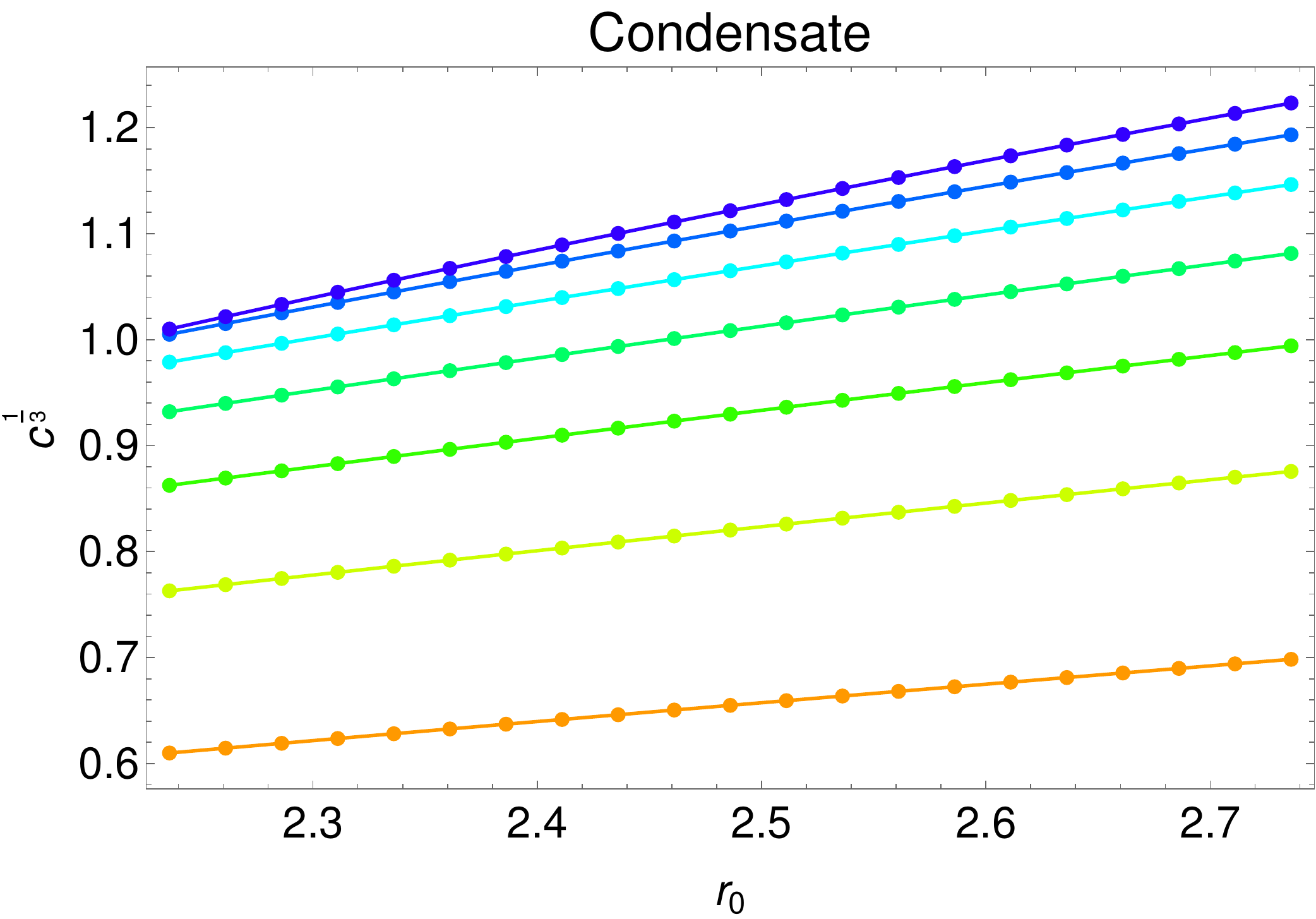}}
    \subfigure{\includegraphics[scale=0.2]{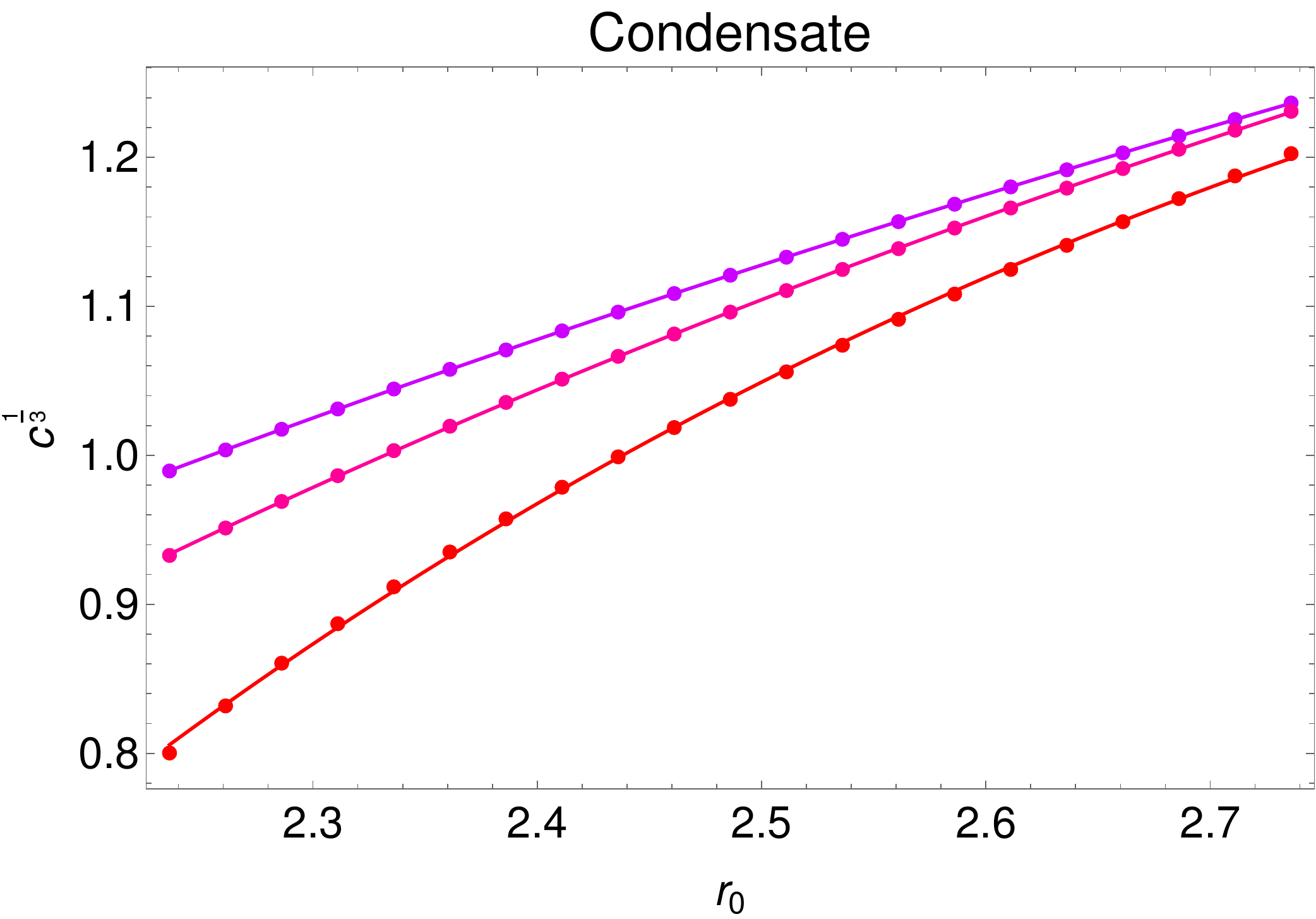}}
    \caption{Condensate of horizon ending branes}
    \label{fig:conBHH}
\end{figure}

In Fig:\eqref{fig:conBHC} shown below, we plot the condensate as a function of temperature for cutoff branes in the black hole background. We find that the condensate is nearly constant at low quark masses (left panel) compared to the larger values. This happens because, as noted earlier, the minimum free energy brane configurations end on the cutoff surface $r=r_m$ at nearly the same angular position for various $r_0$.
\begin{figure}[ht]
    \centering
    \subfigure{\includegraphics[scale=0.2]{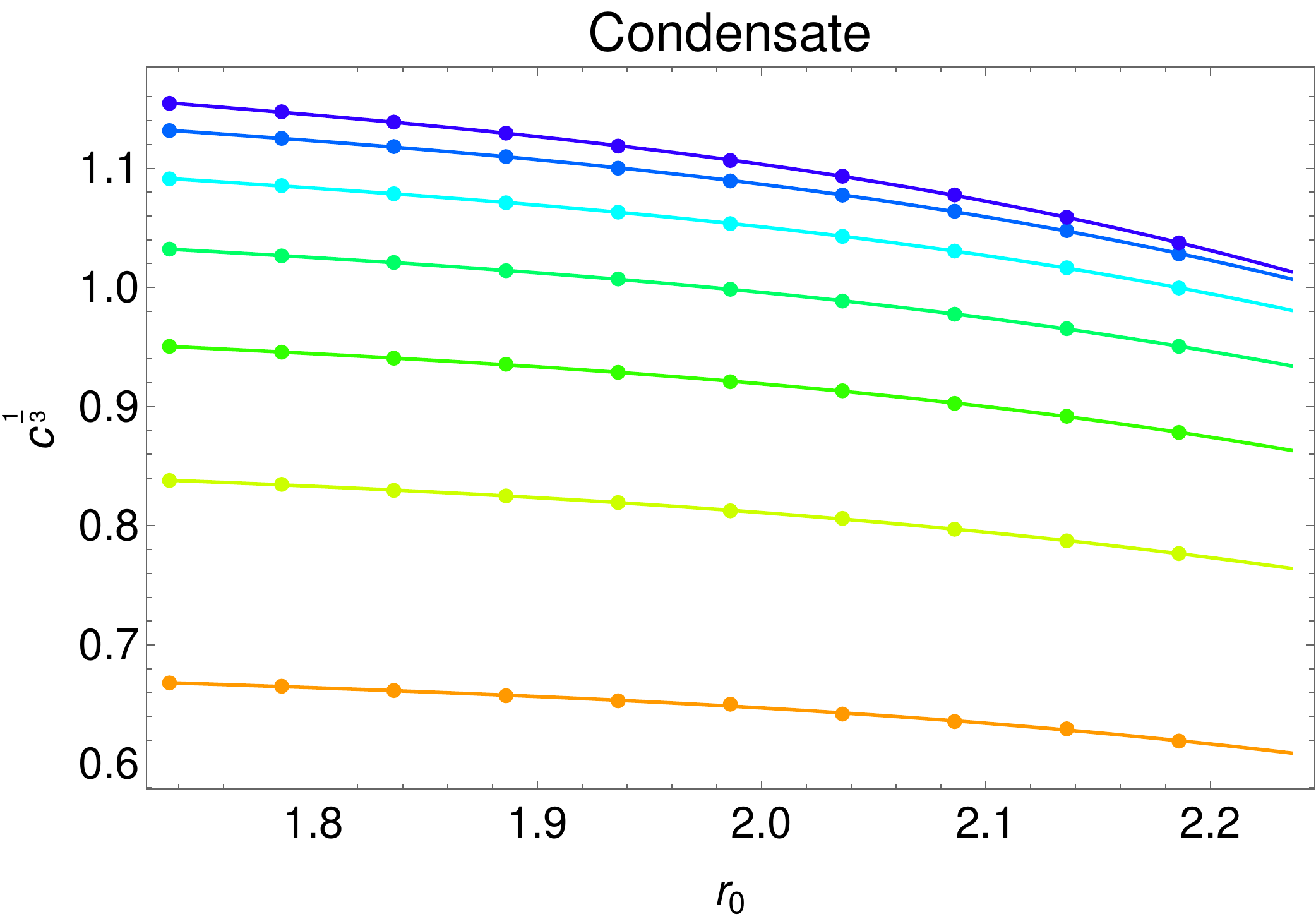}}
    \subfigure{\includegraphics[scale=0.2]{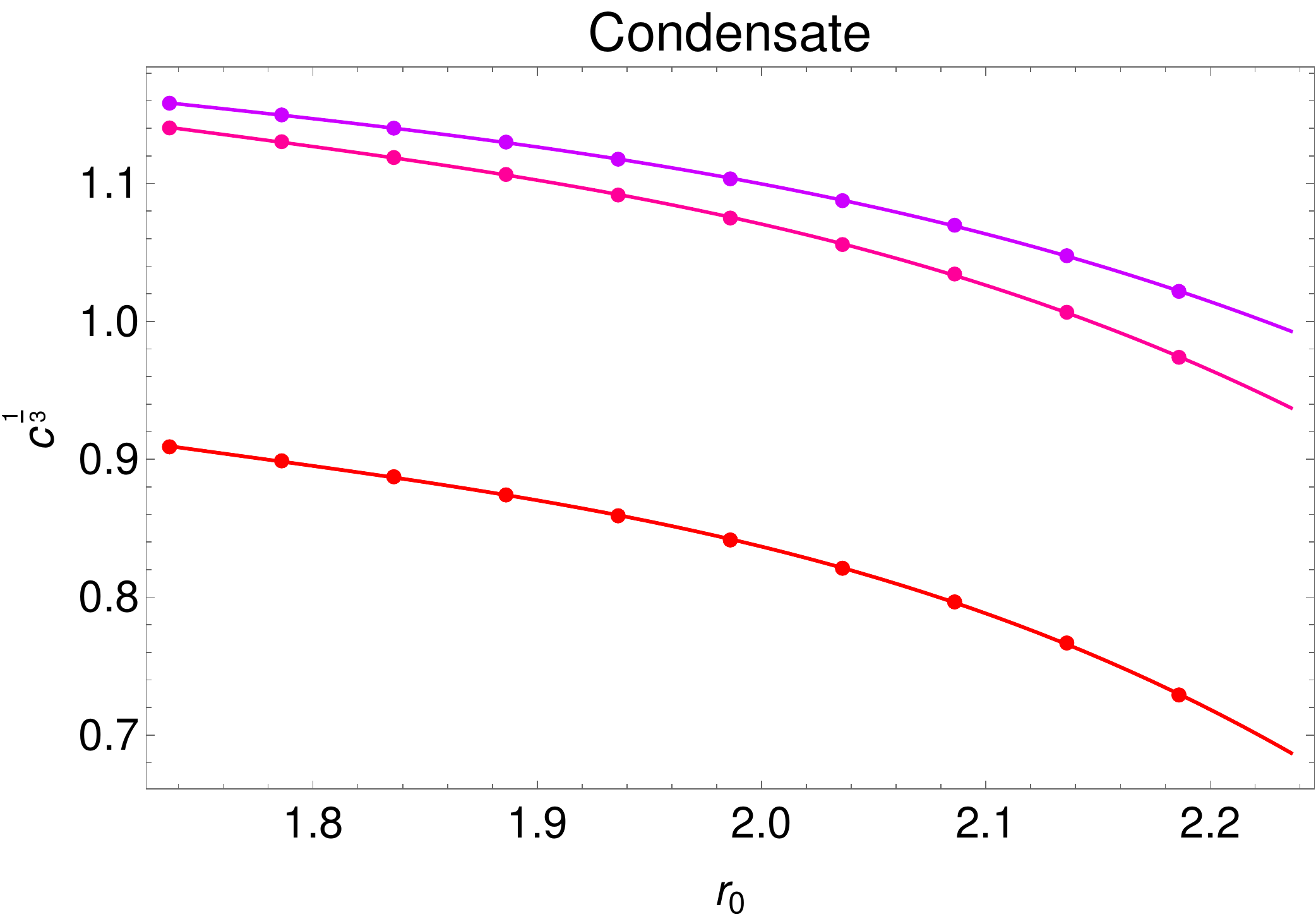}}
    \caption{Condensate for cutoff branes in the Black hole background}
    \label{fig:conBHC}
\end{figure}
However, for the branes ending on the black hole horizon, the condensate always increases with temperature Fig:\eqref{fig:conBHH}. By fitting, we can identify a logarithmic increase with temperature for low quark masses although this is not as clear for larger masses (right panel).

Finally, we plot the condensate as a function of temperature for the straightish branes in the black hole geometry for different values of $m_q$. The least value ($m_q=2.7$) is shown in yellow, while the highest value ($m_q=3.2$) is shown in red.
\begin{figure}[ht]
\centering
 \includegraphics[scale=0.2]{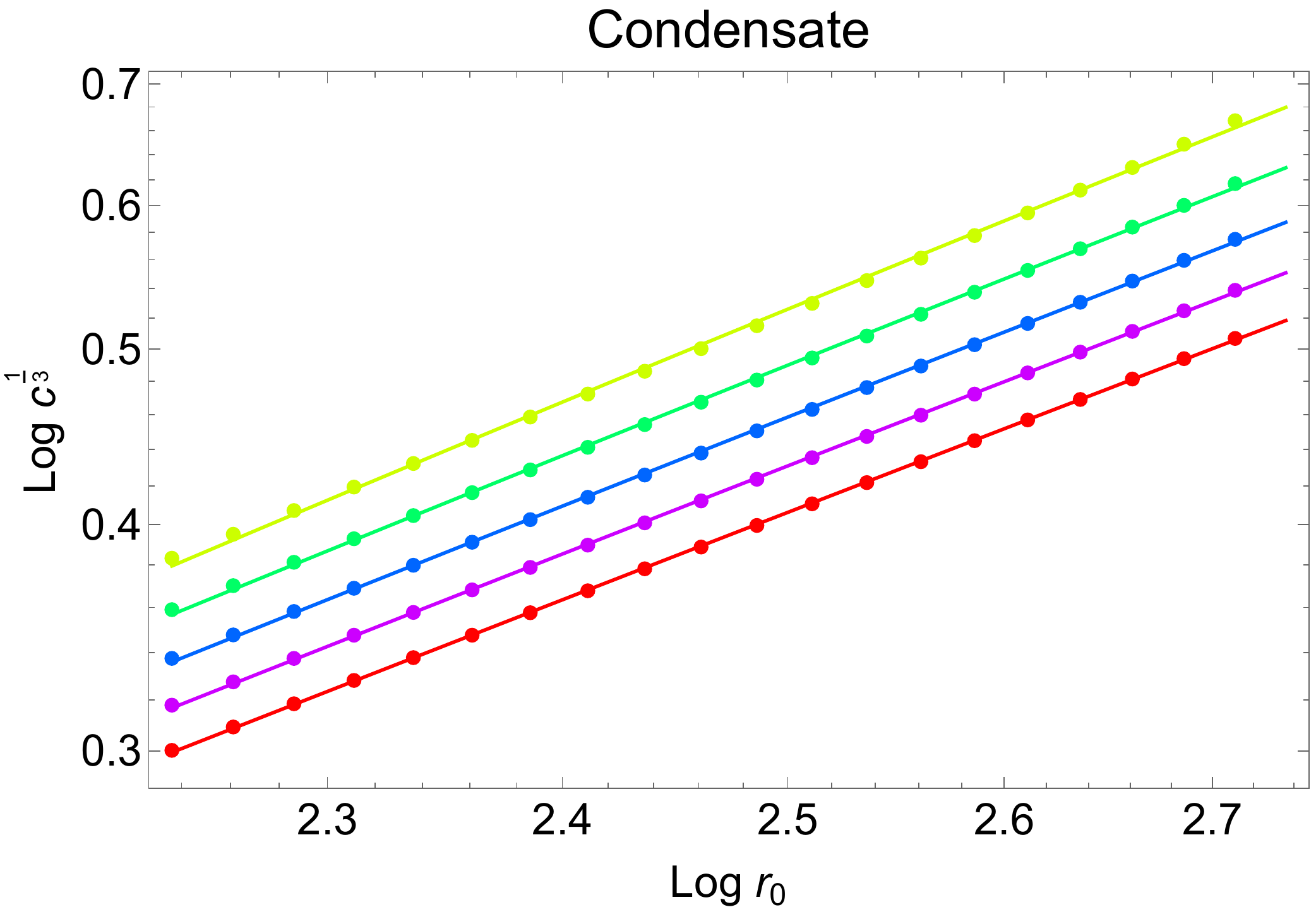}
    \caption{Condensate vs temperature for straightish branes in black hole geometry $\frac{r_m}{r_g}=1.288$.}
    \label{fig:condBHS}
\end{figure}  
The log-log plot shown in  Fig:\eqref{fig:condBHS} clearly shows that the condensate is a power law as a function of the temperature with an approximate exponent $c \approx r_0 ^ {2.6}$.
This behavior of the condensate captures the qualitative difference between the straightish branes in the black hole geometry and the curved branes.

On the other hand, if we plot the behavior of the condensate at fixed temperature as we vary the quark mass \cite{Babington}, we get the following curves. 
\begin{figure}[ht]
\centering
\subfigure[Curved branes]{\includegraphics[scale=0.2]{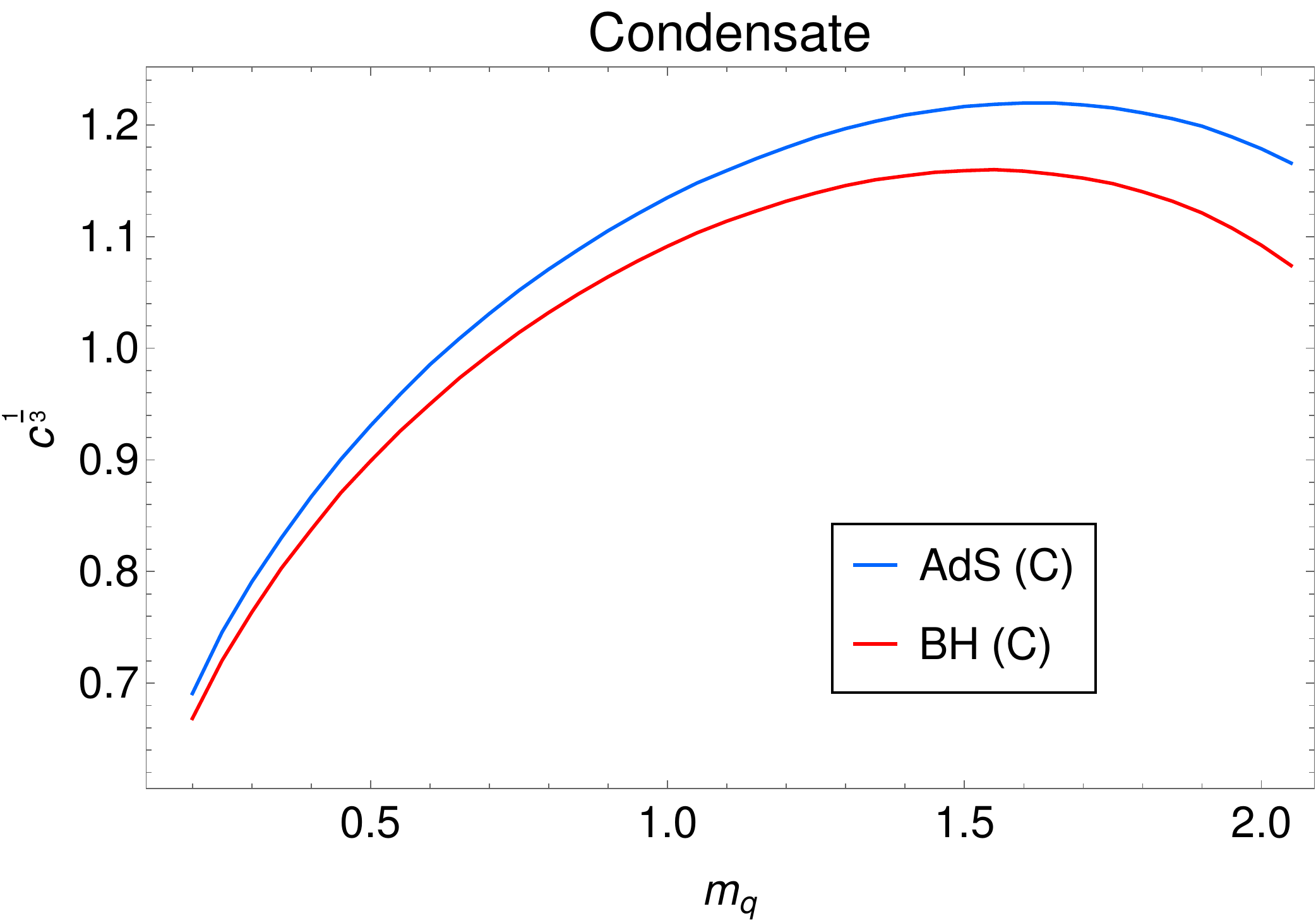}}
\subfigure[Horizon branes]{\includegraphics[scale=0.2]{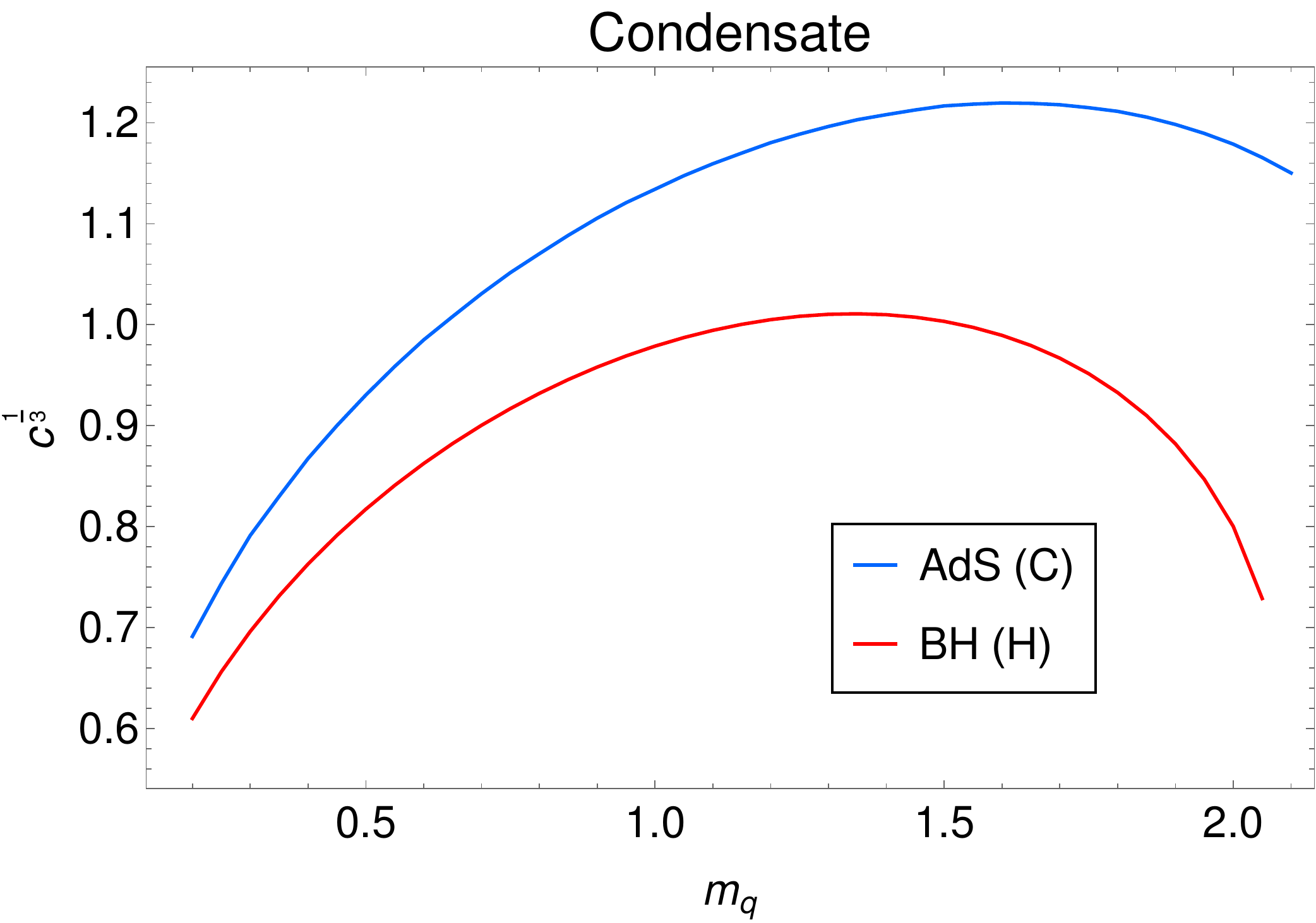}}
\subfigure[Straight branes]{\includegraphics[scale=0.2]{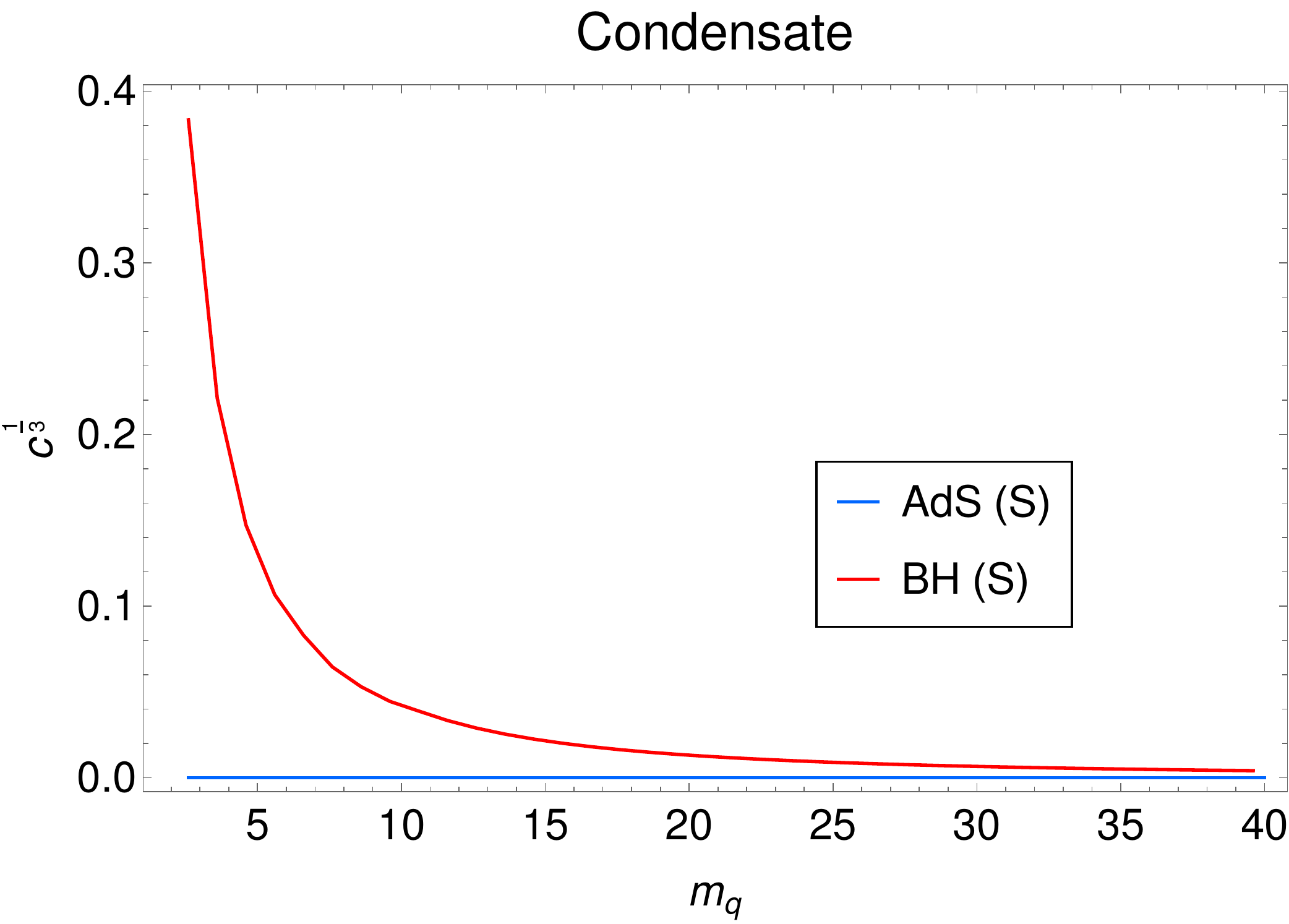}}
\caption{Condensate as a function of Mass}
\label{fig:bhcondM}
\end{figure}
The first two panels of Fig:\eqref{fig:bhcondM} suggests that the dependence of the condensate on the quark mass seems to be qualitatively the same in both AdS and black hole geometry. The condensate vanishes at very small masses as discussed earlier because we minimize the energy as a function of the condensate.

The straightish branes in the black hole geometry, on the other hand, show a qualitatively different behavior with the quark mass, decreasing to zero with large mass as depicted in the last panel. This is understandable from the bulk since for large $m_q$, the brane is far from the black hole's gravity and hence is essentially straight. This behavior is similar to that obtained in the D4-D6 case \cite{Kruczenski}.

It will be interesting to correlate the behavior of the condensate with that of entropy, but we will not discuss this further in this paper.

\subsection{Phase diagram revisited}

In this section, we will see that all the phases of the preceding section are fully described by these two order parameters. 
We have described the different phases (or solutions) in the language of the bulk gravity in Section \ref{PD}. The various phases that have appeared can be summarized by the following list.
\begin{itemize}
    \item Phase \Romannum{1}- Zero entropy, finite condensate describe curved branes in AdS background ending on a cutoff surface. 
    \item Phase \Romannum{2}-  Zero entropy, zero condensate - these correspond to straight branes in the AdS background. 
    \item Phase \Romannum{3}- Finite entropy, finite condensate - brane ending on the IR-cutoff $r_m$ in a black hole background.
    \item Phase \Romannum{4}- Finite entropy, finite logarithmic condensate - this situation corresponds to branes ending on the black hole horizon.
    \item Phase \Romannum{5}- Finite entropy, power law condensate is applicable to straightish branes in the black hole background. 
\end{itemize}

Thus, considering the following figure, for instance, we can describe the phases entirely in boundary terms using the characterization discussed above. 
\begin{figure}[ht]
    \centering
    \includegraphics[scale=0.2]{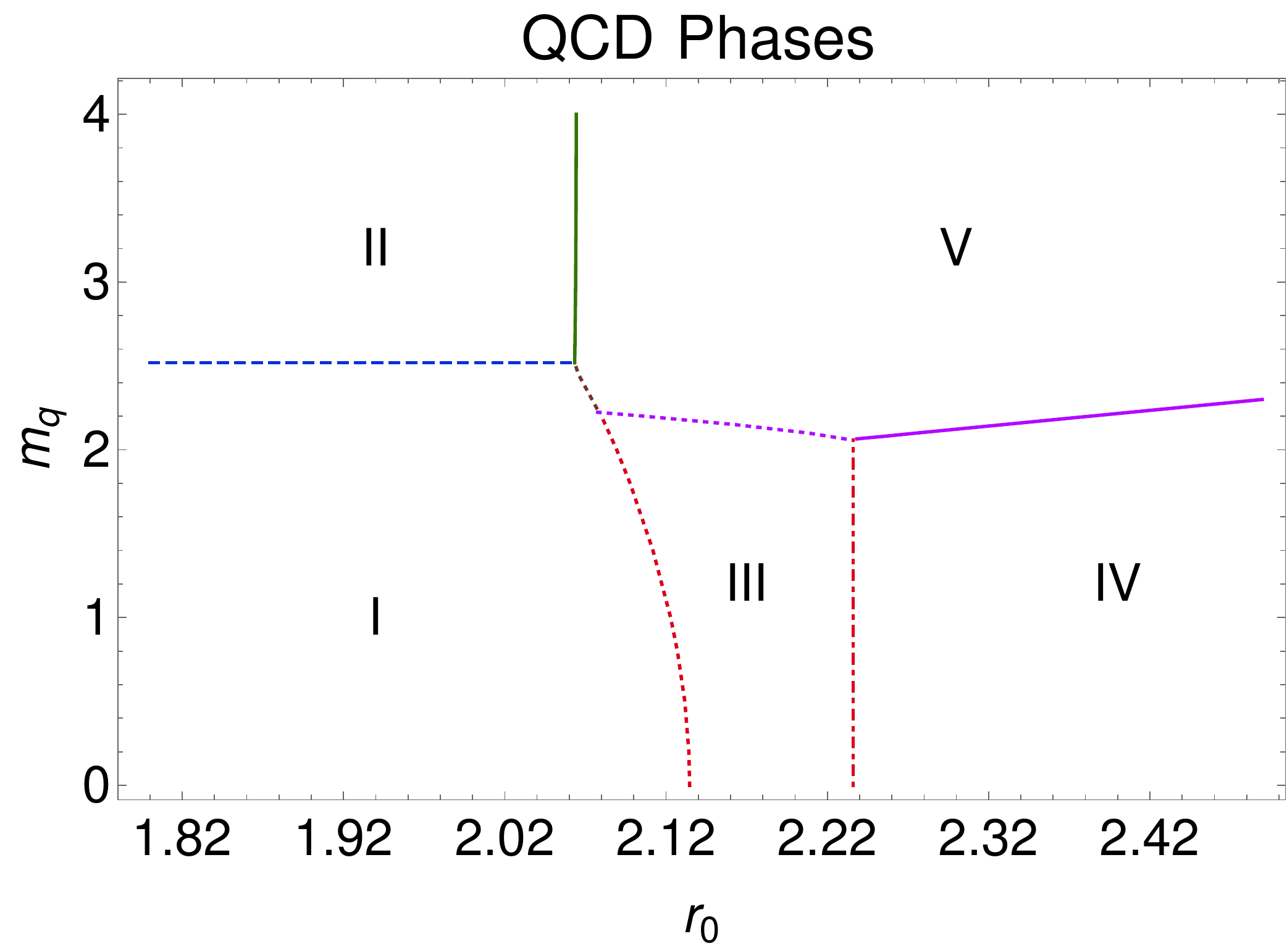}
    \caption{Phase diagram}
    \label{fig:mqvsr0}
\end{figure}
In the phase diagram, as we increase the mass, the nature of ground state changes across the horizontal blue dotted line; the condensate changes from non-zero to zero while the entropy remains zero. Similarly, in crossing the magenta line (at high temperature) we again transit from a logarithmic condensate to a small (and vanishing) condensate at large quark masses as seen in Fig:\eqref{fig:bhcondM}. These transitions involve comparing two different branes in the same gravitational background, and hence the transitions are independent of $b\sim \frac{N_f}{N_c} \l.$

As we increase the temperature, for small fixed $m_q$, the entropy changes from zero to non-zero in crossing the dotted red glue deconfinement line. The condensate remains finite and hardly changes as shown in Fig:\eqref{fig:conBHC} because the entropy of the cutoff branes does not vary significantly with temperature.  As we increase the temperature further, the entropy jumps again, and the quarks become unbound. This transition is shown by the red dotted dashed line in the phase diagram. In the high temperature phase, the condensate {\it increases} with temperature Fig:\eqref{fig:conBHH}. 

From the various phase diagrams shown in Section \ref{PD}, based on the discussions in the preceding section,  it is clear that as we increase the temperature, the entropy always increases for any fixed value of the quark mass. This is in spite of the observation that in the Phase \Romannum{3}, the quarks lead to a decrease in entropy!

We conclude this section with a brief summary of the effect of varying the ratio $\frac{r_m}{r_g}$ and the parameter $b$.
\begin{figure}[ht]
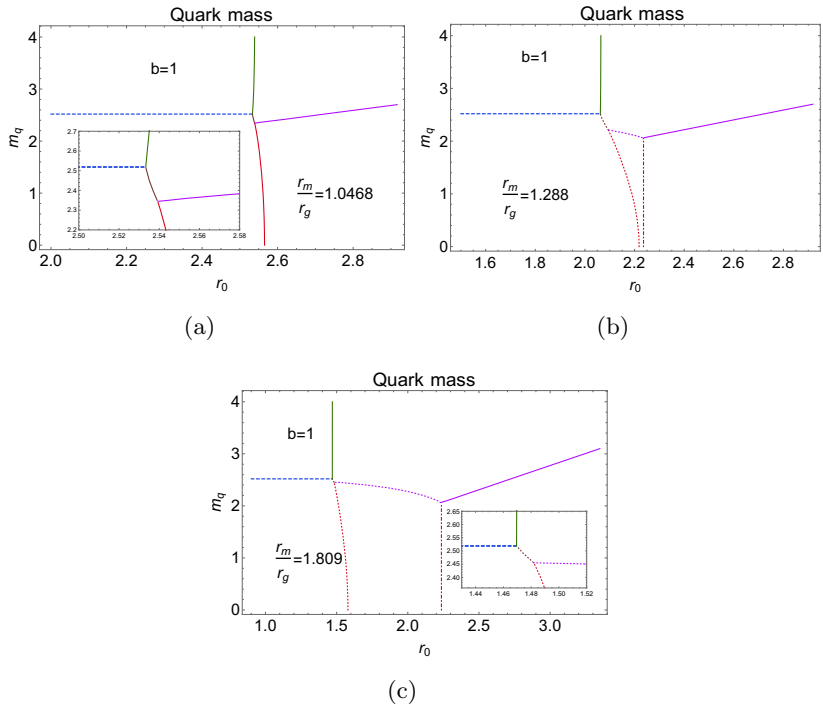

    \centering
   \subfigure[]{\includegraphics[scale=0.25]{FiguresDBI/mqvsr0_b1_rg_1p04.pdf}}
   \subfigure[]{\includegraphics[scale=0.21]{FiguresDBI/mqvsr0_b1_rg_1p28.pdf}}
   \subfigure[]{\includegraphics[scale=0.25]{FiguresDBI/mqvsr0_b1_rg_1p8.pdf}}
    \caption{Phase diagram for varying the ratio $\frac{r_m}{r_g}$  and fixed b}
    \label{fig:Interm}
\end{figure}
Fig:\eqref{fig:Interm} shows the effect of varying the dimensionless ratio $\frac{r_m}{r_g}$ keeping $b=1$ and $r_m=\sqrt{5}$. As we increase the ratio, the temperature required to deconfine the gluons and/or quarks decreases. 
If we interpret the radial direction as an energy scale of the boundary theory, increasing $\frac{r_m}{r_g}$ amounts to keeping a smaller fraction of the quark degrees of freedom. Plausibly, this is the reason for the decrease in the transition temperature.

Finally, we discuss the role played by the parameter $b$ which, in field theory terms, is given by \eqref{b}. Thus, increasing $b$ is equivalent to increasing the coupling between the gluons {\it and} the quarks. Thus, it is unsurprising that the glue deconfinement temperature is unaffected by $b$ since our study can be viewed as being in the quenched quark approximation. At low values of $b$, when the quarks and gluons deconfine separately, the quark deconfinement occurs when $r_0=r_m$. At larger values of $b$, both deconfine together and this transition temperature depends on $b$.

\section{Physical units}\label{PhysicalUnits}
In this section, we describe one method to translate the dimensionless parameters of the preceding into physical units.
The bulk cutoff $r_g$ can be related to the glueball mass spectrum if we study the fluctuations of the metric, dilaton or the other bulk supergravity fields. The Gauge/Gravity duality relates the two point function of bulk fields to the two point functions of boundary operators. The graviton field is related to a spin two color-neutral operator in the boundary theory and by the spectral theorem, the two point function of such an operator will have a pole at the mass of the lightest particle state that can be created by such an operator. 
This was studied by \cite{rinaldi} in a recent paper where the author attempted to fit the known glueball spectrum to hardwall AdS computations in five dimensions (a recent review comparing holographic results to lattice can be found in \cite{Dymarsky:2022ecr}). Using linearized Einstein's equations in thermal AdS, the author fit the computed glueball masses to lattice data which fixed the IR-cutoff $z_g$ in physical units. For Dirichlet and Neumann boundary conditions, the author determined a best fit value of  $\frac{1}{z_g}=250$ MeV and $\frac{1}{z_g}=290$ MeV, respectively as best fit to the glueball spectrum. However, using an average of the lowest glueball masses gives $\frac{1}{z_g} \sim 330$ MeV, $\frac{1}{z_g} \sim 435$ MeV  for Dirichlet and Neumann boundary conditions, respectively. We can use the same numerical values in our 10-D hardwall model by simply considering only s-wave graviton fluctuations on the $S^5$ part of the spacetime.

The cutoff $\r_m$ on the brane world volume coordinate can be determined by computing the mass of mesons similar to the work of \cite{ekss}. The key difference in our case is the presence of the background field $y(\r)$. 
We will outline this calculation below, since it involves features that are unique to the 10-D embedding of the D7-branes. 

\subsection{Vector Mesons}

We start with the DBI action for the D7-brane in $AdS_5\times S^5$ background. 
\be 
 S=-N_f\m_7\int d^8\s \sqrt{-det(P[g]_{\m \n}+2\p \a' F_{\m \n})}.
\ee 
We will study the fluctuations of the Vector field around the zero Vector field background and determine the location of the poles in the two point function which gives us the mass of the particles in the dual field theory. Since the background vector field $A_\n$ is zero, we denote the fluctuations $\delta A_\n$ as $a_\n$ and the field strength as $f$.
The equation of motion is easily obtained as
\begin{equation}\label{eom-vec}
    \partial_\mu (\sqrt{-det P[g]} P[g]^{\mu\sigma}P[g]^{\nu\lambda} f_{\sigma\lambda} ) = 0.
\end{equation}
In this equation, $P[g_{\m\n}]$ depends on the profile of the D-brane $y(\r).$ The vector fields $a_\m$ are functions of all the world volume coordinates. Since we are interested in the lowest vector meson modes, we will assume that $a_\m$ are constant on the $S_3.$ Because we are interested in the {\it vector} mesons, we will consider only fluctuations in $a_\m,  \m=0,1,2,3$
and use Lorentz gauge $\partial_\mu a^\mu=0$ which has the most general Lorentz invariant solution of the form $a_\mu=(q_\mu q_\nu - q^2 g_{\m\n})a^\nu (r)$. 

We can then simplify \eqref{eom-vec} to get,
\begin{equation}
    \rho^3 \sqrt{1+y'^2}\left(\frac{L^2}{\rho^2 +y^2}\right)^2 q^2 a_\n(q,\rho) - \partial_\rho \left(\frac{\rho^3}{\sqrt{1+y'^2}}\partial_\rho a_\n(q,\rho)\right) = 0,
\end{equation}
where, $a_\n(q,\rho)$ is the Fourier Transformation of $a_\n(x,\r)$. In comparison with \cite{ekss} say, the difference is the presence of the profile $y$ of the D7-brane which forms an effective open-string metric for the modes on the world volume. The above equation, for any component $a(\r)$ of vector field $a_\n(\r)$,  defines a Sturm-Liouville problem for the eigenvalues $q^2=-m^2 _\r$ ($m_\r$ is the $\r$-meson mass), with boundary conditions and normalization given by 
\be \del_\r a(\r_m)=0 \qquad a(\r_{UV})=0 \qquad 2\int_{\tilde\r_m} ^{\tilde\L}  d\r \frac{\sqrt{-det P[g]}}{(\r^2+y^2)^2} a(\r)^2=1 .
\ee
It should be noted that both the differential operator and the normalization conditions depend on the shape of the D-brane via both $y(\r)$ as well as $y'(\r).$
Determining the eigenvalues allows us to fix $r_m$ in terms of the $\r-$meson mass in the following manner. The above equation explicitly involves four parameters $m_q,c,r_m$ and $\r_m$ and two other parameters $\l, L$ which are not manifest.

$m_q$ is fixed in terms of the physical quark mass by using $M_q=\frac{m_q\sqrt{\l}}{2\pi L^2}$ provided we know $\l$ and $L.$ The value of the AdS radius $L$ is fixed by the glueball mass via $L=r_g z_g$. In \cite{Hoyos:2016zke}, $\l$ was estimated to be $10.74$ using the form of the free energy at zero temperature and comparing with a gas of quarks. We use this value in what follows, but also point out options. 

In our case, the two flavor branes have been assumed to be coincident. Therefore, we take the physical quark mass to be the average of the up- and down quark masses $M_q=3.55\pm 0.5$.  Using these, the numerical value of $m_q$ is determined to be $m_q=0.06.$ Setting $N_c=3$ and $N_f=2$, the parameter $b=\frac{\l}{4\pi^2}\frac{N_f}{N_c}\approx0.18.$

Knowing $m_q$, and for a given value of $r_m$, the parameter $c$ is fixed by the condition $r^2_m =y(\r_m)^2+\r_m ^2$ at the IR-cutoff. We then vary $\r_m$ to find the minimum energy configuration of D7-brane for the fixed $m_q.$ This still does not fix the parameter $r_m$ which can now be varied until the meson spectrum is suitable.  In our work, we use the mass of the lowest rho-meson to fix $r_m.$

The results of these calculations are as follows. 

For $\frac{1}{z_g}=250$ MeV, the ratio is determined to be $\frac{r_m}{r_g}=1.29062$ for which the conditions described in Section \ref{section0qm} are satisfied. In this case, the relevant phase diagram is approximately the one shown in Fig:\eqref{fig:Intermediate_a} because $b=0.18$. In this phase diagram, we need to set $m_q$ on the y-axis to be equal to $m_q=0.06.$ Thus, we can say that the gluons deconfine first at a temperature $T_{cg}=95.737$ MeV followed by the melting of the quarks at temperature $T_{cq}=102.704$ MeV. 

However, for $\frac{1}{z_g}=290$ MeV, $\frac{r_m}{r_g}=1.11263$ and the the relevant phase diagram changes significantly to Fig:\eqref{PhaseDiagram1_a}. In this case, the quarks and gluons deconfine together at a critical temperature  $T_c=110.403$ MeV.

These conclusions are not robust though. Changing $\l\approx 50$ will make $b\approx 1$ which will lead us to consider the phase diagrams shown in Fig:\eqref{fig:Interm}.
Interestingly, we find that as we decrease the value of bulk IR-cutoff $z_g$, the dimensionless ratio $\frac{r_m}{r_g}$ can even go below unity. This happens if we use the values $\frac{1}{z_g} \sim 330$ MeV, $\frac{1}{z_g} \sim 435$ MeV set by the mean glueball mass \cite{rinaldi}. For the special case of the ratio being unity, we recover the phase transition temperature $T_c\sim122$ MeV found in \cite{Herzog}. 
The numerical value of $\l$ determined in \cite{Hoyos:2016zke} will vary with the compactification manifold (such as Klebanov-Strassler geometries \cite{KS}) which will give different volume factors. More importantly, varying dilaton which models running coupling will also play a significant role \cite{Kaempfer}. 

In our calculation $L$ was fixed by the glueball spectrum - but we can fix $L$ using the meson spectrum or any other dimensionful observable. In the AdS case, this does not change the results because of the scaling symmetry present in the underlying background. 

\subsection{Pseudoscalar meson and GOR relation}

Similarly, we can study the fluctuations of the worldvolume scalar field $\phi$, which is a pseudoscalar particle from the 4D viewpoint (the parity properties are determined from the 10D string theory). This scalar is special since it will be the Goldstone boson of the spontaneously broken $U(1)_R$ rotation symmetry which is an axial $U(1)_A$ symmetry of the boundary 4D theory \cite{Kruczenski:2003uq}. Solutions with quark masses lead to explicit breaking of this rotational symmetry, while it is also possible to find solutions with zero quark masses but nonzero profiles (and hence nonzero condensate). Because of these reasons, we can expect the masses of these fluctuations to satisfy a GOR relation. However, we remind the reader that the minimum energy condition fixes the condensate $c^3$ dynamically for a given $m_q$ and prevents symmetry breaking in the massless limit.  

If we do not impose the minimum energy requirement, for a given $m_q$ (including $m_q=0$), there is a family of solutions with varying condensates $c$. By keeping $c$ fixed, we can study the GOR relation for the breaking of $U(1)_A$:
\be  
M_{\h}^2f_{\h}^2=N_f M_q \s
\ee

The action up to quadratic order for the fluctuations $\delta\phi$ is,
\be\label{actionphi} 
S=-\m_{7}N_f\int d^8\s\sqrt{-det \tilde G}\: \frac{G_{\phi\phi}}{2}\del_{a}\delta\phi\del^a\delta\phi
\ee 
The equation of motion we get from  action \eqref{actionphi} is 
\be 
\del_a(\sqrt{-det\tilde G} G_{\phi\phi}\del^{a}\delta\phi)=0
\ee 
We are interested in the lowest mass modes. Therefore, we will not consider the $S^3$ directions and parametrize the fluctuations as $\delta\phi=e^{-ikx}\varphi(\r)$. We can re-write the equation as 
\be 
\del_{\r}\left(\frac{\r^3 y^2}{\sqrt{1+y'^2}}\del_{\r}\varphi\right)+\frac{\r^3y^2L^4}{r^4}\sqrt{1+y'^2} M^2 \varphi=0
\ee 
where, we have defined the meson mass,
\be 
 M^2=-k^2
\ee
This is an eigenvalue equation that will determine the masses of the meson provided we specify boundary conditions and normalization as
\be \del_\r \varphi(\r_m)=0 \qquad \varphi(\r_{UV})=0 \qquad 2\int_{\tilde\r_m} ^{\tilde\L}  d\r \frac{\sqrt{-det P[g]}}{(\r^2+y^2)^2} \varphi(\r)^2=1 .
\ee
Using the numerical values determined in the previous section, for physical quark mass $M_q=3.55 \pm 0.5$ MeV, the mass of turns out to be $936.93$ MeV and $948.09$ MeV for Dirichlet and Neumann boundary conditions, respectively which compares very well with the experimental mass of   $\h'$-meson which is expected to correspond to the fluctuations of worldvolume field $\phi$ \cite{Kruczenski:2003uq}.

We read off the pion decay constant from the normalization of the kinetic term in the four-dimensional low-energy effective Lagrangian for these fluctuations (which, in principle, depends on the shape of the brane). We compare \eqref{actionphi} with
\be 
S=-\frac{f_\p^2 }{2}\int d^4x \del_\m \delta \phi \del^\m\delta \phi
\ee
and find the decay constant,
\be\label{fpi}
f_\p^2=\m_7 N_f \O_3\int d\r\  \frac{\r^3y^2 L^4}{r^4}\sqrt{1+y'^2}=\frac{N_c N_f \l}{(2\p)^4 L^2}  I.
\ee 
which depends on the quark mass and condensate via the shape of the brane $y.$

Using \eqref{con},\eqref{M},\eqref{fpi}, we can rewrite the GOR relation in terms of dimensionless bulk quantities: 
\be 
q^2 I=\hat{m}_q \hat{c}^3 
\ee  
where $q^2=M_\h^2 L^2$, $\hat{X} L= X$ for $X=\{m_q,c,y,\r...\}$, $I=\int d\hat{\r}\  \frac{\hat{\r}^3\hat{y}^2}{\hat{r}^4}\sqrt{1+\hat{y}'^2}$ is a dimensionless number. We find $I$ for different values of $m_q$ and take the mean to fit the data points. 

\begin{figure}[ht]
    \centering
    \includegraphics[width=6cm]{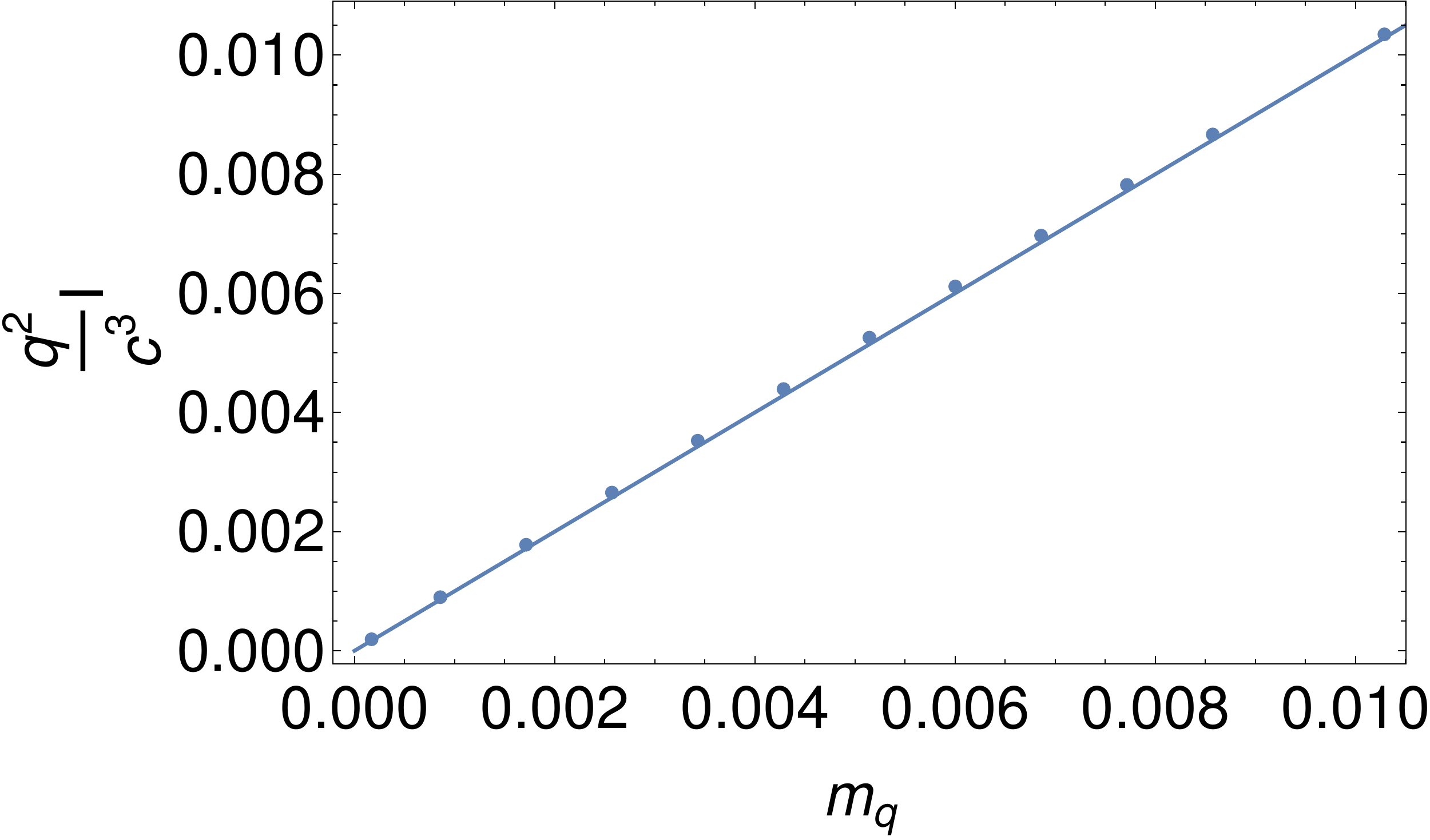}
    \includegraphics[width=6cm]{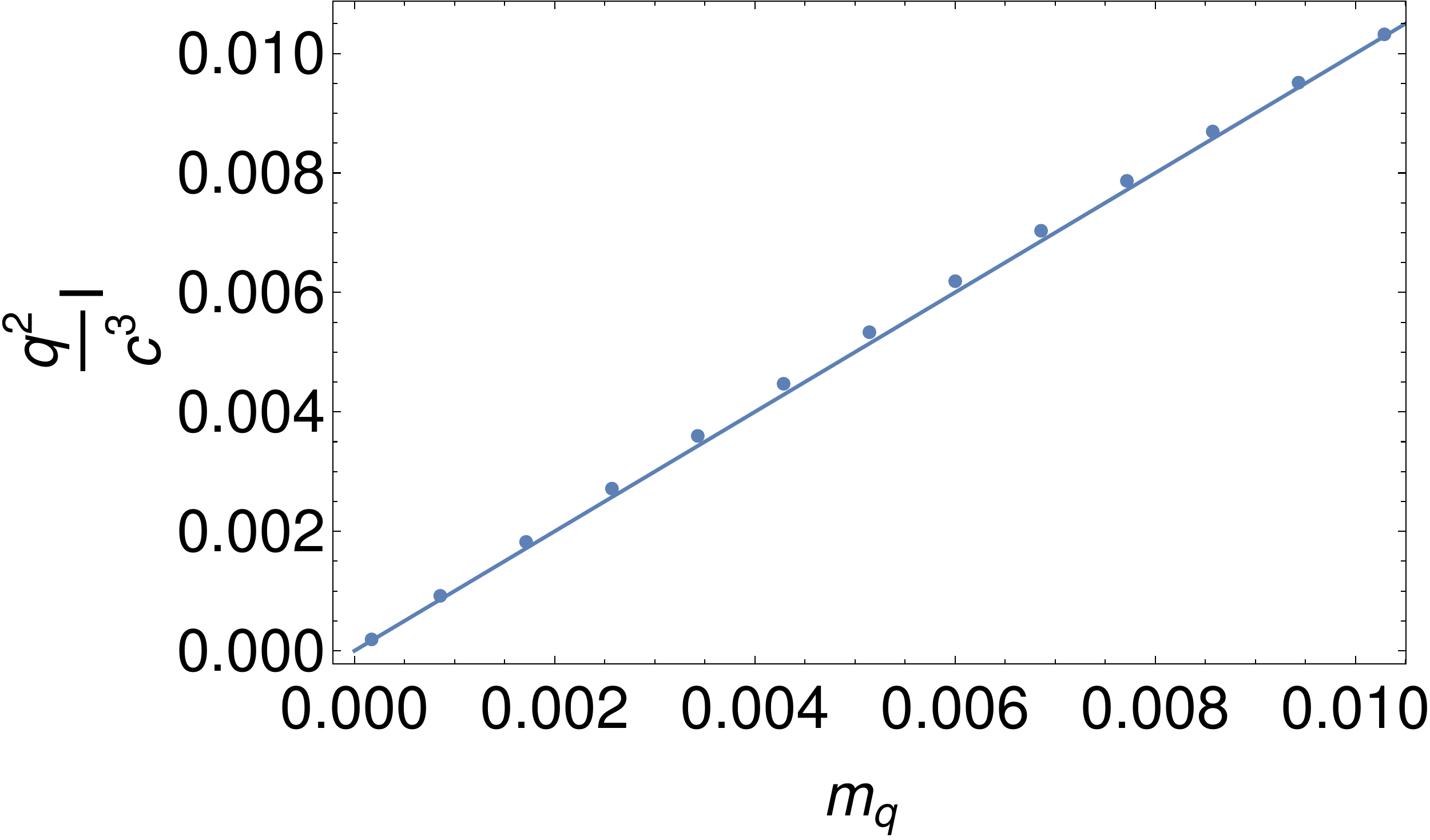}
    \caption{Left panel is for $c=-2, I=0.061$  and right corresponds to $c=-1.5, I=0.011$}
    \label{GOR}
\end{figure}

We numerically solve the eigenvalue problem for fixed $c^3$ to find the meson mass as a function of quark mass $m_q$. The GOR relation between the meson mass and quark mass is satisfied, as can be seen from Fig:\eqref{GOR}.

\section{Reliability estimates}
The computations of the preceding section need to be supplemented by an estimate of the corrections, especially those that are likely to change the predictions substantially. In this section, we show that such an estimate can be performed for the model studied in this paper. 

The gravity approximation we use is valid in the standard low curvature $\l=g_{YM} ^2 N_c>>1$ regime. This is the region of interest in the field theory in any case. Higher derivative corrections are weighted by powers of $\frac{1}{\l}.$ String loop corrections are suppressed by $\frac1{N_c}$ which makes our calculations reliable at large $N_c.$

However, there are other sources of systematic error. Firstly, we work in the probe approximation which is reasonable only 
when the energy sourced by the D7-brane is small compared to the background. We can estimate this by comparing the two sources
\be
|F^{(5)}|^2\ \  vs\ \  2\k ^2 _{10} N_f T_{7} \sqrt{\rm det (P[G]+2\pi \a' F)}
\ee
whose relative contribution to the free energy \ref{0qmFreeEnergy} is weighted by the parameter $b=\frac{N_f}{N_c}\frac{\l}{4\pi^2}.$ Therefore the validity of the probe approximation requires $b<<1.$

Let us consider whether the phase transitions we have identified will be reliable. The glue deconfinement transition which changes the background occurs,
for zero $m_q$, at $T_{cg}=\frac{8^{\frac14}r_g}{\pi L^2(4-b)^{\frac14}}\approx \frac{8^{\frac14}r_g}{4\pi L^2}(1+\frac{b}{16})$  and survive even if $b<<1.$ On the other hand, the quark deconfinement transition which occurs because of a change in the shape of the brane, takes place in a fixed bulk geometry. As we can see from $T_{cq}=\frac{(8r_g ^4+2br_m ^4)^{\frac14}}{\pi L^2(4+b)^{\frac14}}\approx \frac{8^{\frac14}r_g}{4\pi L^2}(1+\frac{b}{16}(-1+\frac{r_m ^4}{r_g ^4})$, it also survives in the probe approximation. The possibility that there are separate transitions depends on the ratio $\frac{r_m ^4}{r_g ^4}$ which is independent of the probe approximations and also of the $\l>>1$ and $N_c>>1$ conditions. 

At finite quark masses and low temperatures, the quark deconfinement transition will involve comparing branes in thermal AdS background.
Thus, this will not be affected by the probe approximation. A similar argument applies to the black hole background. 

The transition that is likely to be most affected is the transition where the background changes. In this case, under the probe approximation, the phase transition will be decided by the free energy difference in the gravity backgrounds. The effect of the probe branes can then be estimated as a shift in various quantities, such as the transition temperature to first order in the parameter $b.$

Secondly, considering the profile of the D7-brane, we see that the derivative $y'$ becomes large near the tip. While the Abelian DBI action remains sensible even for large $y'$, there are higher derivative curvature type terms weighted by $(\a')^2 R^2.$ For these to be small, we get the condition 
\be
\frac{1}{\l} \frac{y_m ^4}{r_m ^4}<<1
\ee
which is satisfied for large $\l$.

There is another sense in which the holographic calculations are likely to remain reliable. The different transitions are all characterized by a change in topology. While $T_c$ and other thermodynamic properties are likely to be corrected, the transitions themselves will not survive only if some other configurations have lower free energy than the ones considered. 

Finally, this model differs from QCD proper in that there is no running coupling, no spontaneous chiral symmetry breaking and it includes additional symmetries ($SO(4)\times SO(2)$ R-symmetry), massless adjoint scalars and fermions together with other extra fundamental degrees of freedom. Corrections from these are unlikely to benign.

\section{Discussion}

In this work, we have presented a rich generalization of the hardwall model in ten dimensions, by introducing a full DBI action in the 10D-\Romannum{2}B supergravity action to model the quark degrees of freedom. Working in ten dimensions allows us the possibilities of changing the compact part of the 10D-spacetime from $S^5$ to other scenarios \cite{Yadav:2020pmk} which are closer to QCD. This potentially enables a systematic exploration of universality classes and features in the phase diagrams. Using the full DBI action has the consequence that the quark degrees of freedom see a different effective geometry given by the open-string metric $K=P[g]$ which depends on the shape of the embedding $y$. This shape degree of freedom allows us to describe the phases geometrically and can motivate searches for other natural brane embeddings including polarized branes as arising from the Dielectric effect \cite{Myers:1999ps} which are likely to be important at finite densities \cite{Kovensky:2021ddl}. Finally, our model is significantly different from all previous work in the introduction of two distinct cutoffs for the DBI and the bulk gravity parts of the action, the sole exception being the recent work \cite{Rebhan}. This is likely to be a key advantage since complex backreaction and Non-Abelian configurations deep in the interior of the bulk can be hidden behind the IR-cutoff. Nevertheless, the effects of this interior geometry can be incorporated by suitable IR boundary conditions which can be fixed by using experiments as illustrated in this work.

In our work, we have shown to handle the IR-cutoff for the brane degrees of freedom including several subtleties. It was pointed out that instead of imposing a cutoff $r_m$, introducing a cutoff on the $\rho$ coordinate leads to a entirely different scenario. Perhaps the difference arises because under scale transformations (RG in the boundary language) the $y-$fields mix with the $\r$-direction. This difference needs to be understood better since it can inform other such holographic constructions as well. 

As discused, $r_g$ leads to a glueball mass $\sim \frac{r_g}{L^2}$ and thus can be related to $\L_{QCD}.$ The parameter $r_m$ decides the masses of the mesons and is related to the condensate $c$ by minimizing the energy. 
In particular, the relation between $c$ and $\L_{QCD}\sim r_g$ is being indirectly determined by by relating the mass of the meson to experiments. Thus we have a handle on exploring the relationship between the condensate scale and $\L_{QCD}$ which can be explored systematically by considering other models which show chiral symmetry breaking in the massless limit. 

The most striking observation that emerged from the introduction of separate cutoffs was that the deconfinement of gluons and the `melting' of mesons can be separated and controlled by the inequality $0<b\leq 4\left(1-2\frac{r_g^4}{r_m^4}\right)$.  The low temperature AdS phase is always characterized by a condensate, but as the temperature increases a first order transition can lead to either zero or non-zero condensate phases depending on the quark mass. Similar observations have appeared earlier in the literature \cite{Kruczenski:2003uq}, \cite{Bak:2004nt}.
In either case, the background involves a black hole - which leads to a perimeter law for the Polyakov loop following standard calculations. Only when the branes intersect the horizon, we can say that the mesons undergo ``melting".
This is reminiscent of \cite{Glozman} who has been arguing that as temperature increases, the deconfinement The second distinct quark deconfinement transition can be tuned to be closeThe second distinct quark deconfinement transition can be tuned to be closetransition in QCD leads to a phase wherein to  gluons are deconfined, but nevertheless, quarks are bound in hadrons.
This separation of scales could be insightful in understanding the relation between the axial anomaly and chiral and deconfinement transition in actual QCD \cite{Choun:2019xyo},\cite{Oh:2019zbr},\cite{Cui:2022vsr}.

A second remarkable feature was the change in the nature of the ground state for large $m_q$ characterized by a vanishing condensate. The ratio $\frac{m_q}{r_m}$ of quark mass to pion mass defines a critical value $m_q ^1$ above which the transition will necessarily occur. This is similar to the chiral transition in that at large masses, a condensate vanishes - in fact, this transition is very similar to that in the Sakai-Sugimoto model once we allow ourselves to ``complete" the brane configurations behind the IR-cutoff. 

The shape field $y$ gives a deep insight into the boundary theory order parameters, the condensate, and the entropy. It allows us to distinguish the quark phases topologically depending on the vanishing of the thermal circle, the $S^3$ or their nonvanishing. Boundary conditions on the slope are natural in this geometric view. Finally, the background shape $y$ also affects the location of the meson poles and spectral density. Another somewhat surprising observation that can be made is that the phase diagrams presented in Section \ref{PD} bear a remarkable similarity to the phase diagrams obtained in \cite{Basu:2016mol, Horowitz:2010jq} which are those of a Holographic superconductor by relating the quark mass of our work and chemical potential in those studies.

By using the procedure of holographic renormalization, we have also been able to characterize the various phases entirely in boundary terms. This allowed us to observe qualitative differences in the various phases in the variation of entropy and the condensate with temperature. In fact, careful consideration of the {\it geometry} of the solutions leads to natural expansions of the free energy as a function of $r_0$ and $m_q$.
For instance, in the {\it subsequent} work \cite{Weldon}, the author presents a particular form for the pressure of QCD - which seems to emerge quite naturally from the brane and bulk contributions.
Fitting the temperature dependence could lead to further understanding in terms of dependence on quark masses and couplings. If we assume the presence of a Fermi surface, we can identify the presence of fermionic (quark) degrees of freedom via linear terms in specific heat. We have not undertaken this exercise in detail in {\em this} work. 
In the intermediate phase where even though the gluons are deconfined, the quarks remain bound in mesons and effectively lead to a decrease in entropy and thereby the specific heat. The coexistence curves between various phases depend, in general, on all the parameters in the model. Our results are likely to persist in IR-complete models since the various D-branes can be characterized topologically by vanishing cycles. 

This study appears to be the first to explore the phase diagrams as they depend on the parameters $b,m_q$ and the ratio $\frac{r_m}{r_g}$, which in the boundary language are related to the 't Hooft coupling, the quark mass, and the ratio of the meson mass to the glueball mass. We have shown that the nature of the ground state changes at large $m_q$ marked by a vanishing condensate. This transition is independent of temperature and independent of the coupling $b$ albeit, in the probe approximation. For the probe approximation to be valid, we require the energy in D-brane embeddings be small (compared to the contribution from the five-form). Hence, the fact that the phase diagrams change upon changing $\frac{r_m}{r_g}$ can be taken to be an illustration that backreaction effects are going to be significant. This can also suggest ways of improving the bulk models.

As discussed in the preceding section, the results are sensitive to the physical observables that are used to fix the IR boundary conditions and parameters. In fact, our work opens the way for a controlled incorporation of multiple features in ten dimensions utilizing the separate hardwall cutoffs can be used in conjunction with physical boundary conditions to hide strong curvature and string coupling regions. 

Firstly, to model the running of the QCD coupling, we can include the dilaton field \cite{Gubser:2008ny} in the background geometry and in the DBI action. In fact, various backgrounds with varying dilaton such as the various soft wall models and the Witten-Sakai-Sugimoto models fall into this class. The dilaton profile is also important to modeling confinement and  obtaining good meson/glueball spectra.  An effective use of the IR-cutoff $r_g$ could aid in separating various length scales (such as the KK scale from $\L_{QCD}$ in the WSS models), strong curvature $\a'$ and strong coupling regions arising deep in the interior of the bulk geometry. Secondly, multiple flavor branes and chiral symmetry can also be included together with a judicious use of the second IR-cutoff $r_m.$ This also allows us to extend this investigation to finite chemical potentials. In the context of the QCD phase diagram at finite densities we may cite \cite{Gubser}, \cite{Ratti}, \cite{Ishii:2019gta}, \cite{Knaute:2017opk}. 
Distributions of polarized branes representing a gas of baryons can be hidden away behind the IR-cutoff in an attempt to avoid the problems associated to Non-Abelian DBI actions. Simultaneous inclusion of both baryon and isospin chemical \cite{Kovensky:2021ddl} potentials are also quite simple via non-Abelian gauge fields in the DBI as is the possibility of different current quark masses especially the strange quark. 
The further advantage of working in ten dimensions is that there is a large body of techniques to find solutions to the gravitational system which can take us beyond the probe approximation. The D7-brane backreaction is controllable to some extent for one (due to linear growth of the dilaton in the UV), and the entire set up, at zero temperature, has a well-defined supersymmetric dual. For instance, \cite{PolGrana} present a fully backreacted D3-D5-D7 geometry presenting, for our purposes, useful metric ansatz for more general explorations.  
We note here that the AdS/CFT correspondence tells us that the energy density of the boundary theory is obtained as the coefficient of a subleading term in the Fefferman-Graham expansion of the bulk metric. Since the branes are being treated in a probe approximation, the energy density obtained from the FG-expansion will not agree with that obtained from thermodynamics $-\frac{\del F}{\del \b}.$ Thus, the probe approximation is not entirely self-consistent and backreaction will contribute significantly \cite{Karch2}. 
Even if not, backreaction effects can be estimated and robustness of conclusions can be tested by stability analysis - because the ten dimensional description should be matched up to a complete string theory.

We can then attempt to fix the various parameters in the above studies in multiple ways. For instance, as in this work, we can fix the parameters at zero temperature and then compute other field theory quantities such as susceptibilities and viscosity at finite density and temperature. Naturally, another possibility is to fix only those observables (both at zero and nonzero temperatures) which are robust against changes in parameters and in the actual bulk action used.

We expect that the ideas presented in this paper and explorations suggested above will lead to insights into the strong coupling physics of gauge theories in general with a better hold on universal features. 

\section{Acknowledgment}
We also wish to acknowledge the use of the supercomputing facility Param at NABI. 
We wish to record our thanks to overleaf.com for their invaluable free service.

\end{document}